\documentclass[aps,twocolumn,superscriptaddress,preprintnumbers]{revtex4}
\usepackage{braket}
\usepackage{subfigure}
\usepackage[latin9]{inputenc}
\usepackage{color}
\usepackage{verbatim}
\usepackage{amsthm,amsmath,amssymb}
\usepackage{mathrsfs}
\usepackage{graphicx}
\usepackage{esint}
\usepackage{mathrsfs}
\usepackage{bm}

\usepackage{hyperref}

\makeatletter
\@ifundefined{textcolor}{}
{%
 \definecolor{BLACK}{gray}{0}
 \definecolor{WHITE}{gray}{1}
 \definecolor{RED}{rgb}{1,0,0}
 \definecolor{GREEN}{rgb}{0,1,0}
 \definecolor{BLUE}{rgb}{0,0,1}
 \definecolor{CYAN}{cmyk}{1,0,0,0}
 \definecolor{MAGENTA}{cmyk}{0,1,0,0}
 \definecolor{YELLOW}{cmyk}{0,0,1,0}
}

\usepackage{mathrsfs}

\draft
\makeatother

\begin{document}
\title{Emergent gapless topological Luttinger liquid}

\author{Sen Niu}
\affiliation{International Center for Quantum Materials and School of Physics, Peking University, Beijing 100871, China}
\affiliation{Collaborative Innovation Center of Quantum Matter, Beijing 100871, China}

\author{Yucheng Wang}
\affiliation{Shenzhen Institute for Quantum Science and Engineering,
Southern University of Science and Technology, Shenzhen 518055, China}
\affiliation{International Center for Quantum Materials and School of Physics, Peking University, Beijing 100871, China}
\affiliation{Collaborative Innovation Center of Quantum Matter, Beijing 100871, China}

\author{Xiong-Jun Liu}
\thanks{Corresponding author: xiongjunliu@pku.edu.cn}
\affiliation{International Center for Quantum Materials and School of Physics, Peking University, Beijing 100871, China}
\affiliation{Collaborative Innovation Center of Quantum Matter, Beijing 100871, China}
\affiliation{Shenzhen Institute for Quantum Science and Engineering,
Southern University of Science and Technology, Shenzhen 518055, China}
\affiliation{CAS Center for Excellence in Topological Quantum Computation, University of Chinese
Academy of Sciences, Beijing 100190, China}

\begin{abstract}
Gapless Luttinger liquid is conventionally viewed as topologically trivial, unless it hosts degenerate ground states and or entanglement spectrum, which necessitates partial bulk degree of freedom to be gapped out. Here we predict an emergent gapless topological Luttinger liquid which is beyond the conventional scenarios and is characterized by the nontrivial many-body bulk spin texture, and propose feasible scheme for experimental observation. We consider a one-dimensional spin-orbit coupled Fermi-Hubbard model with fractional filling, whose low-energy physics is effectively described by a spinless Luttinger liquid and is trivial in the conventional characterization. We show that, as being tuned by the filling factor and interaction strength, the many-body ground state may exhibit nontrivial winding in its bulk spin texture in the projected momentum space, manifesting an emergent topological phase. A topological transition occurs when the projected spin-state at a high symmetry momentum becomes fully mixed one, resulting from competing processes of particle scattering to the lower and higher subbands, for which the spin texture at such momentum point is ill-defined, but the Luttinger liquid keeps gapless through the transition. Surprisingly, at relatively small filling the Luttinger liquid remains topologically nontrivial even at infinitely strong interaction. The results can be generalized to finite temperature which facilitates the real experimental detection. This work shows a novel gapless topological Luttinger liquid whose characterization is beyond the low-energy effective theory, and can be verified based on current experiments.
\end{abstract}
\date{\today }
\maketitle

\section{introduction}

The notion of topological quantum phases revolutionized the characterization of fundamental phases of quantum matter, out of the Landau
theory of symmetry breaking paradigms. In particular, the discoveries of Quantum Hall effect \cite{klitzing1980new,PhysRevLett.49.405} and time reversal invariant topological insulators \cite{PhysRevLett.95.226801,PhysRevLett.95.146802,PhysRevLett.96.106802,bernevig2006quantum,konig2007quantum} have stimulated the extensive investigation and classification of free fermion topological phases~\cite{schnyder2008classification,kitaev2009periodic,ryu2010topological,chiu2016classification}. The gapped topological phases are characterized by nontrivial bulk topological invariants 
and gapless boundary excitations linking to the bulk topology. 
The free-fermion topological states can be extended to correlated topological phases through the adiabatic principle in gapped systems~\cite{schnyder2008classification,kitaev2009periodic,ryu2010topological}. Depending on the existence of long-range or short-range entanglement, the gapped many-body topological states can be classified into topologically ordered phases \cite{wen1990topological} or the symmetry protected topological (SPT) phases \cite{gu2009tensor,PhysRevB.82.155138,pollmann2010entanglement,chen2011classification,schuch2011classifying}. Unlike topological orders which are stable against arbitrary local perturbations, the SPT phases are stable against only symmetry-preserving perturbations.

Aside from the gapped phases, topological states have been also predicted in one dimensional (1D) correlated systems with bulk being partially gapless~\cite{fidkowski2011majorana,cheng2011majorana,bonderson2013quasi,kainaris2015emergent,keselman2015gapless,montorsi2017symmetry,PhysRevB.96.085133,thorngren2020intrinsically,scaffidi2017gapless,parker2018topological,verresen2019gapless}.
These 1D topological phases have gapless bulk excitations while retain part of characteristics of gapped topological phases such as edge modes and degenerate entanglement spectrum. Conventionally, the gapless topological phases can be realized in systems with spin-charge separation or constructed from decorated domain walls. For the systems with spin-charge separation \cite{fidkowski2011majorana,cheng2011majorana,kainaris2015emergent,keselman2015gapless,montorsi2017symmetry,PhysRevB.96.085133,thorngren2020intrinsically}, one degree of freedom is gapped and can be refermionized to a free fermion topological insulator in the continuum limit, while the other degree of freedom is a trivial gapless Luttinger liquid. The second type of gapless phases are constructed by extending the decorated domain wall mechanism for constructing gapped SPT phases~\cite{chen2014symmetry}. 
As is known that one can apply a local unitary twist to entangle two local trivial gapped degrees of freedom to construct gapped SPT phases~\cite{chen2014symmetry}. Similarly, one can use the similar unitary twist to entangle a trivial gapped degree of freedom with a trivial gapless degree of freedom to construct gapless SPT phases~\cite{scaffidi2017gapless,parker2018topological}. We note that the both types of gapless phases require one gapped degree of freedom \cite{note_gapless} and are all characterized by protected edge modes and degenerate entanglement spectrum.

Nevertheless, so far the understanding of the correlated gapless topological phases is still primitive. A natural question is that, whether the phases that are trivial according to the above conventional topological characterization are featureless and all the same? In this work we show that the answer is no by studying an interacting spin-orbit coupled (SOC) fermionic chain at fractional filling, and predict an emergent gapless topological Luttinger liquid characterized by nontrivial many-body spin texture. We uncover that the trivial Luttinger liquid in conventional characterization has two emergent phases classified by the winding numbers defined from many-body spin textures of the correlated ground state. This emergent topological phase predicted here has no gapped degree of freedom, and is beyond the scope of the conventional topological phases featured by edge modes and degenerate entanglement spectrum.

The model of our study without interaction is a 1D AIII class topological insulator \cite{guo2011topological,liu2013manipulating}, which was already realized in experiment~\cite{song2018observation}. In the presence of onsite Hubbard interaction, we observe the nontrivial features by analyzing the two different limits for the strongly repulsive interacting regime. 
First, for the parameter regime with half filling condition, i.e. $\nu=1$, the tensor product many-body ground state
is naturally captured as a trivial Mott insulator. Second, for the system with only a single fermion so that the filling factor $\nu\rightarrow0$, one naively see that the eigenstates of the fermion form the topological band of the noninteracting Hamiltonian, say, the system is nontrivial in the band topology. 

\begin{figure}[t]
\centering
\includegraphics[width=\columnwidth]{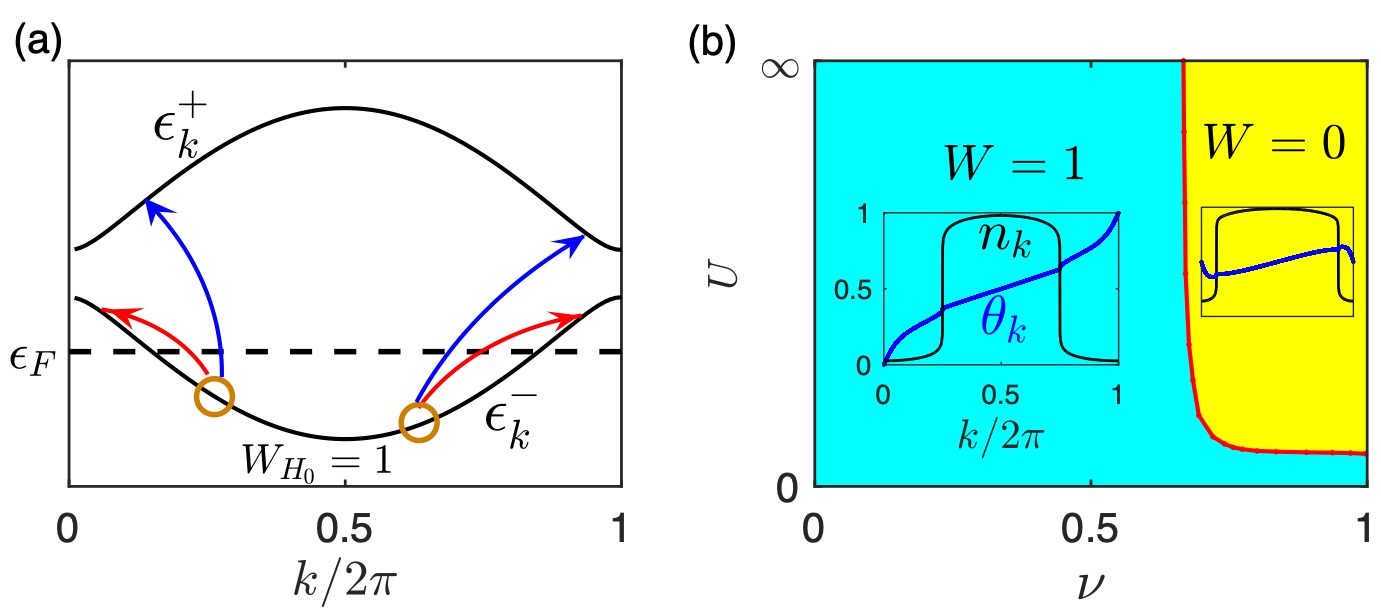}
\caption{Diagrammatic sketches for the many-body ground state and the topological phase diagram. (a), The Hubbard interaction induces scattering between unoccupied states and occupied states. $\epsilon_{k}^{\pm}$ (black curves) denotes dispersion of upper/lower topological bands of noninteracting $H_{0}$ with nontrivial band topology $W_{H_{0}}=\mp1$. $\epsilon_{F}$ (black dashed line) denotes Fermi energy of the noninteracting ground state. (b), The typical phase diagram has two different phases with nontrivial and trivial winding numbers $W$ respectively, tuned by filling factor $\nu$ and interaction $U$. In the insets blue (black) curves denotes typical spin direction $\theta_{k}$ (momentum distribution $n_{k}$) in momentum space.}
\label{fig:sketch}
\end{figure}

While the above trivial and nontrivial limits correspond to different definitions of topology, their physical features are expected to be fundamentally different. It is a nontrivial task to characterize the transition between such two limits, which cannot be captured in the conventional theory. 
First, in the view point of low energy physics, the ground state within $0<\nu<1$ is given from the gapless Luttinger liquid. Changing the Hamiltonian parameters only quantitatively modify the magnitude of Luttinger parameters. Thus the transition cannot be determined from low energy physics. Second, in the conventional characterization of topological phases, the ground state with $\nu<1$ is trivial due to absence of edge modes and degenerate entanglement spectrum (see numerical confirmation in Sec. V). Therefore a new topological characterization is necessary to distinguish physical features at different filling factors. In particular, we uncover that the gapless Luttinger liquid undergoes an emergent topological transition in varying the Hubbard interaction and filling factors, with the topological invariants in different phases defined by the many-body bulk spin texture. The main features are summarized below:
\begin{enumerate}
\item[(i)] The ground state of the Luttinger liquid exhibits nontrivial many-body spin texture in the projected momentum space, which defines a winding number characterizing the emergent global topology of the phase. The winding number cannot be obtained from the low energy physics near Fermi points and, instead, is determined by the competing processes of particle scatterings to the full lower- and higher-energy bands as illustrated by Fig.~\ref{fig:sketch} (a).

\item[(ii)] Two phases with nontrivial and trivial spin textures emerge in the Luttinger liquid
for $0<\nu<1$ [Fig.~\ref{fig:sketch} (b)]. The winding number undergoes a discrete change across topological transition through tuning filling factor $\nu$ or interaction $U$. Surprisingly, at small filling factor the phase remains topologically nontrivial even for infinitely large interaction, a novel feature of the current phase beyond mean-field and perturbation regime    .

\item[(iii)] Due to gapless nature of the quantum state away from half filling, adiabatic connection is not applicable and the winding number
is defined according to the observable many-body spin texture. The stability of this
topological invariant is guaranteed by the fact that the observable spin texture changes continuously with Hamiltonian
parameters without crossing the transition point.
\item[(iv)] The definition of many-body spin texture can be directly generalized to finite temperature. In this case the spin texture and winding number describe properties of Hamiltonian and low energy states. The features of
spin textures and phase diagrams are similar to those of the ground state.
This facilitates the observation in real cold atom experiments
\cite{Wu83,PhysRevLett.121.150401} via the spin-resolved time-of-flight imaging.
\end{enumerate}

The manuscript is structured as follows. In Sec. II, after introducing the model,
we define the bulk spin texture and the winding number which characterize the emergent topological phase in this work. The quantization of the winding number and the underlying physics of the winding number are also discussed. Through both analytical and numerical methods, we investigate the topological patterns of many-body spin textures and phase diagrams at zero temperature in Sec. III and finite temperature in Sec. IV. In Sec V we provide a comparison study on the emergent topological phase featured by topological spin textures and the conventional topological phases, with the ground state in the gapless ($\nu<1$) and gapped ($\nu=1$) regimes being investigated, respectively. Finally, the conclusion and outlook are presented in Sec. VI.

\section{Generic theory: model, definition and formalism}

\subsection{1D topological model with interaction}

We consider a 1D interacting SOC Hamiltonian $H=H_{0}+H_{I}$ that consists of a two
band AIII class topological model $H_{0}$ \cite{liu2013manipulating,song2018observation}
and repulsive on-site Hubbard interaction $H_{I}$. Using the spinor operator $\psi_{i}^{\dagger}=[c_{i,\uparrow}^\dag,c_{i,\downarrow}^\dag]$ the Hamiltonian has the form
\begin{align}
H_{0}= & -t_{0}\sum_{i}( \psi_{i}^{\dagger}\tau_z \psi_{i+1}+h.c.)+\delta\sum_{i}(n_{i,\uparrow}-n_{i,\downarrow}) \notag \\
 & +it_{\rm so}\sum_{i}( \psi_{i}^{\dagger}{\tau}_y \psi_{i+1}+h.c.), \notag \\
H_{I}= & U\sum_{i}n_{i,\uparrow}n_{i,\downarrow}.\label{eq:model}
\end{align}
Here $c_{i,\uparrow}^\dag,c_{i,\downarrow}^\dag$ denotes the fermionic creation operators for spin-up and spin-down, $\tau_{x/y/z}$ denotes Pauli matrices, $(t_{0},t_{\rm so},\delta,U)$ denote the spin conserving hopping, spin flip hopping, Zeeman energy, and on-site repulsive energy, respectively. This Hamiltonian
has been proposed in optical lattice \cite{liu2013manipulating},
and the noninteracting part has been realized experimentally \cite{song2018observation}.

In momentum space, through Fourier transformation the noninteracting Hamiltonian $H_{0}$ is written as
\begin{equation}
H_{0}=-\sum_{k}(h_{k}^{y}\psi_{k}^{\dagger}\tau_{y}\psi_{k}+h_{k}^{z}\psi_{k}^{\dagger}\tau_{z}\psi_{k}),
\end{equation}
here $\psi_{k}^{\dagger}=[c_{k,\uparrow}^\dag,c_{k,\downarrow}^\dag]$ is the Fourier transformation of $\psi_i^\dag$ to momentum space, the Bloch Hamiltonian coefficients $h_{k}^{y/z}$ are $h_{k}^{y}=2t_{\rm so}\sin k, h_{k}^{z}=2t_{0}\cos k-\delta$,
and the length of vector $(h_{k}^{y},h_{k}^{z})$ is denoted as $ h_{k}=\sqrt{(h_{k}^{x})^2+(h_{k}^{y})^2}$. Without loss of generality, in this work we focus on the parameter region $t_{\rm so}>0,0<\delta<2t_{0}$, and take $t_{0}=1$.
The bands of $H_{0}$ are gapped and the lower subband has nontrivial band topology characterized by the nonzero winding number $W_{H_{0}}=1$, defined as \cite{chiu2016classification}
\begin{align}
\theta_{k}^{H_{0}}= & \frac{1}{2\pi}\arctan\frac{h_{k}^{y}}{h_{k}^{z}},\notag \\
W_{H_{0}}= & \int_{FBZ}d\theta_{k}^{H_{0}}. \label{theta_W}
\end{align}
Here $W_{H_{0}}$ characterizes the continuous map from the 1D FBZ to 1D sphere. The $\theta_{k}^{H_{0}}$ is the spin direction of the lower subband in FBZ restricted to the range $[0,1]$. Moreover, the condition that there are two fermi
points in lower band for any $\nu<1$, i.e., the dispersion of each
band is monotone in half FBZ $k\in[0,\pi]$ (such as Fig.~\ref{fig:sketch}(a))
can be satisfied by choosing
\begin{equation}
\frac{\delta}{t_{0}}>  2|\frac{t_{\rm so}^{2}}{t_{0}^{2}}-1|.\label{eq:two_fermi_point_condition}
\end{equation}
In this parameter region, we shall see that the only possible fixed point at $\nu<1$
is the trivial Luttinger liquid in the bosonization language and from the conventional picture. 
In the case with four Fermi points in the lower band, even more
sophisticated bosonization analysis will be needed to reveal Luttinger
liquid properties. Nevertheless, we shall find that the emergent topological phases are similar in both cases.

\subsection{Definition and Formalism: many-body spin texture and winding number}

Now we introduce the many-body spin texture and the corresponding winding number as a topological invariant. In this work, the phases with nontrivial winding number are dubbed \emph{emergent topological phases}. We first define spin operator in momentum space as
\begin{equation}
s_{k}^{x/y/z}=\psi_{k}^{\dagger}\tau_{x/y/z}\psi_{k}.
\end{equation}
The observable momentum space many-body spin texture is
defined via many-body ground state $|\psi_{g}\rangle$
\begin{equation}
\langle s_{k}^{x/y/z}\rangle=\text{Tr}(|\psi_g\rangle\langle \psi_{g}|s_{k}^{x/y/z})/\text{Tr}|\psi_g\rangle\langle \psi_{g}|,
\label{eq:def_texture_groundstate}
\end{equation}
where the trace is performed in momentum space. Note that this quantity can be measured in cold atom experiments
\cite{Wu83,PhysRevLett.121.150401} via the spin-resolved time-of-flight imaging. The observable spin texture can be generalized to finite temperature via many-body density matrix
\begin{equation}
\langle s_{k}^{x/y/z}\rangle=\frac{\text{Tr}[\exp(-\beta K)s_{k}^{x/y/z}]}{\text{Tr}[\exp(-\beta K)]},\label{eq:def_texture}
\end{equation}
here $K=H-\mu\sum_{i,\sigma}n_{i,\sigma}$ is the grand canonical
Hamiltonian, in which $\sigma=\uparrow,\downarrow$ denotes spin direction, $\mu$ is the chemical potential, $\beta$ is the inverse temperature. When $\beta\rightarrow\infty$ at zero temperature, the Eq. (\ref{eq:def_texture}) is reduced to Eq. (\ref{eq:def_texture_groundstate}). The full Hilbert space can be represented as tensor product of four-dimensional subspaces at each momentum. Then the expectation value can also be expressed with the single-momentum reduced density matrix $\rho_{k}$
\begin{align}
\langle s_{k}^{x/y/z}\rangle&=\frac{\text{Tr}[\rho_{k}s_{k}^{x/y/z}]}{\text{Tr}\rho_{k}},\notag \\
\rho_{k}&=\text{Tr}_{\text{FBZ} \setminus k} \exp(-\beta K).
\end{align}
Here the partial trace in the second line of above equations is performed in the full FBZ except for the $k$ momentum point. $\rho_{k}$ is the four dimensional reduced density matrix at the $k$ momentum point, with the bases being $|0\rangle,c_{k,\uparrow}^{\dag}|0\rangle,c_{k,\downarrow}^{\dag}|0\rangle,c_{k,\uparrow}^{\dag}c_{k,\downarrow}^{\dag}|0\rangle$. Note that only the two singly occupied $c_{k,\uparrow}^{\dag}|0\rangle,c_{k,\downarrow}^{\dag}|0\rangle$ configurations have nonzero contribution to spin texture expectation value at $k$ momentum point. When interaction is nonzero, $\rho_{k}$ characterizes a mixed state in general even for ground state.

\begin{figure*}[t]
\centering
\includegraphics[width=6.2in]{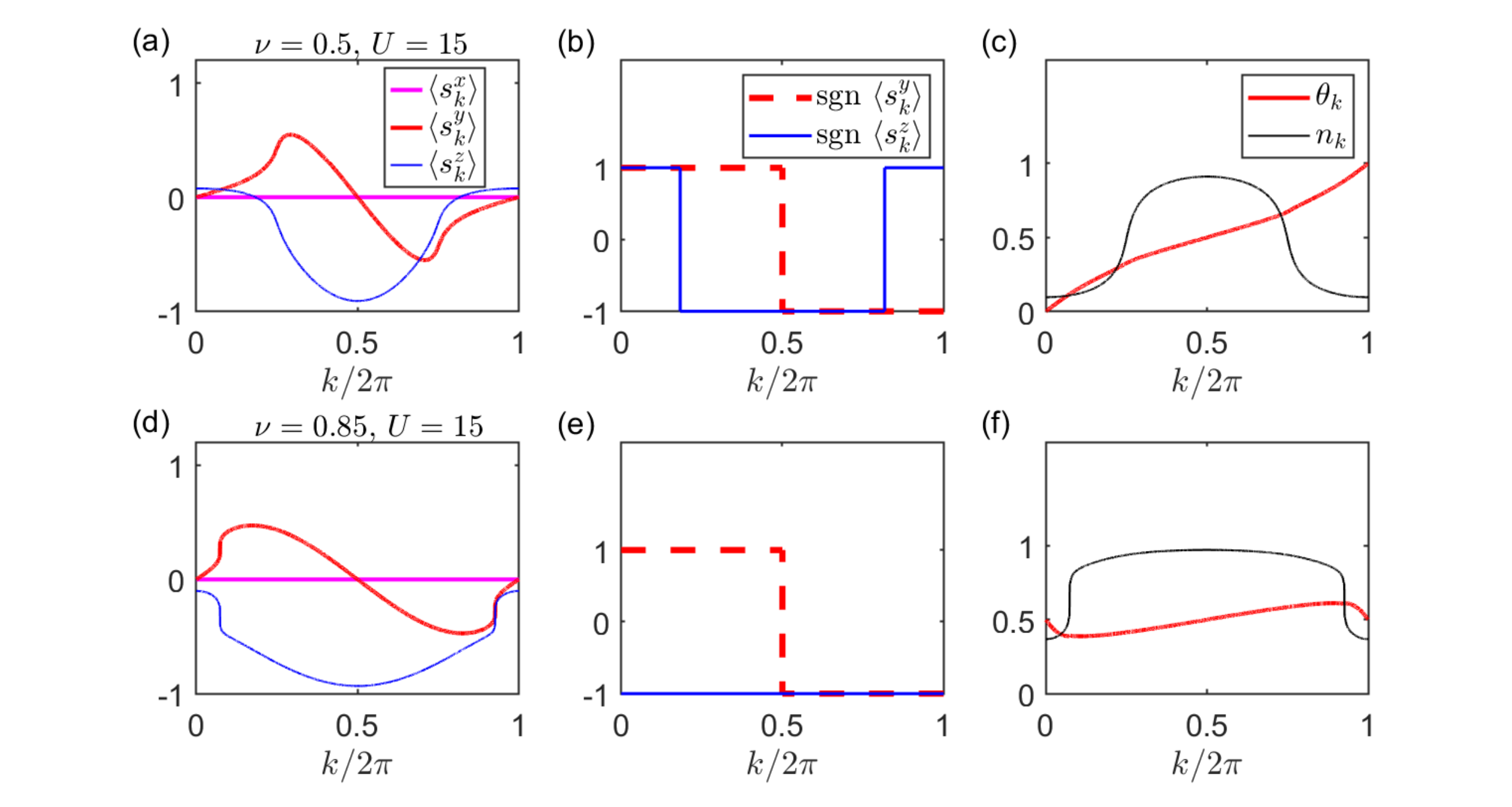}
\caption{The patterns ground state spin textures with different topologies. Parameters are chosen as $(t_{0},t_{so},\delta)=(1,1,0.5)$ and data are obtained from VUMPS simulation. (a)-(c) correspond to $(\nu,\,U)=(0.5,15)$ with nontrivial winding number $W=1$ and (d)-(f) correspond to $(\nu,\,U)=(0.85,15)$ with trivial winding number $W=0$. Due to Hamiltonian symmetries, the spin direction is in $y-z$ plane, with $\langle s_k^{y}\rangle$ and $\langle s_k^{z}\rangle$ being odd and even functions, respectively.}
\label{fig:textures_ob}
\end{figure*}

As an observable, the many-body spin texture has the same symmetry as the Hamiltonian, i.e. it has the spin reflection symmetry $R_{x}$ and parity symmetry $P$ in the presence of Hubbard interaction (for details see Appendix A). One can then show that
the spin textures satisfy
\begin{align}
\langle s_{k}^{x}\rangle =0, \ \langle s_{k}^{y}\rangle =-\langle s_{-k}^{y}\rangle, \ \langle s_{k}^{z}\rangle =\langle s_{-k}^{z}\rangle.
\end{align}
Thus $\langle s_{k}^{y}\rangle$ and $\langle s_{k}^{z}\rangle$ are odd and even functions versus momentum, respectively. Thus at high symmetry $k=0,\pi$ points one has $\langle s_{k=0,\pi}^{y}\rangle=0$.
The observable many-body spin texture can then be represented by the spin length $S_{k} =\langle s_{k}^{y}\rangle^{2}+\langle s_{k}^{z}\rangle^{2}$ at each momentum and the spin direction $\theta_{k}$ defined by
\begin{align}
\theta_{k}=\frac{1}{2\pi}\arctan \frac{\text{Tr}[\rho_{k}s_{k}^{y}]}{\text{Tr}[\rho_{k}s_{k}^{z}]}.
\label{eq:spinlength}
\end{align}
In general the spin length satisfies $S_{k}<1$ for $\nu<1$, implying that due to the interaction the density matrix $\rho_k$ becomes mixed. Away from the transition point, the spin length should be finite at each momentum point.
The winding number of the many-body spin texture is defined as
\begin{equation}
W =\int_{FBZ}d\theta_{k} \label{eq:many_body_winding},
\end{equation}
which characterizes the global topology of the interacting phase. Without interaction the winding number $W$ is equivalent to that of the band topology at half-filling, but not well-defined for $\nu<1$. The correlated phase with Hubbard interaction is far beyond the single-particle counterpart. The topological transition, across which $W$ varies, implies the existence of singularity at certain $k$ point, on which the density matrix $\rho_k$ will be shown to be fully mixed and the spin length $S_k=0$. Two examples of spin textures are shown in Fig.~\ref{fig:textures_ob}, where Fig.~\ref{fig:textures_ob} (a)-(c) correspond to a nontrivial case with winding number $W=1$ and Fig.~\ref{fig:textures_ob} (d)-(f) correspond to a trivial case with zero winding number.

The Luttinger liquid properties will be investigated from the momentum distribution of particle-number density $n_{k}=n_{k,\uparrow}+n_{k\downarrow}=\langle c_{k,\uparrow}^{\dagger}c_{k,\uparrow}\rangle+\langle c_{k,\downarrow}^{\dagger}c_{k,\downarrow}\rangle$.
The total magnetization is given by
\begin{equation}
m=n_{\uparrow}-n_{\downarrow}=\sum_k (n_{k,\uparrow}-n_{k,\downarrow}).
\end{equation}
The physical quantities at $\nu>1$ can be related to those at $\nu<1$ through particle-hole
transformation, thus in this work we stick to the $\nu\le1$ case. The thermodynamic limit of the system is achieved by increasing particle number $N$ and lattice size $L$ with the filling factor $\nu=N/L$ being fixed. In particular, the limit case with $\nu\rightarrow0$ should be regarded as the regime with $L\rightarrow\infty$ while the particle number $N$ is finite.

\subsection{Quantization of the ground state winding number}

Here we present the generic quantization features of the winding
number under the proper conditions satisfied for Luttinger liquid. The first condition is that except for the transition point, the spin length $S_{k}$ is nonzero in FBZ such that spin direction $\theta_{k}$ and the winding number are well-defined. For the Luttinger liquid the momentum
distribution $n_k$ is nonzero at each $k$ in the full FBZ. In momentum space spin rotation
symmetry is broken by $H_{0}$ at each $k$, hence the spin length $S_{k}$ should be
nonzero in FBZ. As to be shown by analytic results, including the weak/large $U$ limit in Sec. III and two exact cases in Appendix D, and numerical results in Sec. III, the $y$ direction spin polarization $\langle s_{k}^{y}\rangle$ of this model in FBZ always has the same
sign as the Bloch Hamiltonian coefficient $h_{k}^{y}=2t_{\rm so}\sin k$. Thus the topological transition must be featured by that the density matrix $\rho_k$ become fully mixed at either $k=0$ or $\pi$, where the spin length is either $\langle s_{k=0}^{z}\rangle=0$
or $\langle s_{k=\pi}^{z}\rangle=0$ by tuning the parameters. Thus the first condition is generally satisfied.

The second condition is that $\theta_{k}$
is continuous in FBZ so that the winding number is stable. 
For $\nu=1$ case with the gapped ground state, from Fourier transformation
and the exponential decay of correlation functions $\langle c_{i,\sigma}^{\dagger}c_{j,\sigma'}\rangle$
\cite{hastings2006spectral} one can show that the observables $\langle s_{k}^{y/z}\rangle$
and $\theta_{k}$ are continuous functions of $k$. For $\nu<1$,
with the ground state being a gapless Luttinger liquid phase, the momentum
distribution at Fermi points has power law singularity instead of fermi
liquid discontinuity \cite{brech1990momentum,karrasch2012luttinger}.
Thus the $\theta_{k}$ is also continuous. We will numerically verify
this power law behaviour by analysing the long distance behaviour of
$\langle c_{i,\sigma}^{\dagger}c_{j\sigma'}\rangle$ in Sec. V. In this way
the second condition is also satisfied.

\subsection{Physics of the winding number}

We introduce the single-particle bases which diagonalize the single-particle Hamiltonian as
\begin{equation}
H_{0}=\sum_{k}(\epsilon_{k}^{+}d_{k,+}^{\dagger}d_{k,+}+\epsilon_{k}^{-}d_{k,-}^{\dagger}d_{k,-}),\label{eq: H0_diag}
\end{equation}
where the lower ($+$) and upper ($-$) band dispersions are
\begin{equation}
\epsilon_{k}^{\pm}=\pm h_{k},
\end{equation}
and single-particle bases
\begin{align}
d_{k,-}^{\dagger} & =\alpha_{k}c_{k,\uparrow}^{\dagger}+\beta_{k}c_{k,\downarrow}^{\dagger},\notag \\
d_{k,+}^{\dagger} & =-\beta_{k}^{*}c_{k,\uparrow}^{\dagger}+\alpha_{k}^{*}c_{k,\downarrow}^{\dagger}.\label{eq: H0_basis}
\end{align}
Here $\alpha_{k}$ and $\beta_{k}$ are not gauge-invariant and can
be chosen as imaginary odd function and real even function in FBZ,
respectively. In this work for $k\neq0,\pi$ (hence $h_{k}\neq h_{k}^{z}$)
we choose
\begin{align}
\alpha_{k}= & ih_{k}^{y}/\sqrt{2h_{k}(h_{k}-h_{k}^{z})},\notag \\
\beta_{k}= & (h_{k}^{z}-h_{k})/\sqrt{2h_{k}(h_{k}-h_{k}^{z})}.
\end{align}
At $k=0$ point we have $(\alpha_{k},\beta_{k})=(-i,0)$ for $0<\delta<2t_{0}$
and $(\alpha_{k},\beta_{k})=(0,-1)$ for $\delta>2t_{0}$, while at
$k=\pi$ point $(\alpha_{k},\beta_{k})=(0,-1)$ for all $\delta>0$.

Due to interaction effect the particle number distributions in both upper and lower subbands are nonzero. To describe the particle distribution between upper and lower subbands, we introduce the rotating-frame spin operator related to the single-particle bases as
\begin{align}
\mathcal{S}_{k}^{y} & =id_{k,-}^{\dagger}d_{k,+}-id_{k,+}^{\dagger}d_{k,-},\notag \\
\mathcal{S}_{k}^{z} & =d_{k,+}^{\dagger}d_{k,+}-d_{k,-}^{\dagger}d_{k,-},\label{eq:basis_change}
\end{align}
which are also useful for perturbation analysis in next section. The spin texture within the true spin up and down bases can be obtained from inverse transformation
\begin{align}
\langle s_{k}^{y}\rangle= & i(\alpha_{k}^{*}\beta_{k}-\beta_{k}^{*}\alpha_{k})\langle\mathcal{S}_{k}^{z}\rangle-(\alpha_{k}^{2}+\beta_{k}^{2})\langle\mathcal{S}_{k}^{y}\rangle, \notag\\
\langle s_{k}^{z}\rangle= & (|\beta_{k}|^{2}-|\alpha_{k}|^{2})\langle\mathcal{S}_{k}^{z}\rangle-2i\alpha_{k}\beta_{k}\langle\mathcal{S}_{k}^{y}\rangle,
\label{eq:inverse_transform}
\end{align}
and $\theta_{k}$ can be expressed using $\langle \mathcal{S}_{k}^{y/z} \rangle$. Thus spin direction $\theta_{k}$ and winding $W$ can be written as summations of contributions from noninteracting $H_0$ and the rotating frame spin direction:
\begin{align}
\theta_{k} & =\theta_{k}^{H_{0}}+\theta_{k}^{rel}, \notag \\
W & =W_{H_{0}}+W_{rel},
\end{align}
where $\theta_{k}^{rel}$ is the rotating frame spin direction in the bases
of upper and lower subbands obtained from
\begin{equation}
\theta_{k}^{rel} =\frac{1}{2\pi}\arctan \frac{\langle \mathcal{S}^{y}\rangle}{\langle \mathcal{S}^{z}\rangle},
\end{equation}
and $W_{rel}$ is the winding number of $\theta_{k}^{rel}$ in FBZ. One can immediately
see that $W=W_{H_{0}}$ if $\mbox{sgn}(\mathcal{S}_{k}^{z})$ is unchanged in full FBZ. Accordingly, if $W\neq W_{H_{0}}$, $\mathcal{S}_{k}^{z}$ must change sign
in the FBZ, indicating that there exists momentum points where the reduced density matrix $\rho_k$ is dominated by the scattering to the states of upper and lower bands due to the Hubbard interaction. This implies that in general the topology emerging in the present gapless Luttinger liquid cannot be characterized by the conventional bosonization treatment, in which only the low-energy physics near fermi points are considered.

\begin{figure*}[t]
\centering
\includegraphics[width=6.2in]{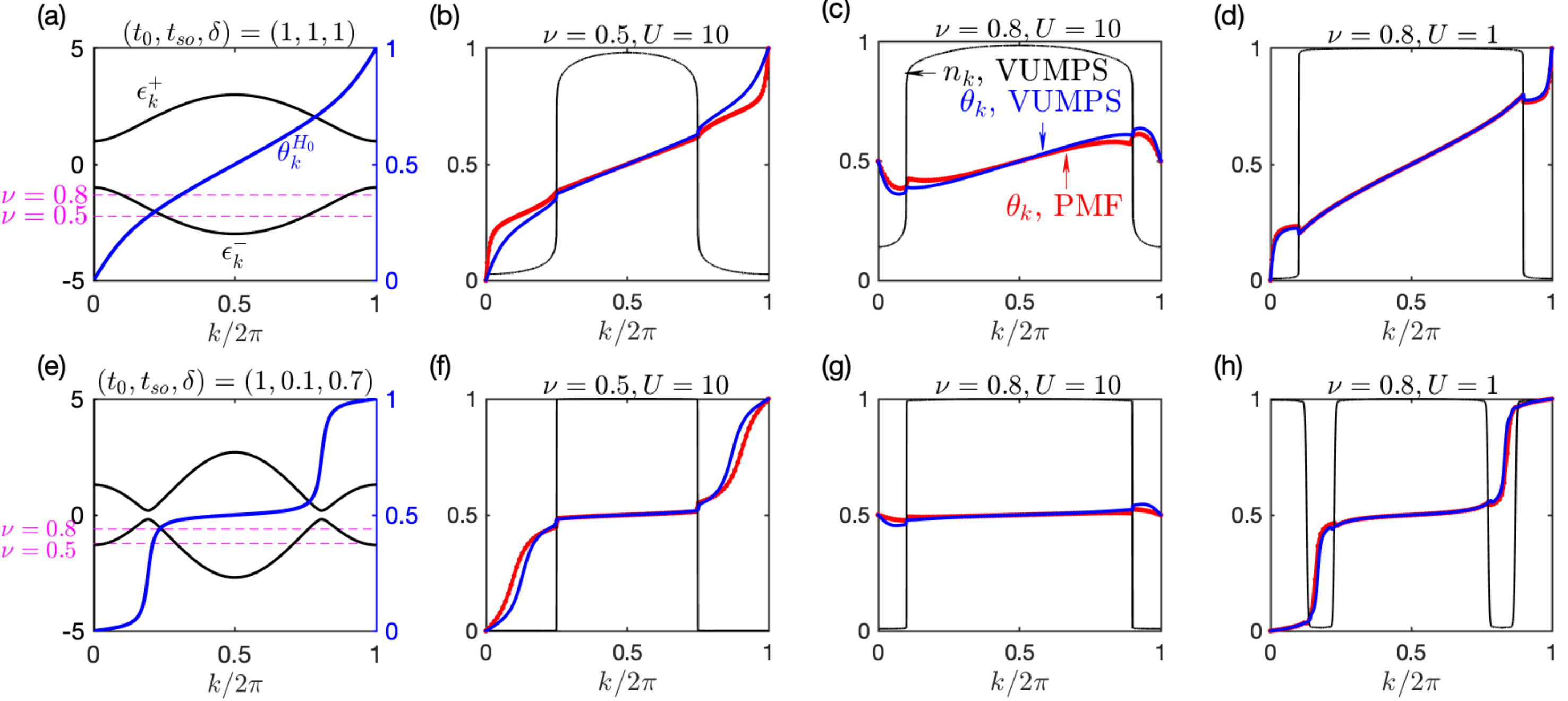}
\caption{Typical ground state spin textures with different topololy tuned by filling factor $\nu$ and interaction strength $U$. Lattice parameters are chosen as $(t_{0},t_{\rm so},\delta)=(1,1,1)$ in (a)-(d) and $(t_{0},t_{\rm so},\delta)=(1,0.1,0.7)$ in (e)-(h). (a) and (e), The dispersions (black curves), spin directions $\theta_{k}^{H_{0}}$ (blue curves), and Fermi levels (purple dashed lines) of the noninteracting Hamiltonian, respectively. (b)-(d) and (f)-(h) show spin directions $\theta_k$ obtained from analytical PMF and numerical VUMPS methods, which agree well even at moderate $U$. Positions of Fermi momenta can be found from the momentum distributions $n_k$, and the effect of interaction induced Fermi sea deformation in case (e)-(h) can be captured by PMF method. In (b), (d), (f), and (h) the spin textures are nontrivial with $W=1$; in (c) and (g) the spin textures are trivial with $W=0$.}
\label{fig:typycal_textures}
\end{figure*}
\section{Ground state properties of spin textures}

In this section we first investigate typical patterns of ground state spin
textures at generic filling factor $\nu$ via both analytical and numerical
methods. We further map out the winding number
phase diagrams which show two gapless phases with nontrivial and trivial spin textures emerge in the Luttinger liquid for $0<\nu<1$. The winding number undergoes a discrete change across the topological transition, and may be driven from nonzero to zero through increasing filling factor or interaction strength. The existence of emergent topology and its transition imply that low energy physics is insufficient to describe the gapless Luttinger liquid, instead the high-energy physics and global features in momentum space should be taken into account for a full characterization.

\subsection{Perturbation calculation of spin textures}

The bosonization method which only takes into account low energy density
fluctuation around Fermi points is not applicable to studying the spin textures in global FBZ. More importantly, at generic filling
$\nu<1$, Fermi points are at lower subband, and excitations to the upper band that affect spin textures are not bosonic due to the absence of Fermi points there. If the upper band degree of freedom is discarded, the particle scattering within lower subband only modifies the momentum distribution, and the global topology of the many-body spin texture will not be affected. Thus we use perturbation
treatment, from which effects of both lower and upper band in full
FBZ need to be taken into account. Although at the Fermi points
there exists divergence problem \cite{giamarchi2003quantum} in finite
order perturbation treatment, in the case relevant to the present study the spin directions can only be flipped
(or changed qualitatively compared with noninteracting states) at $k$ momentum points where the dispersion energies
$\epsilon_{k}^{\pm}$ in upper and lower subbands are both above Fermi level $\epsilon_{F}$,
as seen from numerical results in Fig.~\ref{fig:typycal_textures}.
Thus the divergence problem in perturbation does not affect
the present study in general.

\subsubsection{Ground state from expansion of $U$ }

We develop the perturbed mean-field (PMF) wavefunction method to study ground states at relatively weak $U$, from which qualitatively correct spin textures and winding numbers are obtained. We first apply mean-field (MF) approximation to characterize the effects of band deformation caused by Hubbard interaction, and then use perturbation expansion of $U$ as did in appendix of Ref. \cite{PhysRevB.41.2326} to tackle with the competing processes of particle scattering to the lower and higher subbands.

Note that it is inappropriate to choose the noninteracting Hamiltonian as the unperturbed Hamiltonian since the total energy of unperturbed
eigenstates will cross during increase of $U$. So we choose mean-field
ground state, i.e., the Slater determinant that minimizes the total energy as
the starting point of perturbation expansion. As the positive $\delta$
will be enhanced by Hubbard interaction \cite{liu2013manipulating}, ferromagnetic order parameter is assumed and the MF Hamiltonian is written as
\begin{align}
H_{mf}(M) & =H_{0}-\sum_{i}\frac{UM}{2}(n_{i,\uparrow}-n_{i,\downarrow})+\sum_{i}U\frac{\nu^{2}+M^{2}}{4} \notag \\
 & =\sum_{k}(\epsilon_{k}^{+}d_{k,+}^{\dagger}d_{k,+}+\epsilon_{k}^{-}d_{k,-}^{\dagger}d_{k,-}),
\end{align}
where $M$ is the variational parameter. Note that in this PMF treatment all the notations such as $\epsilon_{k}^{\pm}$, $d_{k,\pm}^{\dagger}$, $\alpha_{k},\beta_{k}$ are defined according to $H_{mf}$ instead of $H_{0}$. The solution of order parameter $M$ can be determined by variational condition
$\frac{\partial E(M)}{\partial M}|_{M=m}=0$, where $E(M)$ is ground state energy of MF Hamiltonian $H_{mf}(M)$
at filling $\nu$, and $m$ is the solution of order parameter $M$ that
satisfies self-consistent condition $m=n_{\uparrow}-n_{\downarrow}$.
If there is more than one local minimum of $E(M)$, one should
choose global minimum of $E(M)$. Since $m<0$, the Zeeman field $\delta$
in $H_{0}$ is corrected to be
\begin{equation}
\tilde{\delta}=\delta-\frac{Um}{2}>\delta.
\end{equation}
From above expression and Eq. (\ref{eq:two_fermi_point_condition})
one can see that the band of MF Hamiltonian will be deformed
by $U$, and there will eventually exist two fermi points at large
$U$ (even if there are four Fermi points at $U=0$), as indicated
by momentum distribution in Fig.~\ref{fig:typycal_textures}.

Now we tackle with the correlation effects of momentum scattering caused by off-diagonal parts of Hubbard interaction. Conceptually, it is incorrect to take the winding number of MF Hamiltonian as the true winding number of the many-body ground state. The reason is that MF ground state wavefunction
\begin{equation}
\psi_{mf}=\prod_{\epsilon_{k}^{-}<\epsilon_{F}}d_{k,-}^{\dagger}|vac\rangle
\end{equation}
with $\epsilon_{F}$ being Fermi energy is a Slater determinant whose
momentum distribution is a step function at $\nu<1$. For $\psi_{mf}$
the spin direction is ill-defined for momentum points outside fermi
sea, and the true ground state with nonzero interaction is a scattering
state with nonzero momentum distribution in full FBZ. To include momentum
scattering effect of the Hubbard interaction, we transform it to
momentum space as
\begin{align}
H_{I}= & \frac{U}{L}\sum_{k,k',q,q'}c_{q\uparrow}^{\dagger}c_{q'\downarrow}^{\dagger}c_{k'\downarrow}c_{k\uparrow}\notag \\
= & \frac{U}{L}\sum_{k,k',q,q'}(\alpha_{q}^{*}d_{q,-}^{\dagger}-\beta_{q}d_{q,+}^{\dagger})(\beta_{q'}^{*}d_{q',-}^{\dagger}+\alpha_{q'}d_{q',+}^{\dagger})\notag \\
 & \times(\beta_{k'}d_{k',-}+\alpha_{k'}^{*}d_{k',+}) (\alpha_{k}d_{k,-}-\beta_{k}^{*}d_{k,+}).\label{eq:HI_k_space}
\end{align}
Here coefficients $\alpha_{k},\beta_{k}$ are determined following Eq. (\ref{eq: H0_basis}) by diagonalizing $H_{mf}$
instead of $H_{0}$, and in the summation of momentum the constraint $\mod (k+k'-q-q',2\pi)=0$ should be satisfied. From this formula one can see interaction induces particle scattering
between lower and upper bands. In the standard Bosonization treatment, the terms containing $d_{k,+}^\dag,d_{k,+}$ will be omitted since Fermi points lie at the lower subband, while in this model they are responsible for transition of winding number and should be kept.

We choose MF Hamiltonian $H_{mf}$ as
unperturbed part and $H-H_{mf}$ that creates two pairs of particle-hole excitations
in MF ground state as the perturbation, and calculate the
spin texture of the PMF wavefunction at (the lowest) second
order of $U$. The spin textures $\langle s_{k}^{y/z}\rangle$ in true spin up/down
bases are obtained from combining Eq. (\ref{eq:basis_change}) with Eq. (\ref{eq:inverse_transform}), in which the single-particle bases are determined from MF Hamiltonian. The perturbation expression of $\langle\mathcal{S}_{k_{0}}^{z}\rangle$
for $k_{0}$ outside Fermi sea has the form
\begin{align}
\langle\mathcal{S}_{k_{0}}^{z}\rangle= &  (\frac{U}{L})^{2}\sum_{q>q'}\frac{|(\alpha_{q}\beta_{q'}-\alpha_{q'}\beta_{q})(\beta_{k'}\alpha_{k_{0}}-\alpha_{k'}\beta_{k_{0}})|^{2}}{(\epsilon_{q}^{-}+\epsilon_{q'}^{-}-\epsilon_{k_{0}}^{+}-\epsilon_{k'}^{+})^{2}} \notag\\
\times & (1-f_{k',+})(1-f_{k_{0},+})f_{q,-}f_{q',-} \notag \\
+ &\cdot\cdot\cdot .\label{eq:sz_k0_formula}
\end{align}
Here $f_{k,\pm}$ denotes the Fermi distribution at $k$ point in
the upper/lower subband of MF ground state. As can be seen from
Fermi distribution functions in the above formula, the four terms
all have two hole excitations in the lower subband below Fermi level $\epsilon_{F}$,
but particle excitations can be either in upper or lower subbands. The details of perturbation calculation and complete expressions of spin textures are shown in Appendix B.

When parameter $\nu$ or $U$ are tuned, the corrected Zeeman
field $\tilde{\delta}$ and the bases of unperturbed eigenstates will
change, as a result the spin direction obtained from perturbation calculation
for momentum points both outside and inside Fermi sea will also change.
In short, in this weak $U$ perturbation theory, correlation effects on the many-body spin texture come from corrected Zeeman field $\tilde{\delta}$ and competing processes of particle scattering to upper/lower subbands. The distinction between winding number of many-body spin texture and winding number of the MF lower subband implies that for some momentum points the single-momentum reduced density matrices are dominated by the scatterings to the states of MF upper subband. From above perturbation formulas one can see that when $\langle s_{k=0}^z \rangle=0$ is satisfied via tuning parameters, the zero spin length at $k=0$ point does not imply that there are no particles
at the $k$ momentum point, instead the single-momentum reduced density matrix
is fully mixed and proportional to identity in the singly occupied sector.

However, the weak $U$ perturbation fails at large $U$ regime. 
In particular, at sufficiently large $U$ and any finite filling factor, the MF Zeeman field
$\tilde{\delta}$ will always exceed critical value. Thus the many-body spin polarization
$\langle s_{k}^{z}\rangle$ obtained from PMF becomes all negative in the full FBZ, leading to an incorrect trivial phase (see details in Appendix B). The actual emergent topological phase uncovered in this work is however always nontrivial for small filling, beyond the MF and perturbation regime. 

\subsubsection{Ground state at infinite $U$ from $t_{\rm so}$ expansion: nontrivial spin texture at small $\nu$ }

In order to correctly evaluate the winding number at large $U$ limit analytically, we consider the Hamiltonian at infinite $U$ limit directly and treat $t_{\rm so}$ as the expansion parameter. Through the analyse of infinite $U$ limit, one will see that at relatively small filling factor the winding number of many-body spin texture remains nontrivial even at infinite $U$. This also shows that the lowest order perturbation of $U$ in PMF method can not correctly capture the competing processes of particle scattering to lower and upper bands when $U$ is very large.

The physical properties of the infinite $U$
model at the low energy (temperature scale $T\ll U$) can be captured
by the Gutzwiller projected Hamiltonian
\begin{equation}
\tilde{H} =P_{G}HP_{G}=\tilde{H}_{t_0}+\tilde{H}_{t_{\rm so}}+\tilde{H}_{\delta},
\end{equation}
where the Gutzwiller projector $P_{G}$ is defined as $P_{G}=\prod_{i}(1-n_{i,\uparrow}n_{i,\downarrow})$
and $\tilde{H}_{t_0},\tilde{H}_{t_{\rm so}},\tilde{H}_{\delta}$ denotes single-particle terms in original Hamiltonian Eq.~(\ref{eq:model}) projected by $P_{G}$. These terms are no more quadratic and $\tilde{H}_{t_0},\tilde{H}_{t_{\rm so}}$ terms induce particle scattering in momentum space.

One can see that at $t_{\rm so}=0,\delta>0$ the many-body ground state is exactly known and simple. The Hamiltonian can be reduced to standard Hubbard model by setting
$t_{\rm so}=\delta=0$ and performing a gauge transformation
$c_{i,\downarrow}^{\dagger}\rightarrow(-1)^{i}c_{i,\downarrow}^{\dagger}.$
For the 1D infinite $U$ Hubbard model where total magnetization
is a good quantum number, it has been shown in Refs. \cite{PhysRevB.40.2719,PhysRevB.41.2326,Hodge2010AnIT}
that the ground states in each sector of total magnetization are degenerate
in thermodynamic limit. Thus the infinite $U$ ground state at $t_{\rm so}=0,\delta>0$
is unique and a fully spin polarized Fermi sea $v_{0}$ defined as
\begin{equation}
v_{0}=\prod_{k\in[\pi-k_{F},\pi+k_{F}]}c_{k,\downarrow}^{\dagger}|vac\rangle,
\end{equation}
where $0<k_{F}<\pi$ is the Fermi momentum with $k_{F}=\nu\pi$, and
$[\pi-k_{F},\pi+k_{F}]$ is the range of Fermi sea in FBZ. As a result,
to obtain the ground state spin texture at $t_{\rm so}>0,\delta>0,U=\infty$
it is reasonable to treat $\tilde{H}_{t_{0}}+\tilde{H}_{\delta}$ as
unperturbed Hamiltonian and treat $\tilde{H}_{t_{\rm so}}$ as perturbation.
Although the wavefunction of eigenstates in infinite $U$ Hubbard
model with $t_{\rm so}=0$ can be obtained by Bethe-Ansatz \cite{PhysRevB.41.2326},
their expressions are too complex to allow the standard perturbation treatment.

Instead of calculating perturbation formula directly, we treat the
small $\tilde{H}_{t_{\rm so}}$ term by means of variational principle.
The exact ground state can be obtained by minimizing energy of wavefunction
in the subspace $\text{span}(\{v_{0},\tilde{H}v_{0},...,(\tilde{H})^{p}v_{0}\})$
if positive integer $p$ is taken to be large enough until convergence. As an approximation,
we choose the lowest order $p=2$ where both $\langle s_{k}^{y}\rangle$
and $\langle s_{k}^{z}\rangle$ can be nonzero in FBZ. The orthonormalized
vectors are denoted as $v_{0},v_{1},v_{2}$ in the subspace. The $\nu$ dependence of $v_{1},v_{2}$ comes from the $\nu$ dependence of $v_{0}$.

The physical meaning of vectors $v_{1},v_{2}$ are clear: they are actually scattering states.
The action of (Gutzwiller projected) spin flip $t_{\rm so}$-term on the spin polarized Fermi sea $v_{0}$ defines the $v_{1}$ state where a spin becomes flipped. Similarly, the superposed actions of the spin flip $t_{\rm so}$ term and spin conserving $t_{0}$ term on $v_{1}$ defines the $v_{2}$ state, where particles further move between lattice sites.

The variational ground state wavefunction $\psi_{g}$ can be written
by an expansion of $t_{\rm so}$ as
\begin{equation}
\psi_{g}\approx v_{0}-\frac{H_{01}}{\Delta_{1}+2\delta}(v_{1}-\frac{H_{12}}{\Delta_{2}+2\delta}v_{2}),
\label{eq:WF_infiniteU}
\end{equation}
In above formula non-negative $\Delta_{1},\Delta_{2}\propto t_{0}$
are $\tilde{H}_{t_{0}}$ energy differences between vectors mentioned above,
and $H_{01},H_{12}$ are Hamiltonian matrix elements in the subspace, satisfying $H_{01}\propto t_{\rm so},H_{12}\propto t_{0}$. The explicit
expressions and detailed derivations of these quantities and $v_{1},v_{2}$
states are given in Appendix C. Since the ground states of infinite
$U$ Hubbard model with $t_{\rm so}=\delta=0$ are highly degenerate,
we stress that the choice of spin polarized Fermi sea
$v_{0}$ as a unperturbed ground state is justified when $\delta$ is sizable compared
with $t_{0}$ and $t_{\rm so}$. The spin textures are evaluated from $\psi_{g}$ in Eq. (\ref{eq:WF_infiniteU}), and below we discuss the obtained spin textures for momentum points inside and outside the Fermi sea, respectively.

For $k$ momentum points inside Fermi sea, the simple expression of $\langle s_{k}^{y/z}\rangle$
at leading order of $t_{\rm so}$ is
\begin{align}
\langle s_{k}^{y}\rangle & \approx\frac{16\pi t_{\rm so}}{\Delta_{1}+2\delta}(1-\nu)\sin k,\notag \\
\langle s_{k}^{z}\rangle & \approx-1+O(t_{\rm so}^{2}).\label{eq:sy_infiniteU_insede_fermi_sea}
\end{align}

\begin{figure}[t]
\includegraphics[width=\columnwidth]{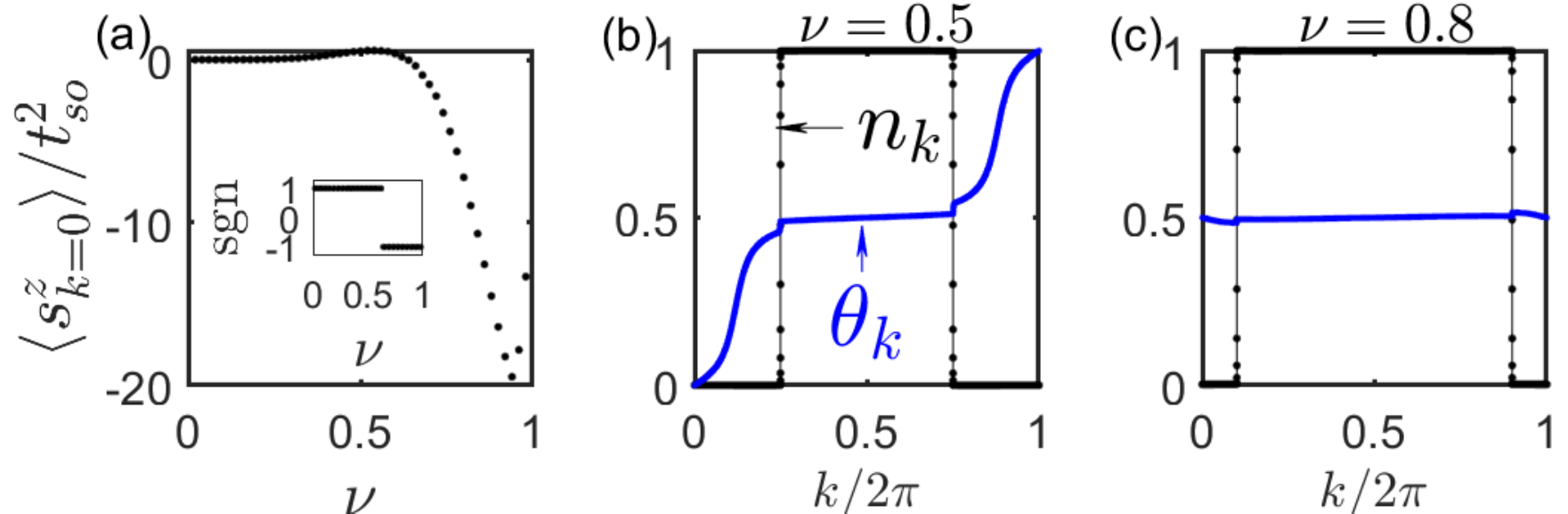}
\caption{At infinite $U$ the ground state still exhibit nontrivial topology below a critical filling factor $\nu_c$. The Hamiltonians in the calculation are projected by Gutzwiller projector. (a), The value of spin
polarization $\langle s_{k=0}^{z}\rangle$ versus $\nu$ from the lowest order $t_{so}$ expansion with parameters $(t_{0},\delta)=(1,1)$ and ansatz in Eq. (\ref{eq:WF_infiniteU}). The inset in (a) shows the sign of spin polarization, indicating a topological transition featured by the spin flip at $k=0$ point. (b)-(c), Spin directions $\theta_{k}$ and momentum distributions $n_{k}$ obtained from VUMPS with parameters $(t_{0},t_{\rm so},\delta)=(1,0.1,1)$, and the topologies are nontrivial and trivial, respectively.}
\label{fig:infiniteU_texture}
\end{figure}

For $k$ momentum points outside Fermi sea, the leading order spin textures $\langle s_{k}^{y/z}\rangle$
from Eq. (\ref{eq:WF_infiniteU}) can only be integrated
numerically in general. In Appendix C, we show that $\langle s_{k}^{y}\rangle$ evaluated from Eq. (\ref{eq:WF_infiniteU}) at generic fillings have the same sign as the Bloch Hamiltonian coefficient $h_k^y$. A special case is the small $\nu$ limit, where
the expression of $\langle s_{k}^{y}\rangle$ in can be simplified to an analytical expression
\begin{equation}
\langle s_{k}^{y}\rangle\approx\frac{64\pi^{4}t_{\rm so}^{3}\nu^{4}}{3(\Delta_{1}+2\delta)^{3}}\sin k.
\end{equation}
The leading order of $\langle s_{k=0}^{z}\rangle$ with $k=0$ outside Fermi sea
obtained from Eq. (\ref{eq:WF_infiniteU})
is at $t_{\rm so}$'s second order and shown in Fig.~\ref{fig:infiniteU_texture}(a).
The sign of $\langle s_{k=0}^{z}\rangle$ changes at finite filling factor, implying existence of a critical filling $\nu_{c}$, below
which the sign of $\langle s_{k=0}^{z}\rangle$ will not change even at arbitrarily large $U$. Therefore the winding number is always nontrivial at small filling factor $\nu<\nu_c$ with arbitrarily strong interaction, as being also confirmed by numerical results in Fig.~\ref{fig:infiniteU_texture}(b)-(c) using the Gutzwiller projected Hamiltonian. In Appendix C, we show behaviours of each components in $\langle s_{k=0}^{z}\rangle$ evaluated from Eq. (\ref{eq:WF_infiniteU}), from which one can see that $v_{1}$ does not contribute to $\langle s_{k}^{z}\rangle$ around $k\approx 0$ at small filling factor, hence the state $v_{2}$ which contributes to the leading order is responsible for existence of finite $\nu_{c}$ at infinite large $U$.

\subsection{Numerical methods for calculation of spin textures}

While being applicable to limit cases, the analytic study shows clearly two different phases tuned by $\nu$ and interaction. The transition between them can be further studied numerically, for which the fermionic Hamiltonian is transformed
to spin model (see Appendix A) using Jordan-Wigner transformation. We apply three methods to extract the ground state properties.
\begin{enumerate}
\item[(i)] The ground state properties such as spin textures and winding number phase diagrams at generic parameters are calculated with variational uniform matrix product state algorithm
(VUMPS) \cite{PhysRevB.97.045145}, from which the spin textures
can be calculated without finite size effects and boundary effects. The filling factor $\nu$
is controlled by fine tuning of chemical potential $\mu$, and for gapless
phases the error of $\nu$ is within $5\times10^{-3}$ in our VUMPS calculation.

\item[(ii)] In order to verify ground state phase diagrams obtained from VUMPS, we also use exact diagonalization (ED) with lattice size $L=6$ to calculate phase diagrams
at very low but non-zero temperature (e.g., $\beta=5$), where filling
$\nu$ is a continuous function of chemical potential $\mu$ even
with small lattice size.

\item[(iii)] We also use finite size matrix product state (MPS) algorithm to investigate the conventional topological characterizations, including energy gaps and entanglement spectra,
where energy variances in our calculation are controlled below $5\times10^{-6}$ under open boundary condition (OBC) and below $5\times10^{-3}$ under periodic boundary condition (PBC).
We use $E_{n}(N)$ to denote $n$'s lowest eigenstate energy in the
sector of total particle number $N$.
\end{enumerate}

Now we discuss in detail about using the finitely correlated VUMPS ansatz to characterize spin textures of the gapless Luttinger liquid. The asymptotic behaviour of momentum distribution and spin textures near Fermi points is determined by the long distance behaviour of single-particle Green's function
\begin{equation}
G_{\sigma,\sigma'}(i,j)=\langle c_{i,\sigma}^{\dagger},c_{j,\sigma'}\rangle.
\end{equation}
In gapless Luttinger liquid phase $G_{\sigma,\sigma'}(i,j)$ exhibits
power law decay in long distance and correspondingly $\langle c_{k,\sigma}^{\dagger}c_{k,\sigma'}\rangle$
has power law singularity near Fermi points. Since connected correlation functions
of an injective VUMPS with fixed bond dimension $\chi$
decays exponentially, the power law characteristics of gapless Luttinger
liquid should be revealed by extrapolation of bond dimension $\chi$
of MPS \cite{karrasch2012luttinger}. For $k$ momentum points away from fermi
points, the momentum distribution and spin textures $\langle s_{k}^{y/z}\rangle$
converges quickly with respect to distance $|i-j|$ in $G_{\sigma,\sigma'}(i,j)$
in Fourier transformation. Therefore for determination of phase diagram
of spin textures, we choose relatively small VUMPS bond dimension $\chi=50$.

\begin{figure*}[t]
\centering
\includegraphics[width=6in]{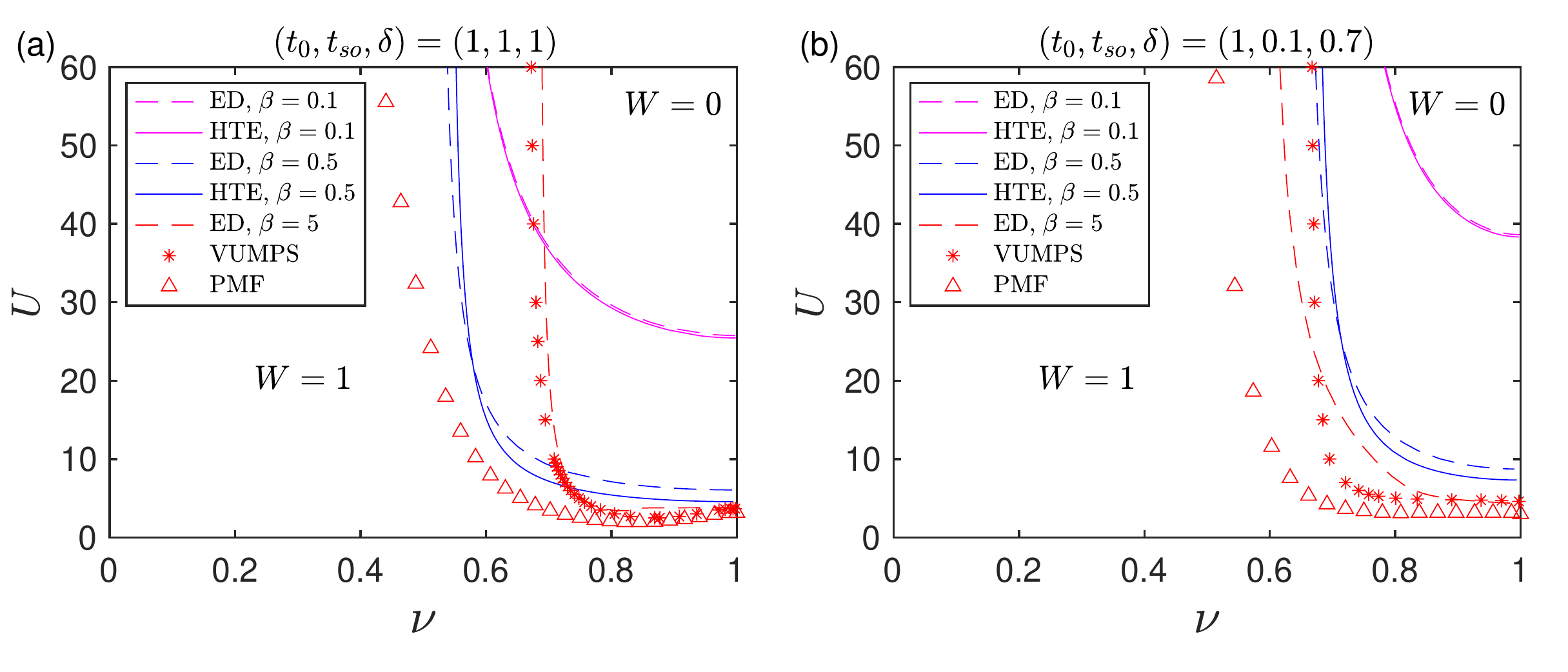}
\caption{Topological phase diagrams at finite $U$. The curves
represent phase boundaries: on the upper right side spin textures are trivial with $W=0$, on the lower left side spin textures are nontrivial with $W=1$. The ground state phase boundaries are obtained from VUMPS and PMF methods; the finite temperature phase boundaries
are obtained from ED and first order HTE.}
\label{fig:finiteU_phase_diagram}
\end{figure*}

\subsection{Many-body spin textures at generic $\nu$ and $U$}

In this subsection we characterize the features of the spin texture patterns at generic filling factor and interaction strength, under the condition of nontrivial band topology $W_{H_0}=1$. One will see that the winding number $W$ can only be $0$ or $1$, depending on spin directions at high symmetry momenta.

Before working on the interacting ground states, let us first review the spin directions $\theta_{k}^{H_{0}}$
of the noninteracting Hamiltonian $H_0$ at half filling. The spin textures of the noninteracting Hamiltonian take the form
\begin{equation}
\langle s_{k}^{y/z} \rangle=h_{k}^{y/z}/h_{k},
\end{equation}
where $h_{k}^{y/z}$ are the Bloch Hamiltonian coefficients.
For $0<\delta<2t_0$ with $W_{H_0}=1$ the spin polarization $\langle s_{k}^{z}\rangle $ has two zero points in FBZ. While the $W_{H_0}$ is trivialized when $\delta>2t_{0}$ with $\langle s_{k}^{z}\rangle$ being polarized in the whole FBZ.

For the interacting regime at half filling condition, the spin texture patterns can also be easily understood. We consider single-particle parameter regime with $0<\delta<2t_{0}$ such that $W_{H_0}=1$. In weak $U$ limit, the slightly modified many-body spin textures still satisfy $W=1$ and have the same features as that of $H_0$. In the large $U$ limit,
we can show that the spin texture becomes trivial and are given by (see Appendix D)
\begin{align}
\langle s_{k}^{y}\rangle= & \frac{4t_{\rm so}}{U+2\delta}\sin k, \notag\\
\langle s_{k}^{z}\rangle= & -\mbox{sgn}(\delta)+O(\frac{1}{U^{2}}).
\end{align}
From above results one can see that the direction of the Zeeman field $\delta$
determines the polarization. The sign of $\langle s_{k}^{y}\rangle$
remains unchanged in FBZ, while large repulsive interaction tends to polarize
$\langle s_{k}^{z}\rangle$ for any $k$ towards the opposite direction of Zeeman field.

The interacting phase for the fractional filling factor $\nu<1$ necessitates careful study. Note that the sign of $\langle s_{k}^{y}\rangle$ remains the same as that of the Bloch Hamiltonian coefficient $h_{k}^{y}$, and the increase of filling factor or interaction strength tends to polarize $\langle s_{k}^{z}\rangle$ towards the opposite direction of Zeeman field. This feature is observed in both analytic calculation (including weak $U$ expansion and $t_{\rm so}$ expansion at infinite
$U$, as studied in the former section, and two exact cases in Appendix D) and numerical VUMPS calculation as depicted in Fig.~\ref{fig:textures_ob}, \ref{fig:typycal_textures} and \ref{fig:infiniteU_texture}. Since $\langle s_{k}^{y}\rangle$ only has only two zero points at high symmetry momenta $k=0,\pi$, we can simplify the determination of the bulk topology by characterizing the winding number $W$, which can only be $0$ or $1$, by the signs of $\langle s_{k}^{z}\rangle$ at $k=0,\pi$~\cite{ZHANG20181385}. Accordingly, the transition of winding
occurs when sign of $\langle s_{k}^{z}\rangle$ at $k=0,\pi$ changes.

The spin polarizations at $k=0,\pi$ points are not given from the low energy physics near Fermi points at generic filling factor. In general the state at $k=\pi$ lies far below the Fermi energy, and its spin polarization cannot be reversed by scattering. Thus we focus on the spin polarization at $k=0$, with
two cases being illustrated in Fig.~\ref{fig:typycal_textures}(a)-(d) and
(e)-(h), corresponding to two different band structures of $H_0$ distinguished by Eq. (\ref{eq:two_fermi_point_condition}).
In the former case of simple band structure as shown in Fig.~\ref{fig:typycal_textures}(a)-(d), the competing particle type excitations to upper and lower subbands determines the spin polarization at $k=0$. In the latter case of non-monotone band structure as shown in Fig.~\ref{fig:typycal_textures}(e)-(h), the MF band structure will eventually be deformed by large $U$ to satisfy Eq. (\ref{eq:two_fermi_point_condition}) due to renormalized $\tilde{\delta}$, as can be seen from momentum distribution $n_{k}$ in Fig.~\ref{fig:typycal_textures}(e)-(h). We have checked that the transition between the two band structures is typically ahead of transition of winding number. This explains why the phase diagrams for emergent topology are similar in two cases, as discussed further below.


\begin{figure*}[t]
\centering
\includegraphics[width=6in]{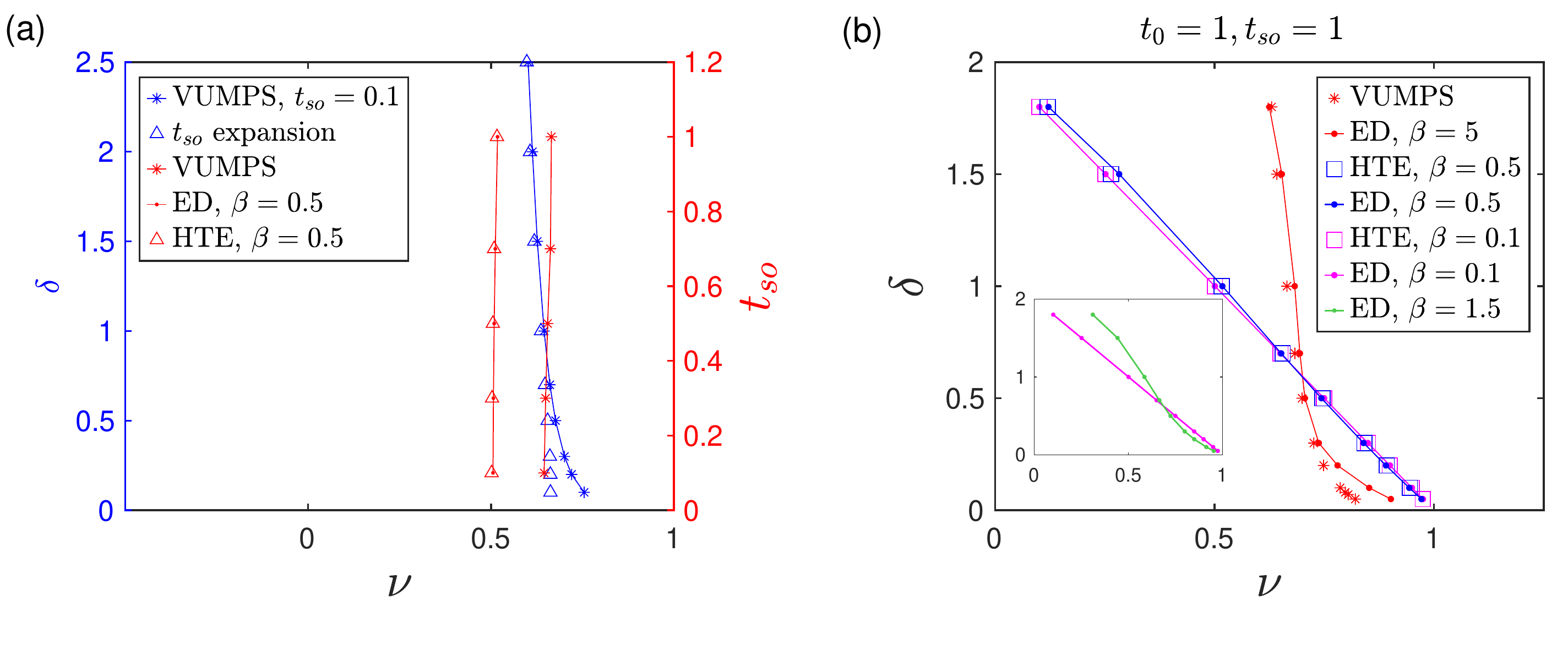}
\caption{Topological phase diagrams at infinite $U$.
Hamiltonians in the calculation are projected by Gutzwiller projector. The
curves denote phase boundaries and winding numbers are $W=1/0$ at
left/right sides of phase boundaries. VUMPS and $t_{\rm so}$ expansion
results corresponds to ground state. HTE (up to third order)
and ED results correspond to finite temperature. (a), Blue $\delta-\nu$ curves show existence
of finite $\nu_{c}$ at ground state, which is confirmed by the analytic
lowest order $t_{\rm so}$ expansion and numerical VUMPS results. Red $t_{\rm so}-\nu$ curves
with $\delta=1$ show that the magnitude of $t_{\rm so}$ has little effect on magnitude
of $\nu_{c}$. (b), Critical $\nu_{c}$ versus $\delta$ at different temperatures.}
\label{fig:phase_diagram_infiniteU}
\end{figure*}

\subsection{Ground state Phase diagrams and transition of the winding number}

We show finite $U$ phase diagrams of winding number in Fig.~\ref{fig:finiteU_phase_diagram}.
One can see that the winding number captures the effects of tuning filling factor and interaction strength on the many-body ground state. The features of phase diagrams are below. First, the {\em trivial} Luttinger liquid in conventional topology characterization has two emergent phases with winding numbers $W=1$ and $W=0$ respectively. Second, the topological transition can be tuned by increasing either filling factor or interaction strength. Third, at small filling factor the ground state remains nontrivial even at infinite $U$. The phase diagrams regarding the two different noninteracting band structures are given in (a) and (b), respectively.

The infinite $U$ phase diagrams are shown numerically in Fig.~\ref{fig:phase_diagram_infiniteU}.
We confirmed existence of nonzero $\nu_{c}$ by
calculating Gutzwiller projected Hamiltonian, and show how the critical
value $\nu_{c}$ depends on $t_{\rm so}$ and $\delta$. One can see that $\nu_{c}$
almost do not depend on $t_{\rm so}$. While increase of $\delta$ tends to decrease $\nu_{c}$,
since $\delta$ tends to polarize spin along $z$ direction and trivialize the winding number. It is noteworthy that when
$\delta\rightarrow0$ the $\nu_{c}$ tends to $1$ as shown by VUMPS
results in Fig.~\ref{fig:phase_diagram_infiniteU}(a)-(b),
which is consistent with the flat band $t_{0}=t_{\rm so},\delta=0$ case where
the spin texture remains nontrivial with arbitrary $U$ as proved in
Appendix D. This again confirms our previous statement that the $t_{\rm so}$ expansion necessitates a finite $\delta$ to correctly obtain
$\nu_{c}$ as shown in Fig.~\ref{fig:phase_diagram_infiniteU}(a), otherwise the
fully spin polarized Fermi sea is not a good ansatz for the $t_{\rm so}$
expansion. Nevertheless, the $t_{\rm so}$ expansion perfectly shows the existence of nontrivial many-body spin textures in the infinite large $U$ limit.

\begin{figure}[t]
\centering
\includegraphics[width=\columnwidth]{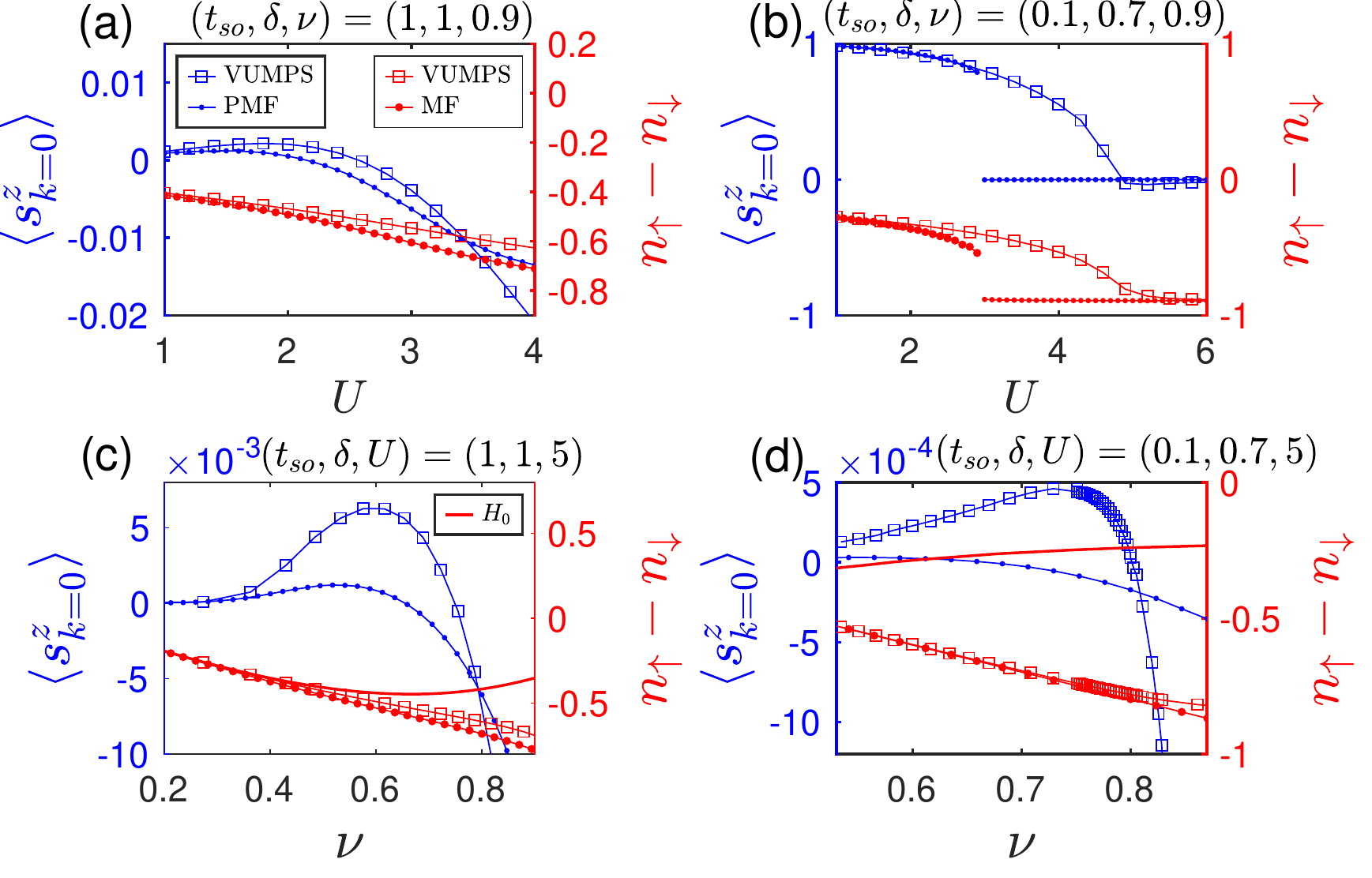}
\caption{Ground state spin polarization $\langle s_{k=0}^{z}\rangle$ and magnetization
$n_{\uparrow}-n_{\downarrow}$ versus $U$ or $\nu$ around the topological transition at fractional filling regime. Typically the winding number transits when spin polarization (length) at $k=0$ crosses zero continuously. Blue and red curves denote spin polarization and magnetization respectively. The red solid lines in (c) and (d) denote magnetization of noninteracting ground states. Note that in (b) the discontinuity of $\langle s_{k=0}^{z}\rangle$ given by the PMF method at large $\nu$ is artificial due to the artificial discontinuity of the MF solution.}
\label{fig:sz_mz_transition}
\end{figure}

To see how the topological transition occurs in gapless regime,
we plot the evolution of magnetization $m=n_{\uparrow}-n_{\downarrow}$
and spin polarization $\langle s_{k=0}^{z}\rangle$ versus parameters $(\nu,U)$ in Fig.~\ref{fig:sz_mz_transition}.
We find a novel phenomenon that within gapless regime the $\langle s_{k=0}^{z}\rangle$ crosses zero continuously with increase of magnetization, which is in sharp contrast to the gapped half filling case where $\langle s_{k=0}^{z}\rangle$ jumps suddenly. For example, consider the noninteracting band insulator at the ctritical point $\delta=2t_{0}$. The band gap closes exactly at the $k=0$ point such that the $k=0$ point is just at the fermi point, leading to a sudden jump of spin polarization from $\langle s_{k=0}^{z}\rangle=1$ to $\langle s_{k=0}^{z}\rangle=-1$.

We explain why in gapless regime the winding number transits continuously through the PMF picture. This relies on the fact that the MF parameter $\tilde{\delta}$ in general varies continuously and the excitation energies to $k=0$ point which is away from Fermi points with finite energy. Being precise, the winding number is determined by spin polarization
$\langle s_{k=0}^{z}\rangle$, which is related to the rotating frame spin direction
$\langle\mathcal{S}_{k=0}^{z}\rangle$ defined in Eq. (\ref{eq:basis_change}) by
\begin{align}
\langle s_{k=0}^{z}\rangle & =(|\beta_{k=0}|^{2}-|\alpha_{k=0}|^{2})\langle\mathcal{S}_{k=0}^{z}\rangle \notag\\
 & =\begin{cases}
\langle\mathcal{S}_{k=0}^{z}\rangle, & \tilde{\delta}>2t_{0}\\
-\langle\mathcal{S}_{k=0}^{z}\rangle, & 0<\tilde{\delta}<2t_{0}
\end{cases}.
\end{align}
Here $\langle\mathcal{S}_{k=0}^{z}\rangle$ measures the weight difference
between particle excitations at MF upper/lower subbands. From Eq. (\ref{eq:sz_k0_formula}) one can easily find $\langle s_{k=0}^{z}\rangle$
is a continuous function as long as Fermi points are away from $k=0$ (at the special point with $\tilde{\delta}=2t_{0}$, both $|\beta_{k=0}|^{2}-|\alpha_{k=0}|^{2}$ and $\langle\mathcal{S}_{k=0}^{z}\rangle$
change signs, so $\langle s_{k=0}^{z}\rangle$ is still continuous). With continuous change of Hamiltonian parameters, the MF bands deforms continuously and the value of $\langle s_{k=0}^{z}\rangle$ crosses zero continuously, showing the sharp difference from the gapped phases. Note that at large $\nu$ the PMF method may give discontinuous observables [Fig.~\ref{fig:sz_mz_transition}(b)], which is artificial (more details seen in Appendix B).

The features of topological transition in the present gapless phase are exceptional. First, the spin texture is a continuous function of system parameters, and the change of topological number is associated with continuous spin flip at high symmetry momentum. Second, at the topological transition, the spin length at $k=0$ vanishes and the winding number becomes ill-defined. Accordingly, the reduced density matrix $\rho_{k=0}$ in the singly occupied subspace (projected spin-state) is fully mixed identity matrix. Third, through the topological transition the ground state is always gapless Luttinger liquid at fractional filling. These features result from the gapless nature of Luttinger liquid and distinguishes sharply from those in conventional topological characterizations.

\section{Finite-temperature many-body spin textures}

In this section we show that the emergent topological phase and phase transition can be generalized to finite temperature regime, which facilitates the real experimental detection. We use the
grand canonical ensemble defined in Eq.~(\ref{eq:def_texture}).
The winding number of spin texture now implies the properties of the many-body Hamiltonian (or low energy eigenstates) instead of the ground state. In general the behaviour of observables at high and low (zero) temperatures can be different. Nevertheless, we show that both the trivial and nontrivial finite-temperature phases appear in the phase diagram, and the topological transition of winding number can be tuned by either the filling factor or interaction strength. Compared to ground state, the particle occupation at momentum points outside Fermi sea can be enhanced at appropriate finite temperature, thus the experimental measurement of the many-body spin textures would be easier.

Similar to the ground state, the winding number also satisfies the quantization conditions at finite temperature. First, the spin length is nonzero in the whole FBZ at finite temperature, as one will see that the sign of $\langle s_{k}^{y}\rangle$ is also the same as that of the Bloch Hamiltonian coefficient $h_{k}^{y}$, and for the current study the zero spin length is only possible at
$k=0$ point at the topological transition point. Further, the continuity of spin textures in FBZ is also satisfied at finite temperature. As shown in \cite{karrasch2012luttinger}, the power law behaviours of Green's function $G_{\sigma,\sigma'}(i,j)$ at zero temperature
Luttinger liquid are cut off by finite temperature and replaced by exponential decay, and the momentum distribution is linear at Fermi
points. Therefore, the winding number at finite temperature is still well-defined and quantized.

To calculate many-body spin texture and determine the winding number at finite temperature, we use high temperature linked cluster expansion (HTE) method introduced
in Refs. \cite{PhysRevB.43.8549,PhysRevE.89.063301} which is valid at high temperature $T$ (small $\beta$). Besides, we shall also use ED (with lattice size $L=6$), which can be applied to arbitrary temperature
although limited to small size, to confirm the HTE results at high temperature.

\subsection{Spin textures at high temperature: HTE method and features of patterns}

We now briefly introduce the HTE method \cite{PhysRevB.43.8549,PhysRevE.89.063301}. In the HTE method we treat the on-site terms
\begin{align}
\tilde{H}_{0}= U\sum_{i}n_{i\uparrow}n_{i\downarrow}+\delta\sum_{i}(n_{i,\uparrow}-n_{i,\downarrow})-\mu\sum_{i\sigma}n_{i\sigma}
\end{align}
are treated as unperturbed, and the hopping terms \begin{align}
\tilde{H}_{1}= & \sum_{i} \psi_{i}^{\dagger}(it_{\rm so}\tau_{y}-t_{0}\tau_z) \psi_{i+1} +h.c.,
\end{align}
as perturbations. The partition function of unperturbed Hamiltonian for each site is independent, and they
are connected by perturbation terms with the expansion parameters being $t_{0}\beta$ and $t_{\rm so}\beta$. Thus the finite-order perturbation expansion is a natural approximation at high temperature.
We define spin-dependent chemical potential $\mu_{\sigma}=\mu-\sigma\delta$ as well as four single-site weights, $(1,x=e^{\beta\mu_{\uparrow}},y=e^{\beta\mu_{\downarrow}},d=e^{\beta(2\mu-U)})$
which correspond to four configurations $|0\rangle$, $|\uparrow\rangle$, $|\downarrow\rangle$, $|\uparrow\downarrow\rangle$, respectively.
Spin texture up to first order expansion has the form
\begin{align}
\langle s_{k}^{y}\rangle= & \frac{4t_{\rm so}\sin k}{(1+x+y+d)^{2}} [\frac{e^{\beta\mu_{\bar{\sigma}}}-e^{\beta\mu_{\sigma}}}{\mu_{\bar{\sigma}}-\mu_{\sigma}}+\cdot\cdot\cdot],\notag \\
\langle s_{k}^{z}\rangle= & \frac{x-y}{1+x+y+d}
+  \frac{2t_{0}\cos k[\beta e^{-\beta U}xy(x+y)+\cdot\cdot\cdot]}{(1+x+y+d)^{2}}.\label{eq:HTE_1st_sim}
\end{align}
Being different from that of $t_0,\delta$, and $U$, the effect of the parameter $t_{\rm so}$ on $\langle s_{k}^{z}\rangle$ is at least second order and is irrelevant at high temperature.

The qualitative features of spin texture patterns at finite temperature is the same as those of the ground state given in Sec. III. 
Therefore, the spin direction is well defined and the winding number at finite temperature is again determined by signs of $\langle s_{k}^{z}\rangle$ at $k=0,\pi$. In particular, the projected spin-polarization $\langle s_{k=\pi}^{z}\rangle$ is unchanged by scatterings even at finite temperature, and the winding number is then determined by the sign of $\langle s_{k=0}^{z}\rangle$. The complete analytical expressions for spin textures and details of above analysis are presented in Appendix E.

\subsection{Finite temperature critical points of winding number}

Instead of showing the technical details, we present the generic results of critical transitions in three limit cases, which reveal the essential features of the system with finite temperature. Then we show the finite-temperature phase diagram.

%


{\em Infinite temperature limit: $\beta\ll\frac{1}{U}\ll1$}.---In this regime, the temperature is much larger than any other parameters including $U$. Thus terms in Eq.~(\ref{eq:HTE_1st_sim}) can be expanded in terms of $\beta\delta,\beta U$ that
\begin{align}
e^{\beta\delta} & \approx1+\beta\delta+(\beta\delta)^{2}/2,\notag\\
e^{-\beta U} & \approx1-\beta U+(\beta U)^{2}/2.
\end{align}
We keep the terms up to the first order of $\beta$ to study the critical condition $\langle s_{k=0}^{z}\rangle=0$ and obtain
\begin{equation}
2t_{0}=\delta,
\end{equation}
which is the same as the critical point for the noninteracting band. This is reasonable since in the high-temperature limit, the interacting effects are no longer relevant. The winding number should also equal that of the noninteracting band: $W=W_{H_0}=1$.


{\em Infinite interaction limit: $\frac{1}{U}\ll\beta\ll1$}.--When $U$ is strong compared with any other parameters including
high temperature $T=1/\beta$, 
the first order expansion of the spin texture takes the form
\begin{align}
\langle s_{k}^{y}\rangle= & 4t_{\rm so}\sin k\frac{1}{(1+x+y)^{2}}\frac{e^{\beta\mu_{\bar{\sigma}}}-e^{\beta\mu_{\sigma}}}{\mu_{\bar{\sigma}}-\mu_{\sigma}},\notag\\
\langle s_{k}^{z}\rangle= & \frac{x-y}{1+x+y}+2t_{0}\cos k\frac{\beta(x+y)}{(1+x+y)^{2}},
\end{align}
which give the exact transition point as
\begin{equation}
2t_{0}(1-\nu)=\delta.
\end{equation}
This result shows that the increase of filling factor tends to trivializes spin texture.
If $\nu$ is close to $1$, only when $\delta$ is very small the winding number is nontrivial at high temperature and strong $U$ limit. On the other hand, if $\nu$ is close to $0$, the interaction effect is largely suppressed at high temperature even at infinite $U$.

At intermediate temperature, the dependence of $\nu_{c}$ on parameters $\beta$, $t_{\rm so}$, and $\delta$ is shown in numerically
in Fig.~\ref{fig:phase_diagram_infiniteU} up to third order HTE. It can be found that $\nu_{c}$ has no evident $t_{\rm so}$ dependence. Further,
the critical filling factor $\nu_{c}$ is a monotonically decreasing function of $\delta$. On the other hand, whether dependence of $\nu_{c}$ on temperature relies on magnitude of $\delta$. At small $\delta$ the critical $\nu_{c}$ is a decreasing function of $\beta$, while at large $\delta$ the critical $\nu_{c}$ is a increasing function of $\beta$. This is due to the fact that for small $\delta$, which corresponds to relatively large single-particle band gap, the major effect of temperature suppresses the scattering, while for large $\delta$ the temperature effect mainly enhances the higher band population.


{\em Half filling case: $\nu=1$, $\beta\ll1$}.--The half filling condition $\nu=1$ implies existence of particle-hole symmetry, i.e., $U=2\mu$. In this case, through expanding the critical condition $\langle s_{k=0}^{z}\rangle=0$ at high temperature, one obtains the simplified equation for topological transition as
\begin{equation}
2t_{0}[\beta U+e^{\beta U/2}(1-e^{-\beta U})]-\delta(\beta U)(1+e^{\beta U/2})=0.
\end{equation}
If all other parameters are fixed, the solution of the product $\beta U$ is also fixed, leading to the critical $U_{c}$ satisfying
\begin{equation}
U_{c}\propto\frac{1}{\beta}.
\end{equation}
This means at high temperature the critical $U_{c}$ for winding number is proportional to inverse of $\beta$, which implies that at half filling the temperature effect mainly suppresses the scattering induced by interaction.

\subsection{Finite temperature Phase diagrams}

The finite-$T$ phase diagrams are shown numerically in Fig.~\ref{fig:finiteU_phase_diagram} Fig.~\ref{fig:phase_diagram_infiniteU}, which agree with the analytic results obtained in limiting cases. For example, when $T=1/\beta$ is the largest energy scale among
all parameters, the winding number is always nontrivial for $\delta<2t_0$. Thus the phase boundary curves
at high $T$ limit moves upwards towards larger $U$. 
On the one hand, the nearly horizontal phase boundary $U_{c}$ near half filling moves
upwards with increasing temperature, which agrees with the result that
$U_{c}\propto1/\beta$ at $\nu=1$ and high temperature.
Finally, the vertical phase boundary $\nu_{c}$ at
infinite $U$ may either increases or decreases with $\beta$,
depending on magnitude $\delta$ as seen in Fig.~\ref{fig:phase_diagram_infiniteU}(b).
The features capture the tendency of the $U-\nu$ phase boundary curves when temperature is changed.
We emphasize that the zero-temperature and finite-temperature phase diagrams have resemblance, and both have nontrivial and trivial phases, which shall facilitate the detection in real experiments.


\begin{figure*}[t]
\centering
\includegraphics[width=6.5in]{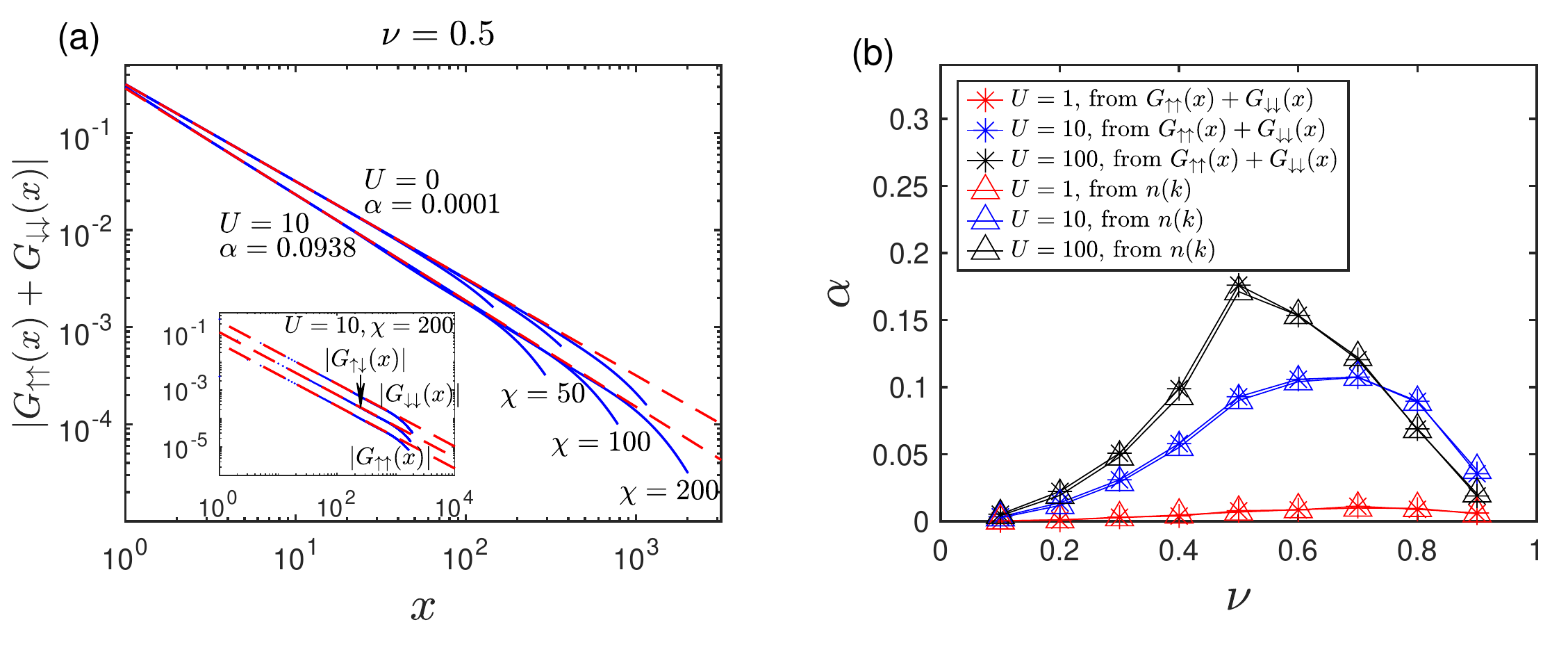}
\caption{Luttinger liquid features of ground state at fractional filling factor. Parameters are chosen as $(t_{0},t_{\rm so},\delta)=(1,1,1)$. (a), Long distance power law decay of single-particle Green's functions, which show that the single-particle excitation is gapless. Blue curves show Green's functions $G_{\sigma\sigma'}(x)=\langle c_{i+x,\sigma}^{\dagger}c_{i,\sigma'}\rangle$ obtained from VUMPS method via extrapolation of bond dimension $\chi$. The anomalous dimension $\alpha$ is extracted from power law fitting $G(x\gg1)\sim\frac{1}{x^{1+\alpha}}$. The red dashed lines correspond to fitting of power law function $\frac{1}{x^{1+\alpha}}$.
(b), Positive anomalous dimension $\alpha$ at different filling factor $\nu$
and interaction $U>0$ extracted from fitting $G(x\gg1)\sim\frac{1}{x^{1+\alpha}}$
and $|n(k)-n(k_{F})|\sim|k-k_{F}|^{\alpha}$, which indicates that the observable spin texture at Fermi points has power law singularities instead of fermi-liquid like discontinuity.}
\label{fig:luttinger_parameter}
\end{figure*}

\section{Comparison with conventional topological phases}
Now we provide a sharp comparison between the present emergent gapless topological phases obtained at fractional filling and the conventional topological phases at gapped ($\nu=1$) regime. In the latter regime, the winding number of the many-body spin texture can capture the topology of a conventional gapped SPT phase. 

\subsection{The gapless $\nu<1$ regime: Luttinger liquid properties and absence of ground state degeneracy}

We first show that the emergent gapless topological Luttinger liquid phase
characterized by the many-body spin texture is indeed beyond the topological characterization theory applicable for the previously studied 1D gapless topological phases~\cite{fidkowski2011majorana,cheng2011majorana,bonderson2013quasi,kainaris2015emergent,keselman2015gapless,montorsi2017symmetry,PhysRevB.96.085133,thorngren2020intrinsically,scaffidi2017gapless,parker2018topological,verresen2019gapless}.
In particular, for the conventional gapless topological phases, the topological characteristics can be summarized by the existence of topological gapped degree of freedom, protected ground state degeneracy and degenerate bipartite entanglement spectrum. In contrast, the present Luttinger liquid is gapless for all degree of freedom and the ground state has no conventional topological characteristics, but exhibits emergent nontrivial topology.


\begin{figure*}[t]
\centering
\includegraphics[width=6.5in]{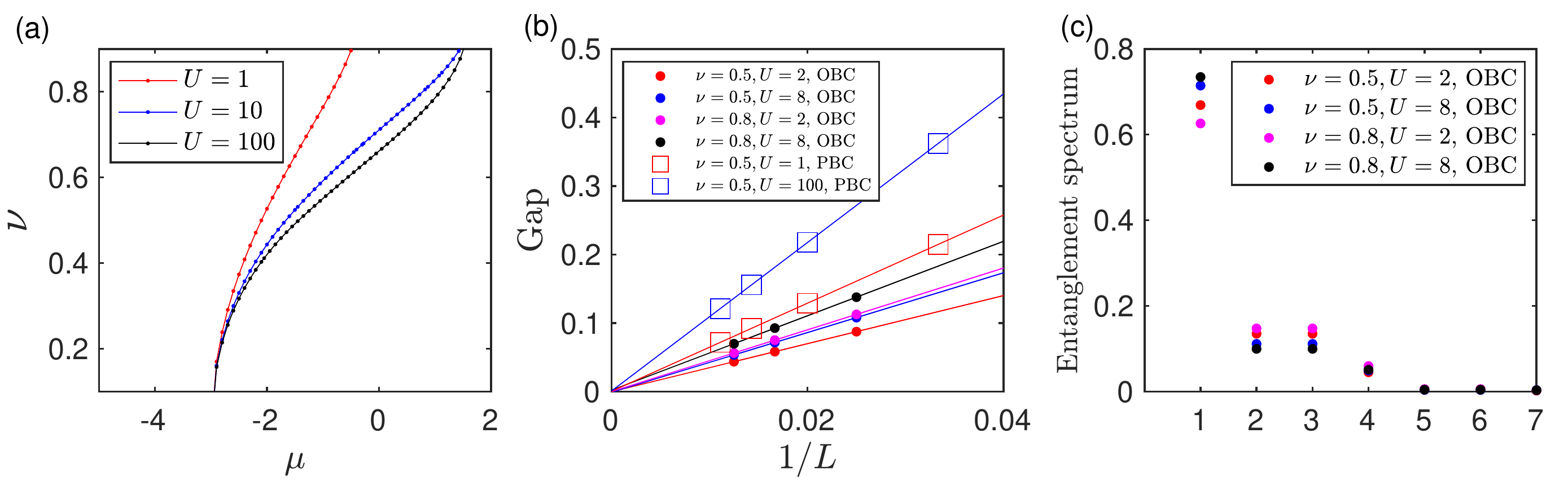}
\caption{The ground states at $\nu<1$ have no conventional
topological characteristics. Parameters are chosen as $(t_{0},t_{\rm so},\delta)=(1,1,1)$. (a), The filling factor $\nu$ versus chemical potential $\mu$ obtained from VUMPS method with $\chi=50$. Absence of filling factor's plateau indicates that single-particle excitation is always gapless. (b), Finite size scaling of excitation gaps $E_{e}(N)=E_{1}(N)-E_{0}(N)$ (circle markers) under OBC and single-particle gaps $E_{sp}=E_{0}(N+1)+E_{0}(N-1)-2E_{0}(N)$ (square markers) under PBC obtained from finite size MPS simulation, which show that the bulk single-particle excitation is gapless and there is no ground state degeneracy under OBC, respectively. (c), Non-degenerate ground state half chain entanglement spectrum in descending order at the bond between site $39$ and site $40$ with lattice size $L=80$ obtained from finite size MPS.}
\label{fig:gaps_ES}
\end{figure*}

\subsubsection{The gapless Luttinger liquid ground state}

We confirm that the ground state is always the gapless Luttinger liquid at fractional filling, including the quarter filling $\nu=0.5$ case, in which the Umklapp scattering exists.
This is similar to the case of the $t-V_{1}-V_{2}$ spinless fermion models studied in Refs. \cite{schmitteckert2004charge,duan2011bond,mishra2011phase,CDW_2020}, where in the limit of large nearest-neighbour interaction $V_1$ and large next-nearest-neighbour interaction $V_2$ the Luttinger liquid phase is still stable due to frustration between $V_1$ and $V_2$. Specifically, our model can also be mapped to a spinless fermion model through projecting out the upper subband, and we expect the mechanisms for stability of the Luttinger liquid phases are similar.

The Luttinger liquid behaviour of the ground states can be studied from the single-particle Green's function $G_{\sigma\sigma'}(x)=\langle c_{i+x,\sigma}^{\dagger}c_{i,\sigma'}\rangle$ at $\nu=0.5$ which exhibit power law decay
\begin{equation}
G(x\gg1)\sim\frac{1}{x^{1+\alpha}}.
\end{equation}
The exponent $\alpha$ characterizes the anomalous dimension \cite{karrasch2012luttinger}
in Luttinger liquid physics and is shown in Fig.~\ref{fig:luttinger_parameter}(a). As fitted
from $G_{\sigma\sigma'}(x)$ with different spin configurations $\sigma\sigma'$, the exponent $\alpha$
is nearly the same for all spin components in our numerical results (see inset). The power law decay behaviour of Green's function is also verified at generic parameters, with $\alpha$ being plotted in Fig.~\ref{fig:luttinger_parameter}(b) and extracted from both Green's function $G(x)$ and momentum distribution
\begin{equation}
|n(k)-n(k_{F})|\sim|k-k_{F}|^{\alpha},
\end{equation}
respectively. The obtained range $0<\alpha<1$ for $U>0$ indicates that the ground state behaves as a standard Luttinger liquid \cite{karrasch2012luttinger} with power law singularities, instead of fermi-liquid discontinuity at Fermi points.

The second evidence of no gap opening is that no translation symmetry breaking (charge density wave) solution can be found in numerical results of two-site unit-cell VUMPS algorithm. We plot the filling factor $\nu$ versus chemical potential $\mu$ obtained
from VUMPS in Fig.~\ref{fig:gaps_ES}(a), where absence of $\nu$'s
plateau at $\nu=0.5$ also indicate the ground state is gapless even
at very large $U$. We calculate single-particle gap defined as
\begin{equation}
E_{sp}(N)=E_{0}(N+1)+E_{0}(N-1)-2E_{0}(N)
\end{equation}
under PBC using finite size MPS as shown
in Fig.~\ref{fig:gaps_ES}(b), and confirm its scaling behaviour
\begin{equation}
E_{sp}(N)\sim\frac{1}{L},
\end{equation}
the same as metallic system with fixed filling factor.

With the power law decay of Green's function and gaplessness of ground state verified, one can conclude the ground state at any fractional filling $\nu<1$ and arbitrary finite Hubbard interaction strength is a gapless Luttinger liquid. 
Accordingly, the emergent topological phase predicted here cannot be determined by the low energy physics near Fermi points.

\subsubsection{Absence of degeneracy of ground state and entanglement spectrum}


Because of no gapped degree of freedom, we expect that there is no ground state degeneracy (gapless edge excitation) or entanglement spectrum degeneracy at fractional filling factor $\nu<1$. We provide the results numerically. The excitation gap $E_{e}(N)$ in a fixed particle number sector defined as
\begin{equation}
E_{e}(N)=E_{1}(N)-E_{0}(N)
\end{equation}
under OBC are shown in Fig.~\ref{fig:gaps_ES}(b).
For both parameter regimes with $W=1$ and $W=0$ in phase diagram
Fig.~\ref{fig:finiteU_phase_diagram}(a) the excitation gaps $E_{e}(N)$
all scales as
\begin{equation}
E_{sp}(N)\sim\frac{1}{L},
\end{equation}
indicating no topological ground state degeneracy. The half
chain entanglement spectra without two-fold degeneracy are also
shown in Fig.~\ref{fig:gaps_ES}(c) in descending order.

\begin{figure}[t]
\centering
\includegraphics[width=\columnwidth]{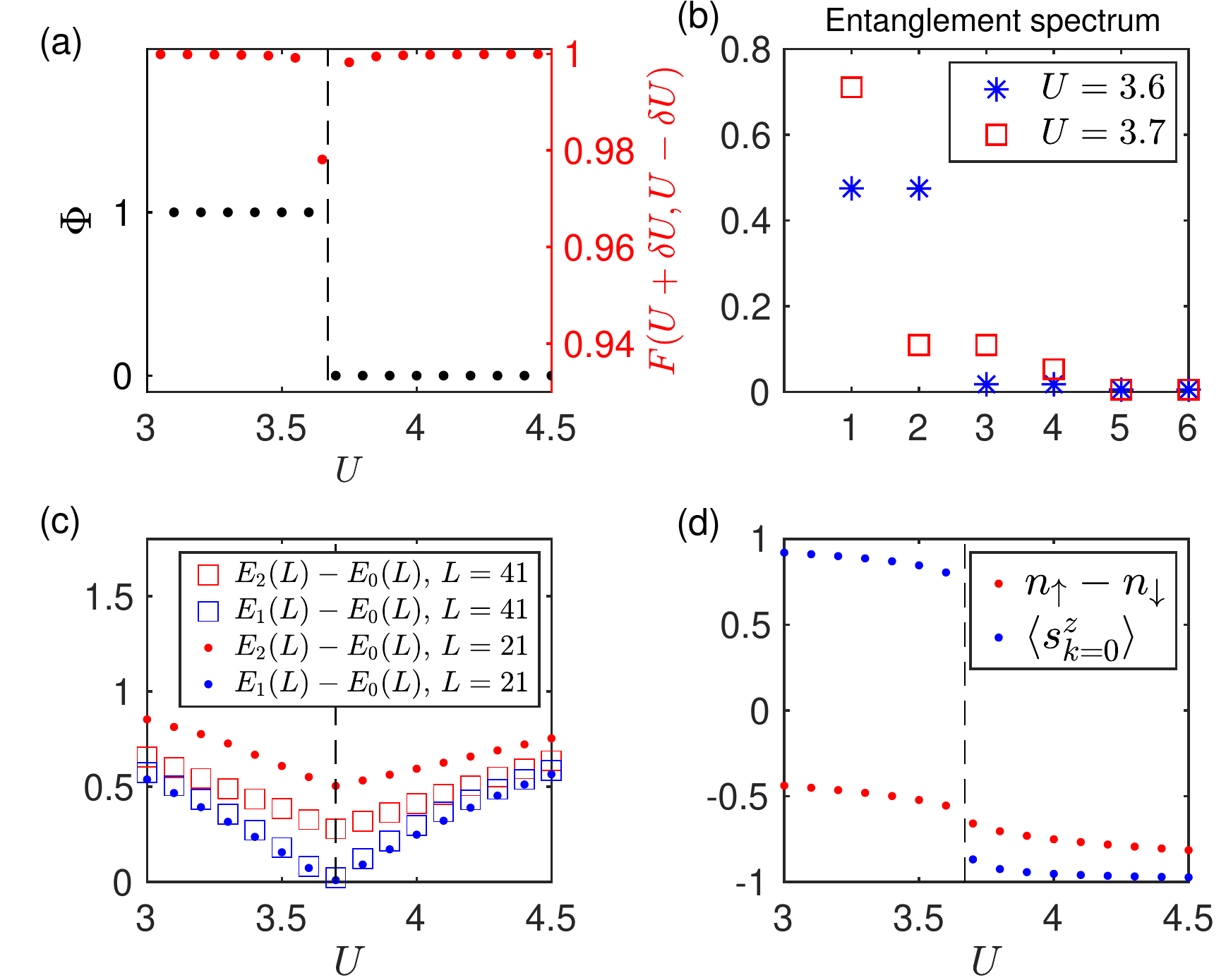}
\caption{In the gapped regime (half filling), the winding number can capture the conventional gapped SPT phase. Parameters are chosen as $(t_{0},t_{\rm so},\delta)=(1,1,1)$. The transition points are marked by black vertical dashed lines in (a), (c), and (d). (a), VUMPS results of SPT topological invariant $\Phi$ and averaged fidelity $F(U+\delta U,U-\delta U)$ as functions of $U$. The sudden change of $\Phi$ together with the fidelity dip imply a conventional SPT transition point (gap closing point) at $3.6<U_{c}<3.7$. (b), Nontrivial (trivial) entanglement spectrum before (after) gap closing point obtained from VUMPS. (c), Finite size scaling of excitation gap with fixed particle number under PBC. (d), Ground state spin polarization $\langle s_{k=0}^{z}\rangle$ (blue dots) and total magnetization $n_{\uparrow}-n_{\downarrow}$ (red dots) obtained from VUMPS. The sudden change in the sign of spin polarization indicates transition of winding number.}
\label{fig:fidelity_ES_gap}
\end{figure}

\subsection{The gapped $\nu=1$ regime: a concrete case study}

Finally we show that the half-filling regime gives the conventional gapped SPT phase. With a specific parameter condition we determine the SPT transition point, which is also the transition point of winding number.

We first introduce the gapped SPT phase in conventional characterization of this model. At half filling $\nu=1$ the ground state at $U=0$ is a gapped free fermion topological insulator, and the gap is still open at generic $U\neq0$ before topological transition. With interaction the topology can be characterized by the conventional gapped SPT framework, and the classification of SPT phases depends on which symmetry is considered. For example, the chiral symmetry is known to protect 1D fermionic topological phase and lead to $Z_4$ classification \cite{liu2013manipulating,z4,z4_2}. In this work, we consider another symmetry, i.e., the parity symmetry $P$ which can also protect the nontrivial gapped SPT phase. From the parity symmetry $P$ and the unique gapped ground state condition we derive a topological invariant $\Phi$ similar to that derived from inversion symmetry in Haldane phase \cite{pollmann2010entanglement}. Specifically, starting from the ground state MPS matrix $A^s$ and through parity symmetry invariant property of injective MPS (as a unique gapped ground state) \cite{perez2008string} one can find
\begin{align}
(\sum_{s'}u^{s,s'}A^{s'})^{T}&=e^{i\phi}R^{-\frac{1}{2}}\tilde{L}^{-1}A^s\tilde{L}R^{\frac{1}{2}},\notag\\
[\tilde{L},R^{\frac{1}{2}}]&=0,\notag\\
\tilde{L}\tilde{L}^{*}&=e^{-i\Phi},
\end{align}
where $u$ is an on-site unitary acting on physical Hilbert space, $\tilde{L}$ is a unitary acting on virtual bond of MPS, and $R$ is a positive-definite matrix acting on virtual bond whose eigenvalues correspond to bipartite entanglement spectrum. One can further obtain $\Phi=0$ or $\pi$ as a SPT topological invariant, where $\Phi=\pi$ corresponds to the nontrivial phase with $\tilde{L}^*=-\tilde{L}$ and  $\Phi=0$ corresponds to the trivial phase with $\tilde{L}^*=\tilde{L}$. Since the SPT topological invariant $\Phi$ only takes discrete values, without gap closing the value of $\Phi$ will not change. Most notably, the nontrivial $\Phi=\pi$ implies even-fold entanglement spectrum degeneracy due to $[\tilde{L},R^{\frac{1}{2}}]=0$ and the condition that $\tilde{L}$ is a skew-symmetric unitary matrix. Thus without gap closing the entanglement spectrum keeps even-fold degeneracy in the nontrivial phase. More details can be found in Appendix F.

For a specific parameter regime $(t_{0},t_{\rm so},\delta)=(1,1,1)$, we determine the SPT phase transition point (gap closing point) numerically. To detect the transition point, we calculate topological invariant $\Phi$, entanglement spectrum, excitation gap as well as averaged fidelity \cite{PhysRevLett.100.080601}
\begin{align}
F(U_{1},U_{2})= & \lim_{L\rightarrow+\infty}|\langle\psi(U_{1})|\psi(U_{2})\rangle|^{\frac{1}{L}},
\end{align}
where $|\psi(U)\rangle$ denotes VUMPS ground state with interaction $U$.
We increase $U$ from $U=0$ towards the large $U$ trivial phase (in large $U$ limit the ground state is spin polarized product state). Through the SPT topological invariant $\Phi$ and dip of fidelity, the SPT transition point is determined to be within $3.6<U_c<3.7$ as illustrated by VUMPS results in Fig.~\ref{fig:fidelity_ES_gap} (a). The degeneracy of entanglement spectrum in Fig.~\ref{fig:fidelity_ES_gap} (b) agrees with prediction of the topological invariant $\Phi$. The finite size scaling of excitation gap in Fig.~\ref{fig:fidelity_ES_gap} (c) confirms that the transition of SPT phases is due to closing of the gap.

Moreover, we study winding number transition of many-body spin texture in the same parameter regime $(t_{0},t_{\rm so},\delta)=(1,1,1)$. In Fig.~\ref{fig:fidelity_ES_gap} (d) one can see that the magnetization $n_{\uparrow}-n_{\downarrow}$ increases with $U$ continuously, while the spin polarization $\langle s_{k=0}^{z}\rangle$ has a sudden change from a positive value to a negative value. Therefore, the transition point for winding number of many-body spin texture coincides with the SPT transition point (gap closing point) in this study.

\section{Conclusion and discussion}

We have predicted an emergent gapless topological Luttinger liquid characterized by nontrivial many-body spin texture, which is a new paradigm of gapless topological phase beyond the conventional topological characterizations relying on partially gapped bulk degree of freedom and ground state degeneracy. 
We considered the 1D AIII class topological Hubbard model with fractional filling factor and found that the ground state, which has no gapped degree of freedom and is viewed as a trivial gapless phase in conventional characterization, exhibits intrinsic topological winding in its many-body spin texture. The topological transition is shown by tuning the interaction and filling factor, with a phase diagram being obtained, and the results have been extended to finite temperature regime. The existence of nontrivial topology and topological transition implies that low energy physics is insufficient to describe the gapless Luttinger liquid, instead the high-energy physics and global features in momentum space should be taken into account for a full characterization. Our results expand the theoretical framework for Luttinger liquids and gapless topological phases, and show that the high-energy physics may enrich quantum phases in the gapless systems.

With the novel new topological phenomena predicted in the bulk, a desirable open question is whether
the bulk winding number of the gapless Luttinger liquid is related to edge properties. While it has been shown that in gapless regime there is no ground state degeneracy in the case with nonzero winding number, we conjecture that the winding number in gapless regime
is related to degenerate high-energy edge excitations. For the noninteracting or mean-field Hamiltonian this conjecture is true: at $\nu<1$ and with nontrivial band topology, one can see that there exists high energy single-particle edge states above the Fermi level.
Nevertheless, with interaction the winding number also combines the correlated effects of particle-particle
scattering, and the concept of single-particle eigenstate is no longer applicable. One possible
way to confirm (or detect) edge excitations is to add one particle to the edge of the
many-body ground state, and observe the peak of spectral function obtained from
the return probability during time evolution.

\section*{ACKNOWLEDGMENTS}
This work was supported by National Natural Science Foundation of China (No. 11825401 and 11761161003), the Open Project of Shenzhen
Institute of Quantum Science and Engineering (Grant No.SIQSE202003), and the Strategic
Priority Research Program of Chinese Academy of Science (Grant No. XDB28000000).

\noindent

\renewcommand{\thesection}{S-\arabic{section}}
\setcounter{section}{0}  
\renewcommand{\theequation}{S\arabic{equation}}
\setcounter{equation}{0}  
\renewcommand{\thefigure}{S\arabic{figure}}
\setcounter{figure}{0}  

\indent

\section*{\Large\bf APPENDIX}

\subsection{Symmetries of Hamiltonian and spin texture}
Here we introduce the symmetries of the Hamiltonian to determine the symmetries of observable many-body spin texture. We use spin reflection symmetry $R_{x}$ to show that the momentum space spin direction is always in $y-z$ plane, and use parity symmetry $P$ to show that $\langle s_{k}^{y/z}\rangle$ are odd/even functions in momentum space, respectively. In the present work we don't consider spontaneous breaking of these symmetries, so the expectation values of spin texture are defined from ensemble average of all ground states if ground states are degenerate.
\subsubsection{Spin reflection symmetry $R_{x}$}

In the Bloch Hamiltonian of $H_0$ only $\tau_y$ and $\tau_z$ appeared, so we define the second quantized spin reflection operator $R_{x}$ that commute with $s_{k}^{y/z}$ and anticommute with $s_{k}^{x}$. The spin operators in position and momentum spaces are defined as:
\begin{align}
s_{i}^{x/y/z} & =\psi_i^{\dagger}\tau_{x/y/z}\psi_i,\notag\\
s_{k}^{x/y/z} & =\psi_k^{\dagger}\tau_{x/y/z}\psi_k,
\end{align}
and the time-reversal symmetry is defined as
\begin{align}
T & =e^{-i\pi/2\sum_{i}s_{i}^{y}}K=e^{-i\pi/2\sum_{k}s_{k}^{y}}K.
\end{align}
Consider the spin operators transformed by $T$:
\begin{align}
Ts_{i}^{x/y/z}T^{-1} & =-s_{i}^{x/y/z},\notag\\
Ts_{k}^{x/y/z}T^{-1} & =-s_{-k}^{x/y/z}.
\end{align}
If we combine $T$ with bond-centered inversion symmetry $I$ that
satisfies
\begin{align}
Ic_{i,\sigma}^{\dagger}c_{j,\sigma}I^{-1} & =c_{-i+1,\sigma}^{\dagger}c_{-j+1,\sigma},\notag\\
Ic_{i,\sigma}^{\dagger}c_{j,\bar{\sigma}}I^{-1} & =c_{-i+1,\sigma}^{\dagger}c_{-j+1,\bar{\sigma}},
\end{align}
we obtain
\begin{equation}
ITs_{k}^{x/y/z}T^{-1}I^{-1}=-s_{k}^{x/y/z}.
\end{equation}
Thus the spin reflection symmetry $R_{x}$ is defined as
\begin{equation}
R_{x}=e^{-i\pi/2\sum_{i}s_{i}^{x}}IT.
\end{equation}
$R^{x}$ is a symmetry of the Hamiltonian and satisfies
\begin{align}
R_{x}HR_{x}^{-1} & =H,\notag\\
R_{x}s_{k}^{x}R_{x}^{-1} & =-s_{k}^{x},\notag\\
R_{x}s_{k}^{y}R_{x}^{-1} & =s_{k}^{y},\notag\\
R_{x}s_{k}^{z}R_{x}^{-1} & =s_{k}^{z}.
\end{align}
Thus the momentum space spin polarization in $x$ direction vanishes
\begin{equation}
\langle s_{k}^{x}\rangle=0,
\end{equation}
and spin polarization is always in $y-z$ plane. We point out this result can also be derived from the chiral (sublattice) symmetry. Since the effects of these symmetries on spin textures are the same, we only present the spin reflection symmetry $R_{x}$.

\subsubsection{Parity symmetry $P$}

The Hamiltonian also satisfies the parity symmetry $PHP^{-1}=H$, where $P$ is a combination of an on-site unitary and inversion symmetries
\begin{align}
P & =(-1)^{\sum_{i}n_{i,\downarrow}}I=(-1)^{\sum_{k}n_{k,\downarrow}}I,
\end{align}
and satisfies
\begin{align}
Pc_{i,\sigma}^{\dagger}c_{j,\sigma}P^{-1} & =c_{-i+1,\sigma}^{\dagger}c_{-j+1,\sigma},\notag\\
Pc_{i,\sigma}^{\dagger}c_{j,\bar{\sigma}}P^{-1} & =-c_{-i+1,\sigma}^{\dagger}c_{-j+1,\bar{\sigma}}.
\end{align}
It's easy to check that the expectation value of spin textures satisfy
\begin{align}
\langle s_{k}^{y}\rangle & =\langle P^{-1}s_{k}^{y}P\rangle=-\langle s_{-k}^{y}\rangle,\notag\\
\langle s_{k}^{z}\rangle & =\langle P^{-1}s_{k}^{z}P\rangle=\langle s_{-k}^{z}\rangle.
\end{align}
Thus at high symmetry momenta $\langle s_{k=0,\pi}^{y}\rangle=0$.

\subsubsection{Hamiltonian and parity operator in Jordan-Wigner representation}

For numerical calculation of physical quantities and analysis of gapped SPT phase
in Sec. V. B, it is necessary to write the fermionic Hamiltonian and
parity operator $P$ in Jordan-Wigner representation. The Jordan-Wigner
transformation is defined as
\begin{align}
d_{j} & =(\prod_{j'<j}\sigma{}_{j'}^{z})\frac{\sigma_{j}^{x}+i\sigma_{j}^{y}}{2},\notag\\
d_{j}^{\dagger} & =\frac{\sigma_{j}^{x}-i\sigma_{j}^{y}}{2}(\prod_{j'<j}\sigma{}_{j'}^{z}),
\end{align}
where $\sigma^{x/y/z}$ denotes Pauli matrices which commute at different sites, $d_{j}^{\dagger}$ denotes spinless fermions and is related to
original fermion operators via
\begin{align}
c_{j,\uparrow}^{\dagger} & =d_{2j-1}^{\dagger},\notag\\
c_{j,\downarrow}^{\dagger} & =d_{2j}^{\dagger}.
\end{align}
The Hamiltonian written with the $d_{j}^{\dagger}$ fermions has the
form
\begin{align}
H= & -\sum_{j}[(t+t_{0})d_{2j-1}^{\dagger}d_{2j+1}+(t-t_{0})d_{2j}^{\dagger}d_{2j+2})+h.c.]\notag\\
 & +t_{\rm so}[\sum(d_{2j-1}^{\dagger}d_{2j+2}-d_{2j}^{\dagger}d_{2j+1})+h.c.]\notag\\
 & +\delta\sum_{j}(n_{2j-1}-n_{2j})\notag\\
 & +U\sum_{j}n_{2j-1}n_{2j}.
\end{align}
Define $\sigma_{j}^{\pm}=\frac{\sigma_{j}^{x}\pm i\sigma_{j}^{y}}{2}$,
then we write the Hamiltonian with tensor product of local Pauli matrix
as
\begin{align}
H= & \sum_{j}[(-t-t_{0})\sigma_{2j-1}^{-}\sigma_{2j}^{z}\sigma_{2j+1}^{+}\notag\\
 & +(-t+t_{0})\sigma_{2j}^{-}\sigma_{2j+1}^{z}\sigma_{2j+2}^{+})+h.c.]\notag\\
 & +t_{\rm so}[\sum_{j}(\sigma_{2j-1}^{-}\sigma_{2j}^{z}\sigma_{2j+1}^{z}\sigma_{2j+2}^{+}-\sigma_{2j}^{-}\sigma_{2j+1}^{+})+h.c.]\notag\\
 & +\delta\sum_{j}(n_{2j-1}-n_{2j})+U\sum_{j}n_{2j-1}n_{2j}.
\end{align}
The form of parity operator $P$ in Jordan-Wigner representation has
the form
\begin{align}
P & =\tilde{I}(\otimes_{i}u_{i}),
\end{align}
where $u_{i}=\text{diag}([1,1,-1,1])$ acts on $4$-dimensional local
Hilbert space of site $i$, with the local bases being $[|0\rangle,|\uparrow\rangle,|\downarrow\rangle,|\uparrow\downarrow\rangle]$.
The $\tilde{I}$ is the inversion operator in Jordan-Wigner representation
that satisfy
\begin{align}
\tilde{I}(\otimes_{i}M_{i})\tilde{I}^{-1}= & \tilde{I}(\cdot\cdot\cdot\otimes M_{i-1}\otimes M_{i}\otimes M_{i+1}\otimes\cdot\cdot\cdot)\tilde{I}^{-1}\notag\\
 & =\cdot\cdot\cdot\otimes M_{-i+1}\otimes M_{-i}\otimes M_{-i-1}\otimes\cdot\cdot\cdot,
\end{align}
where $M_{i}$ is any local operator on $4$-dimensional local Hilbert
space of site $i$. From the relation
\begin{align}
u^{2} & =id\otimes id,\notag\\
u(\sigma^{-}\otimes\sigma^{z}) & =-(\sigma^{-}\otimes id)u,\notag\\
u(\sigma^{+}\otimes\sigma^{z}) & =-(\sigma^{+}\otimes id)u,\notag\\
u(\sigma^{z}\otimes\sigma^{-}) & =(id\otimes\sigma^{-})u,\notag\\
u(\sigma^{z}\otimes\sigma^{+}) & =(id\otimes\sigma^{+})u,\notag\\
u(\sigma^{z}\otimes\sigma^{z}) & =(\sigma^{z}\otimes\sigma^{z})u,\notag\\
u(id\otimes id) & =(id\otimes id)u,
\end{align}
we find
\begin{align}
Pc_{i,\sigma}^{\dagger}c_{i\pm1,\sigma}P^{-1} & =c_{-i+1,\sigma}^{\dagger}c_{-i+1\mp1,\sigma},\notag\\
Pc_{i,\bar{\sigma}}^{\dagger}c_{i\pm1,\sigma}P^{-1} & =-c_{-i+1,\bar{\sigma}}^{\dagger}c_{-i+1\mp1,\sigma},
\end{align}
therefore we have verified $P=\tilde{I}(\otimes_{i}u_{i})$ is indeed
the correct form of parity operator $P$ in Jordan-Wigner representation.

\subsection{Details of the PMF approximation at weak $U$}

\subsubsection{Derivations of spin textures}

In this part we treat MF Hamiltonian $H_{mf}=  \sum_{k}(\epsilon_{k}^{+}d_{k,+}^{\dagger}d_{k,+}+\epsilon_{k}^{-}d_{k,-}^{\dagger}d_{k,-})$ as the unperturbed Hamiltonian, and treat $V=H-H_{mf}=U\sum_{i}n_{i,\uparrow}n_{i,\downarrow}  +\sum_{i}\frac{Um}{2}(n_{i\uparrow}-n_{i\downarrow})-\sum_{i}U\frac{\nu^{2}+m^{2}}{4}$ as the perturbation. Since for $k_{0}$ points outside Fermi sea the unperturbed MF ground state has no particle occupation, the leading order contribution to spin texture observables is at order $O(U^{2})$ order, which requires first order perturbed many-body wavefunction. In standard perturbation theory the perturbed
many-body ground state wavefunction has the form
\begin{widetext}
\begin{align}
|n\rangle= & |n^{(0)}\rangle+\sum_{m\neq n}\frac{V_{mn}|m^{(0)}\rangle}{(E_{n}^{(0)}-E_{m}^{(0)})}-\sum_{m\neq n}\frac{V_{nn}V_{mn}|m^{(0)}\rangle}{(E_{n}^{(0)}-E_{m}^{(0)})^{2}}
 +\sum_{m\neq n}\sum_{l\neq n}\frac{V_{ml}V_{ln}|m^{(0)}\rangle}{(E_{n}^{(0)}-E_{m}^{(0)})(E_{n}^{(0)}-E_{l}^{(0)})} \notag\\
 &-\sum_{l\neq n}\frac{V_{nl}V_{ln}|n^{(0)}\rangle}{(E_{n}^{(0)}-E_{l}^{(0)})^{2}},\label{eq:perturbation_psi}
\end{align}
where $|m\rangle$ denotes the perturbed eigenstates, $|m^{(0)}\rangle$
denotes unperturbed eigenstates (in Fock space), $E_{m}^{(0)}$ denotes the unperturbed
energy (of $H_{mf}$), $V_{ml}=\langle m^{(0)}|V|l^{(0)}\rangle$ denotes the matrix element
of perturbation term $V=H-H_{mf}$.

Since the unperturbed $H_{mf}$ can be diagonalized in mpmentum space, we write the MF correction of Zeeman term and Hubbard interaction to momentum space as
\begin{align}
U\sum_{i}n_{i,\uparrow}n_{i,\downarrow}= & \frac{U}{L}\sum_{k,q,q'}c_{q\uparrow}^{\dagger}c_{q'\downarrow}^{\dagger}c_{-k+q+q'\downarrow}c_{k\uparrow}\notag\\
= & \frac{U}{L}\sum_{k,q,q'}(\alpha_{q}^{*}d_{q,-}^{\dagger}-\beta_{q}d_{q,+}^{\dagger})(\beta_{q'}^{*}d_{q',-}^{\dagger}+\alpha_{q'}d_{q',+}^{\dagger})\notag\\
 & \times(\beta_{-k+q+q'}d_{-k+q+q',-}+\alpha_{-k+q+q'}^{*}d_{-k+q+q',+})\notag\\
 & \times(\alpha_{k}d_{k,-}-\beta_{k}^{*}d_{k,+}),\notag\\
\sum_{i}\frac{Um}{2}(n_{i\uparrow}-n_{i\downarrow})= & \frac{Um}{2}\sum_{k}\{(\beta_{k}\beta_{k}^{*}-\alpha_{k}\alpha_{k}^{*})(d_{k,+}^{\dagger}d_{k,+}-d_{k,-}^{\dagger}d_{k,-})\notag\\
 & +2\alpha_{k}\beta_{k}(d_{k,-}^{\dagger}d_{k,+}-d_{k,+}^{\dagger}d_{k,-})\}.
 \label{eq:HI_quad}
\end{align}
Note that in above formula $d_{k,\pm}^{\dagger},\alpha_{k},\beta_{k}$
are determined by diagonalizing $H_{mf}$ with $\tilde{\delta}=\delta-Um/2$.
Here $\alpha_{k}$ and $\beta_{k}$ have be chosen as imaginary odd
function and real even function in FBZ, respectively.

With the many-body perturbed ground state obtained from Eq. (\ref{eq:perturbation_psi}), it is convenient to first evaluate the relative spin textures $\langle\mathcal{S}_{k_{0}}^{y/z}\rangle$ in the rotating frame defined in Eq. (\ref{eq:basis_change}). The expectation values $\langle\mathcal{S}_{k_{0}}^{y/z}\rangle$
from $0$th to $2$nd order are calculated from
\begin{align}
\langle\mathcal{S}_{k_{0}}^{y/z}\rangle_{0th} & =\langle n^{(0)}|\mathcal{S}_{k_{0}}^{y/z}|n^{(0)}\rangle,\notag\\
\langle\mathcal{S}_{k_{0}}^{y/z}\rangle_{1st} & =\sum_{m\neq n}\frac{V_{mn}^{*}\langle m^{(0)}|\mathcal{S}_{k_{0}}^{y/z}|n^{(0)}\rangle}{(E_{n}^{(0)}-E_{m}^{(0)})}+h.c.,\notag\\
\langle\mathcal{S}_{k_{0}}^{y/z}\rangle_{2nd} & =\sum_{m\neq n}\frac{V_{nm}V_{mn}}{(E_{n}^{(0)}-E_{m}^{(0)})^{2}}(\langle m{}^{(0)}|\mathcal{S}_{k_{0}}^{y/z}|m^{(0)}\rangle-\langle n^{(0)}|\mathcal{S}_{k_{0}}^{y/z}|n^{(0)}\rangle)\notag\\
 & +\sum_{m'\neq n}\sum_{m\neq n,m\neq m'}\frac{V_{m'n}^{*}V_{mn}}{(E_{n}^{(0)}-E_{m'}^{(0)})(E_{n}^{(0)}-E_{m}^{(0)})}\langle m'^{(0)}|\mathcal{S}_{k_{0}}^{y/z}|m^{(0)}\rangle \notag\\
 & +\{\sum_{m\neq n}\sum_{l\neq n}\frac{V_{ml}^{*}V_{ln}^{*}}{(E_{n}^{(0)}-E_{m}^{(0)})(E_{n}^{(0)}-E_{l}^{(0)})}-\sum_{m\neq n}\frac{V_{nn}^{*}V_{mn}^{*}}{(E_{n}^{(0)}-E_{m}^{(0)})^{2}}\}\langle m^{(0)}|\mathcal{S}_{k_{0}}^{y/z}|n^{(0)}\rangle+h.c..
\end{align}

The zeroth order expectation value is simply
\begin{align}
\langle\mathcal{S}_{k_{0}}^{y}\rangle_{0th}= & 0,\notag\\
\langle\mathcal{S}_{k_{0}}^{z}\rangle_{0th}= & -f_{k_{0},-}.
\end{align}
where $f_{k,\pm}$ is the Fermi distribution of MF ground
state in upper/lower subbands, and $f_{k,+}=0$ is satisfied in the global FBZ.

The first order expectation value vanishes
\begin{equation}
\langle\mathcal{S}_{k_{0}}^{y}\rangle_{1st}=\langle\mathcal{S}_{k_{0}}^{z}\rangle_{1st}=0
\end{equation}
due to the fact that $V_{mn}^{*}\langle m^{(0)}|\mathcal{S}_{k_{0}}^{y/z}|n^{(0)}\rangle$
vanishes for both $\mathcal{S}_{k_{0}}^{z}$ and $\mathcal{S}_{k_{0}}^{y}$. To show $V_{mn}^{*}\langle m^{(0)}|\mathcal{S}_{k_{0}}^{z}|n^{(0)}\rangle=0$, only $m=n$ terms need to be considered and one can find $V_{nn}=0$ ($n$ labels the ground state) due to MF self-consistent condition. To show that $V_{mn}^{*}\langle m^{(0)}|\mathcal{S}_{k_{0}}^{y}|n^{(0)}\rangle$, one can take out the quadratic contributions $H_{I}^{quad}$ from Hubbard interaction that create single
particle excitations of the form $d_{k_{0},+}^{\dagger}d_{k_{0},-}|n^{(0)}\rangle$, i.e.,
\begin{align}
H_{I}^{quad}= & \frac{U}{L}\sum_{q,k_{0}}((\alpha_{q}^{*}\alpha_{q}-\beta_{q}^{*}\beta_{q})\alpha_{k_{0}}\beta_{k_{0}}d_{k_{0},+}^{\dagger}d_{k_{0},-}n_{q,-}+h.c.\notag\\
= & \frac{U}{L}\sum_{q,k_{0}}((\alpha_{q}^{*}\alpha_{q}-\beta_{q}^{*}\beta_{q})\alpha_{k_{0}}\beta_{k_{0}}d_{k_{0},+}^{\dagger}d_{k_{0},-}f_{q,-}+h.c.\notag\\
= & 2Um\sum_{k_{0}}\alpha_{k_{0}}\beta_{k_{0}}d_{k_{0},+}^{\dagger}d_{k_{0},-}+h.c..
\end{align}
Note that contributions of the $H_{I}^{quad}$ term cancels with contributions of the Zeeman correction
term in Eq. (\ref{eq:HI_quad}) when evaluating $V_{mn}$.

The second order contributions of $\langle\mathcal{S}_{k_{0}}^{y}\rangle_{2nd}$ and $\langle\mathcal{S}_{k_{0}}^{z}\rangle_{2nd}$ are also obtained after tedious but straightforward calculation. The expression for $\langle\mathcal{S}_{k_{0}}^{y}\rangle_{2nd}$ has the form
\begin{align}
\langle\mathcal{S}_{k_{0}}^{y}\rangle_{2nd}= & -i\langle d_{k_{0},+}^{\dagger}d_{k_{0},-}\rangle_{2nd}+h.c.,\notag\\
\langle d_{k_{0},+}^{\dagger}d_{k_{0},-}\rangle_{2nd}= & (\frac{U}{L})^{2}\sum_{q>q'}\frac{|\alpha_{q}\beta_{q'}-\beta_{q}\alpha_{q'}|^{2}\{\alpha_{k_{0}}^{*}\alpha_{-k_{0}+q+q'}+\beta_{k_{0}}^{*}\beta_{-k_{0}+q+q'}\}(\beta_{k_{0}}^{*}\alpha_{-k_{0}+q+q'}^{*}-\alpha_{k_{0}}^{*}\beta_{-k_{0}+q+q'}^{*})}{(\epsilon_{k_{0}}^{+}-\epsilon_{k_{0}}^{-})(\epsilon_{q}^{-}+\epsilon_{q'}^{-}-\epsilon_{k_{0}}^{-}-\epsilon_{-k_{0}+q+q'}^{-})} \notag\\
 & \times(1-f_{q,-})(1-f_{q',-})f_{-k_{0}+q+q',-}f_{k_{0},-}.\notag\\
+ & (\frac{U}{L})^{2}\sum_{q>q'}\frac{|\alpha_{q}\beta_{q'}-\beta_{q}\alpha_{q'}|^{2}(\beta_{k_{0}}^{*}\alpha_{-k_{0}+q+q'}^{*}-\alpha_{k_{0}}^{*}\beta_{-k_{0}+q+q'}^{*})\{\alpha_{k_{0}}^{*}\alpha_{-k_{0}+q+q'}+\beta_{k_{0}}^{*}\beta_{-k_{0}+q+q'}\}}{(\epsilon_{k_{0}}^{+}-\epsilon_{k_{0}}^{-})(\epsilon_{q}^{+}+\epsilon_{q'}^{+}-\epsilon_{k_{0}}^{-}-\epsilon_{-k_{0}+q+q'}^{-})}\notag\\
 & \times(1-f_{q,+})(1-f_{q',+})f_{-k_{0}+q+q',-}f_{k_{0},-}\notag\\
+ & (\frac{U}{L})^{2}\sum_{q,q'}\frac{|\alpha_{q,}^{*}\alpha_{q'}+\beta_{q,}\beta_{q'}^{*}|^{2}(\beta_{k_{0}}^{*}\alpha_{-k_{0}+q+q}^{*}-\alpha_{k_{0}}^{*}\beta_{-k_{0}+q+q}^{*})\{\alpha_{k_{0}}^{*}\alpha_{-k_{0}+q+q'}+\beta_{k_{0}}^{*}\beta_{-k_{0}+q+q'}\}}{(\epsilon_{k_{0}}^{+}-\epsilon_{k_{0}}^{-})(\epsilon_{q}^{+}+\epsilon_{q'}^{-}-\epsilon_{k_{0}}^{-}-\epsilon_{-k_{0}+q+q'}^{-})}\notag\\
 & \times(1-f_{q,+})(1-f_{q',-})f_{-k_{0}+q+q',-}f_{k_{0},-}\notag\\
+ & (\frac{U}{L})^{2}\sum_{q>q'}\frac{|\alpha_{q}\beta_{q'}-\beta_{q}\alpha_{q'}|^{2}(\alpha_{-k_{0}+q+q'}^{*}\beta_{k_{0}}^{*}-\beta_{-k_{0}+q+q'}^{*}\alpha_{k_{0}}^{*})\{\alpha_{-k_{0}+q+q',}\alpha_{k_{0}}^{*}+\beta_{-k_{0}+q+q',}^{*}\beta_{k_{0}}\}}{(\epsilon_{k_{0}}^{+}-\epsilon_{k_{0}}^{-})(\epsilon_{k_{0}}^{+}+\epsilon_{-k_{0}+q+q'}^{-}-\epsilon_{q}^{-}-\epsilon_{q'}^{-})}\notag\\
 & \times(1-f_{k_{0},+})f_{k_{0},-}(1-f_{-k_{0}+q+q',-})f_{q,-}f_{q',-}\notag\\
- & (\frac{U}{L})^{2}\sum_{q>q'}\frac{|\alpha_{q}\beta_{q'}-\beta_{q}\alpha_{q'}|^{2}(\beta_{k_{0}}^{*}\alpha_{-k_{0}+q+q'}^{*}-\alpha_{k_{0}}^{*}\beta_{-k_{0}+q+q'}^{*})\{\alpha_{k_{0}}^{*}\alpha_{-k_{0}+q+q'}+\beta_{k_{0}}^{*}\beta_{-k_{0}+q+q'}\}}{(\epsilon_{k_{0}}^{+}-\epsilon_{k_{0}}^{-})(\epsilon_{k_{0}}^{+}+\epsilon_{-k_{0}+q+q'}^{+}-\epsilon_{q}^{-}-\epsilon_{q'}^{-})}\notag\\
 & \times(1-f_{k_{0},+})f_{k_{0},-}(1-f_{-k_{0}+q+q',+})f_{q,-}f_{q',-}\notag\\
+ & (\frac{U}{L})^{2}\sum_{q>q'}\frac{|\alpha_{q}\beta_{q'}-\beta_{q}\alpha_{q'}|^{2}\{\alpha_{-k_{0}+q+q'}\alpha_{k_{0}}^{*}+\beta_{-k_{0}+q+q'}\beta_{k_{0}}^{*}\}(\alpha_{-k_{0}+q+q',}^{*}\beta_{k_{0}}^{*}-\beta_{-k_{0}+q+q'}^{*}\alpha_{k_{0}}^{*})}{(\epsilon_{q}^{-}+\epsilon_{q'}^{-}-\epsilon_{k_{0}}^{-}-\epsilon_{-k_{0}+q+q'}^{-})(\epsilon_{q}^{-}+\epsilon_{q'}^{-}-\epsilon_{k_{0}}^{+}-\epsilon_{-k_{0}+q+q'}^{-})}\notag\\
 & \times(1-f_{k_{0},-})(1-f_{-k_{0}+q+q',-})f_{q,-}f_{q',-}\notag\\
- & (\frac{U}{L})^{2}\sum_{q>q'}\frac{|\alpha_{q}\beta_{q'}-\beta_{q}\alpha_{q'}|^{2}\{\alpha_{-k_{0}+q+q'}\alpha_{k_{0}}^{*}+\beta_{-k_{0}+q+q'}\beta_{k_{0}}^{*}\}(\alpha_{-k_{0}+q+q',}^{*}\beta_{k_{0}}^{*}-\beta_{-k_{0}+q+q'}^{*}\alpha_{k_{0}}^{*})}{(\epsilon_{q}^{-}+\epsilon_{q'}^{-}-\epsilon_{k_{0}}^{-}-\epsilon_{-k_{0}+q+q'}^{+})(\epsilon_{q}^{-}+\epsilon_{q'}^{-}-\epsilon_{k_{0}}^{+}-\epsilon_{-k_{0}+q+q'}^{+})}\notag\\
 & \times(1-f_{k_{0},-})(1-f_{-k_{0}+q+q',+})f_{q,-}f_{q',-}.
\end{align}
In the above expression of $\langle d_{k_{0},+}^{\dagger}d_{k_{0},-}\rangle_{2nd}$, the first five terms come from perturbation formula
\begin{equation}
 \sum_{m\neq n}\sum_{l\neq n}\frac{V_{ml}^{*}V_{ln}^{*}}{(E_{n}^{(0)}-E_{m}^{(0)})(E_{n}^{(0)}-E_{l}^{(0)})}\langle m^{(0)}|\mathcal{S}_{k_{0}}^{y/z}|n^{(0)}\rangle
\end{equation}
with $k_0$ inside Fermi sea, while the last two terms come from perturbation formula
\begin{align}
\sum_{m'\neq n}\sum_{m\neq n,m\neq m'}\frac{V_{m'n}^{*}V_{mn}}{(E_{n}^{(0)}-E_{m'}^{(0)})(E_{n}^{(0)}-E_{m}^{(0)})}\langle m'^{(0)}|\mathcal{S}_{k_{0}}^{y/z}|m^{(0)}\rangle
\end{align}
with $k_0$ outside Fermi sea. The expression for $\langle\mathcal{S}_{k_{0}}^{z}\rangle_{2nd}$ has the form
\begin{align}
\langle\mathcal{S}_{k_{0}}^{z}\rangle_{2nd}= & (\frac{U}{L})^{2}\sum_{q,q'}\frac{|(\alpha_{q}^{*}\alpha_{q'}+\beta_{q}^{*}\beta_{q'})(\beta_{-k_{0}+q+q'}\alpha_{k_{0}}-\alpha_{-k_{0}+q+q'}\beta_{k_{0}})|^{2}(1-f_{q,-})(1-f_{q',+})f_{-k_{0}+q+q',-}f_{k_{0},-}}{(\epsilon_{q}^{-}+\epsilon_{q'}^{+}-\epsilon_{k_{0}}^{-}-\epsilon_{-k_{0}+q+q'}^{-})^{2}}\notag\\
+ & (\frac{U}{L})^{2}\sum_{q>q'}\frac{|(\alpha_{q}\beta_{q'}-\alpha_{q'}\beta_{q})(\beta_{-k_{0}+q+q'}\alpha_{k_{0}}-\alpha_{-k_{0}+q+q',}\beta_{k_{0}})|^{2}(1-f_{q,-})(1-f_{q',-})f_{-k_{0}+q+q',-}f_{k_{0},-}}{(\epsilon_{q}^{-}+\epsilon_{q'}^{-}-\epsilon_{k_{0}}^{-}-\epsilon_{-k_{0}+q+q'}^{-})^{2}}\notag\\
+ & (\frac{U}{L})^{2}\sum_{q>q'}\frac{|(\alpha_{q}\beta_{q'}-\alpha_{q'}\beta_{q})(\beta_{-k_{0}+q+q'}\alpha_{k_{0}}-\alpha_{-k_{0}+q+q',}\beta_{k_{0}})|^{2}(1-f_{q,+})(1-f_{q',+})f_{-k_{0}+q+q',-}f_{k_{0},-}}{(\epsilon_{q}^{+}+\epsilon_{q'}^{+}-\epsilon_{k_{0}}^{-}-\epsilon_{-k_{0}+q+q'}^{-})^{2}}\notag\\
+ & (\frac{U}{L})^{2}\sum_{q>q'}\frac{|(\alpha_{q}\beta_{q'}-\alpha_{q'}\beta_{q})(\beta_{-k_{0}+q+q'}\alpha_{k_{0}}-\alpha_{-k_{0}+q+q',}\beta_{k_{0}})|^{2}(1-f_{-k_{0}+q+q',+})(1-f_{k_{0},+})f_{q,-}f_{q',-}}{(\epsilon_{q}^{-}+\epsilon_{q'}^{-}-\epsilon_{k_{0}}^{+}-\epsilon_{-k_{0}+q+q'}^{+})^{2}}\notag\\
- & (\frac{U}{L})^{2}\sum_{q>q'}\frac{|(\alpha_{q}\beta_{q'}-\alpha_{q'}\beta_{q})(\beta_{-k_{0}+q+q'}\alpha_{k_{0}}-\alpha_{-k_{0}+q+q',}\beta_{k_{0}})|^{2}(1-f_{-k_{0}+q+q',-})(1-f_{k_{0},-})f_{q,-}f_{q',-}}{(\epsilon_{q}^{-}+\epsilon_{q'}^{-}-\epsilon_{k_{0}}^{-}-\epsilon_{-k_{0}+q+q'}^{-})^{2}}\notag\\
- & (\frac{U}{L})^{2}\sum_{q>q'}\frac{|(\alpha_{q}\beta_{q'}-\alpha_{q'}\beta_{q})(\alpha_{-k_{0}+q+q'}^{*}\alpha_{k_{0}}+\beta_{-k_{0}+q+q'}^{*}\beta_{k_{0}})|^{2}(1-f_{-k_{0}+q+q',+})(1-f_{k_{0},-})f_{q,-}f_{q',-}}{(\epsilon_{q}^{-}+\epsilon_{q'}^{-}-\epsilon_{k_{0}}^{-}-\epsilon_{-k_{0}+q+q'}^{+})^{2}}\notag\\
+ & (\frac{U}{L})^{2}\sum_{q>q'}\frac{|(\alpha_{q}\beta_{q'}-\alpha_{q'}\beta_{q})(\alpha_{-k_{0}+q+q'}^{*}\alpha_{k_{0}}+\beta_{-k_{0}+q+q'}^{*}\beta_{k_{0}})|^{2}(1-f_{-k_{0}+q+q',-})(1-f_{k_{0},+})f_{q,-}f_{q',-}}{(\epsilon_{q}^{-}+\epsilon_{q'}^{-}-\epsilon_{k_{0}}^{+}-\epsilon_{-k_{0}+q+q'}^{-})^{2}}.
\end{align}
In above expression for $\langle\mathcal{S}_{k_{0}}^{z}\rangle_{2nd}$ all terms come from perturbation formula
\begin{equation}
\sum_{m\neq n}\frac{V_{nm}V_{mn}}{(E_{n}^{(0)}-E_{m}^{(0)})^{2}}(\langle m{}^{(0)}|\mathcal{S}_{k_{0}}^{y/z}|m^{(0)}\rangle-\langle n^{(0)}|\mathcal{S}_{k_{0}}^{y/z}|n^{(0)}\rangle)\notag\\,
\end{equation}
and the last four terms contribute to the case where $k_0$ is outside Fermi sea. When we perform numerical integral using above perturbation expressions, we slightly
modify the dispersion $\epsilon_{k,-}$ to avoid the divergence at
Fermi points by slightly modify the dispersion
\begin{equation}
\epsilon_{k,-}\rightarrow\epsilon_{k,-}-\delta_{\epsilon}f_{k,-},
\end{equation}
where $\delta_{\epsilon}=0.01t_{0}$ is used. This approximation has
little effects on spin directions of momentum points whose energy are away
from Fermi level $\epsilon_{F}$. Finally, the spin textures $\langle s_{k}^{y/z}\rangle $
in original spin up/down bases are obtained by the inverse transformation
\begin{align}
\langle s_{k}^{y}\rangle= & i(\alpha_{k}^{*}\beta_{k}-\beta_{k}^{*}\alpha_{k})\langle\mathcal{S}_{k}^{z}\rangle-(\alpha_{k}^{2}+\beta_{k}^{2})\langle\mathcal{S}_{k}^{y}\rangle, \notag\\
\langle s_{k}^{z}\rangle= & (|\beta_{k}|^{2}-|\alpha_{k}|^{2})\langle\mathcal{S}_{k}^{z}\rangle-2i\alpha_{k}\beta_{k}\langle\mathcal{S}_{k}^{y}\rangle,
\end{align}

\end{widetext}

\subsubsection{Detailed analysis of PMF results}

\begin{figure}[t]
\centering
\includegraphics[width=\columnwidth]{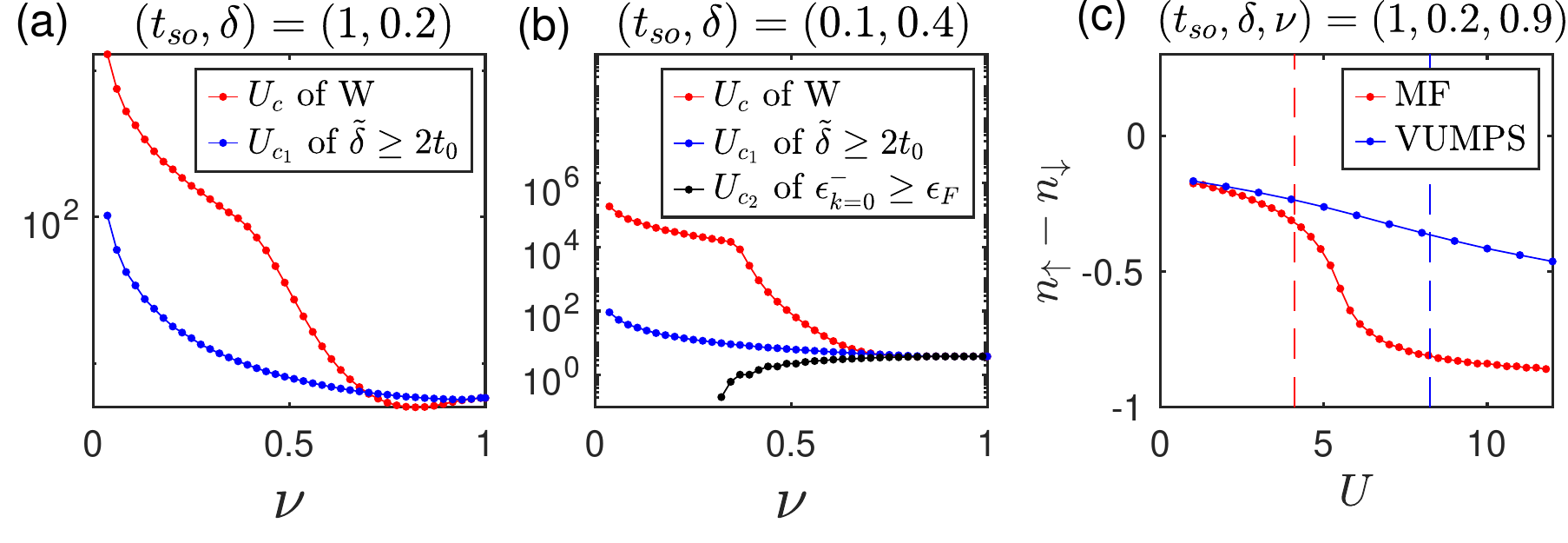}
\caption{Properties and weaknesses of the MF and PMF solutions for topological transitions. (a)-(b), In PMF results, the critical $U$ for topological transition exists at at any filling factor $\nu$, which is qualitatively incorrect in large $U$ limit at small $\nu$. Red curves correspond to critical value $U_{c}$
of the winding number $W$ obtained from PMF, blue curves
correspond to critical value $U_{c_{1}}$ at which the MF renormalized Zeeman field $\tilde{\delta}$ exceeds $2t_{0}$, black curve corresponds to critical value $U_{c_{2}}$ at which MF dispersion energy $\epsilon_{k=0}^{-}$ becomes above Fermi level $\epsilon_{F}$. These critical $U$ values are in general different. (c) shows magnetization $n_{\uparrow}-n_{\downarrow}$ is overestimated
in MF solution compared to the accurate VUMPS result. }
\label{fig:analyse_MF}
\end{figure}

Here we show that the value of winding number defined from the many-body scattering state
wavefunction is different from that of MF Hamiltonian band topology.
We denote $U_{c}$ as transition point of the winding number, and
denote $U_{c_{1}}$ as another transition point where the renormalized
Zeeman energy $\tilde{\delta}$ exceeds $2t_{0}$. Fig.~\ref{fig:analyse_MF}(a)-(b)
shows that $U_{c}$ and $U_{c_{1}}$ are in general different. We also show that typically the MF bands are deformed to satisfy two-fermi-points condition in Eq. (\ref{eq:two_fermi_point_condition}) before the topological transition occurs. To see this we also
denote $U_{c_{2}}$ as the critical value at which the MF lower subband
energy $\epsilon_{k=0}^{-}$ becomes higher than $\epsilon_{F}$ (thus Eq. (\ref{eq:two_fermi_point_condition}) is satisfied by MF Hamiltonian).
In Fig.~\ref{fig:analyse_MF}(b) one can see $U_{c_{2}}$ is typically not
larger than $U_{c_{1}},U_{c}$. Thus when we analysis transition of winding number, we can focus on the two Fermi points case where Eq. (\ref{eq:two_fermi_point_condition}) is satisfied .

\begin{figure}[t]
\centering
\includegraphics[width=\columnwidth]{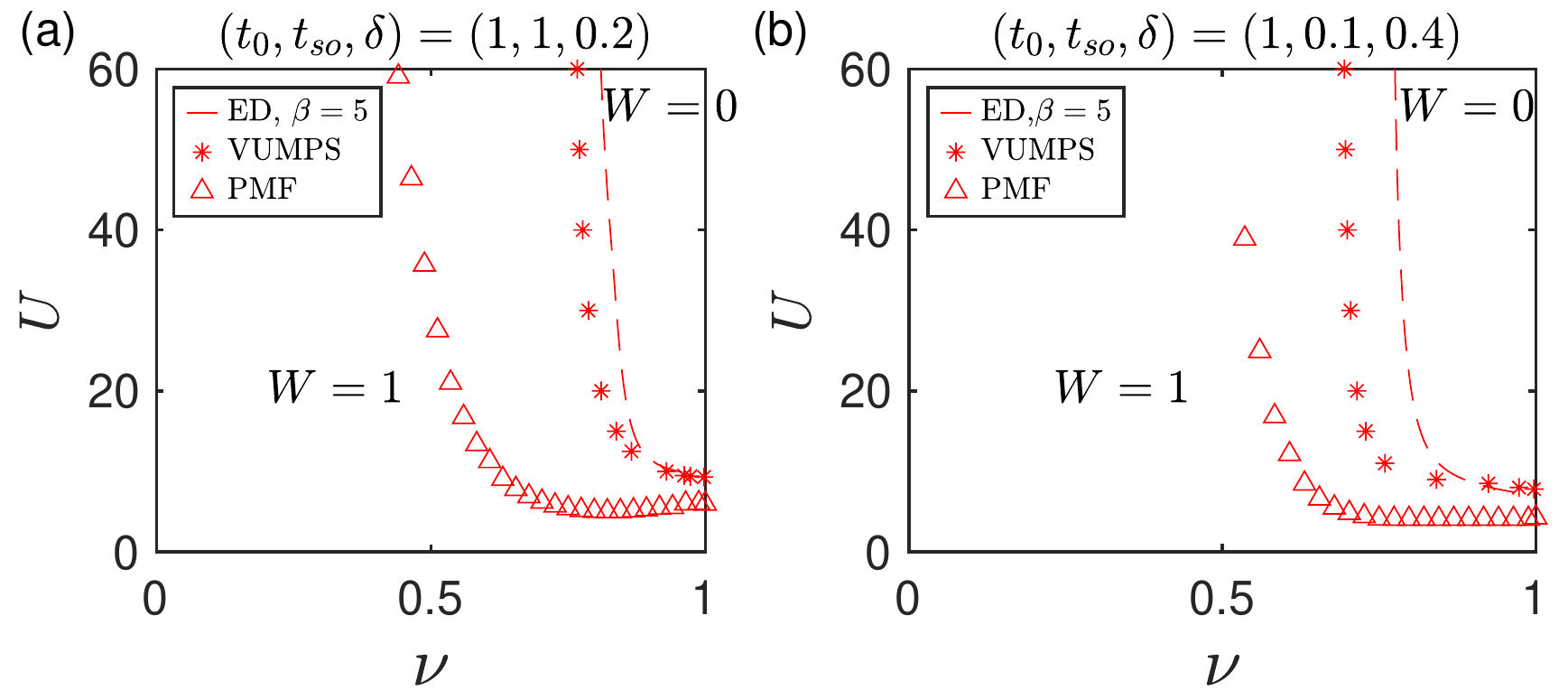}
\caption{Ground state topological phase diagrams at parameter regimes where transitions occur at relatively larger $U$. Parameters are chosen as $(t_{0},t_{\rm so},\delta)=(1,1,0.2)$
in (a) and $(t_{0},t_{\rm so},\delta)=(1,0.1,0.4)$ in (b). The ground state phase boundaries are obtained from VUMPS
and PMF wavefunction. The finite but very low temperature $\beta=5$ phase boundaries are obtained from ED and agrees with VUMPS results well.}
\label{fig:phase_diagram_large_Uc}
\end{figure}

\begin{figure}[t]
\centering
\includegraphics[width=\columnwidth]{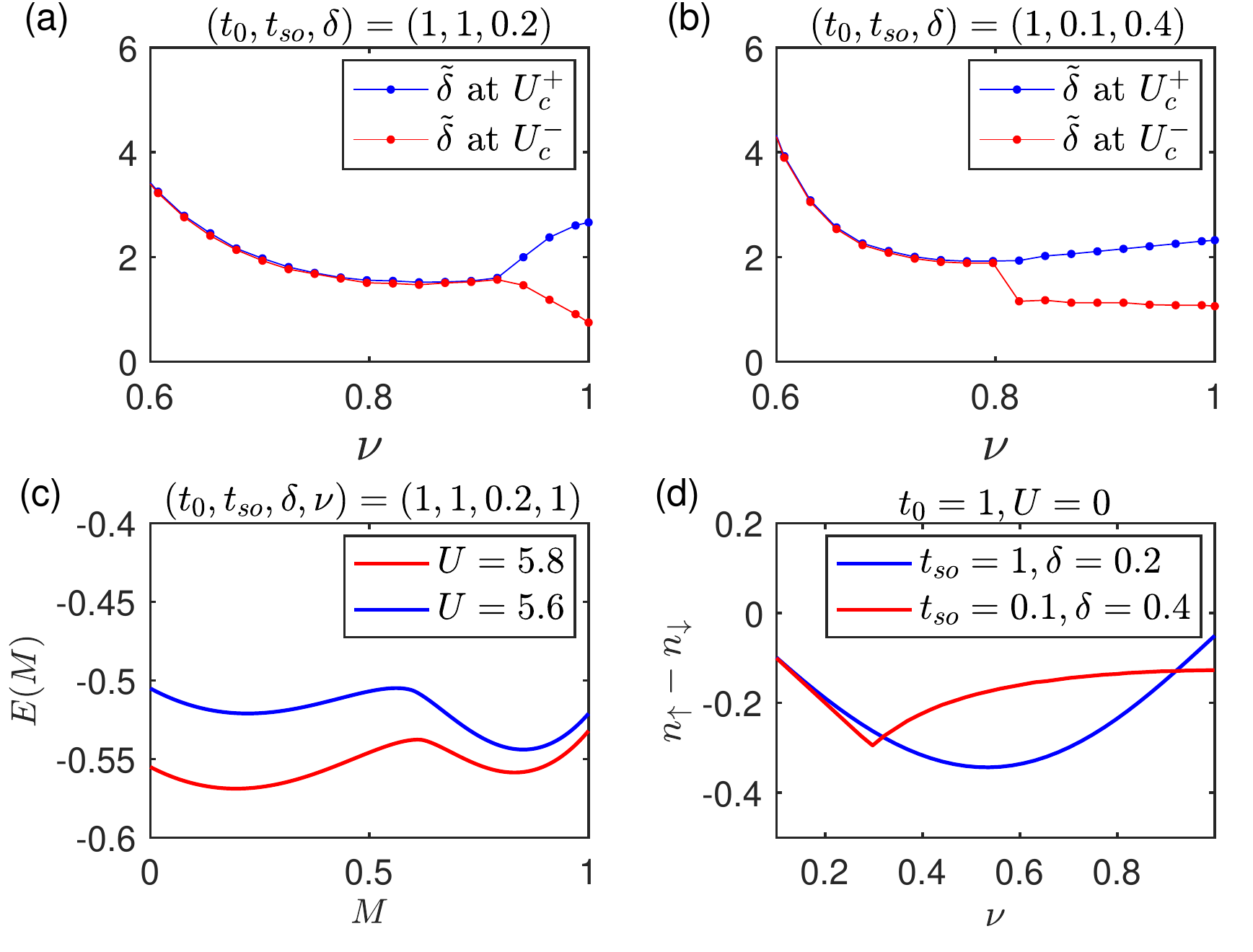}
\caption{Detailed analysis of PMF solutions for ground states. (a)-(b), The renormalized
Zeeman energy $\tilde{\delta}$ in MF solution at critical value $U_{c}^{\pm}=U_{c}+0^{\pm}$,
where $U_{c}$ denotes transition point of winding number $W$ in PMF solution. At large $\nu$, the different quantities of $\tilde{\delta}$ at $U_{c}^{+}$ and $U_{c}^{-}$ indicates that in the MF solution
$\tilde{\delta}$ has an artificial discontinuity at $U_{c}$. (c), The MF ground state energy $E(M)$ may have two local minima, leading
to an artificial discontinuity of $\tilde{\delta}$. (d), With increase of $\nu$, the magnetization of noninteracting ground state may decrease quickly or slowly, depending on the band structure.}
\label{fig:analyse_MF_2}
\end{figure}

Although the PMF treatment provides qualitative understanding of the
winding number and its transition, we should clarify several weaknesses
of the analytical PMF treatment. (i) Such a finite order $U$ expansion is
incorrect at large $U$ limit by definition. From the correct results given by
$t_{\rm so}$ expansion at infinite $U$ and numerical VUMPS results in
Fig.~\ref{fig:infiniteU_texture} and \ref{fig:phase_diagram_infiniteU},
one can see that at small $\nu$ the winding $W=1$ is unchanged by
infinite large $U$. However, in the large $U$ limit the PMF method gives incorrect result that the winding at any $\nu$ will become trivial, as long as $U$ and thus
$\tilde{\delta}=\delta-Um/2$ are large enough as seen in Fig.~\ref{fig:analyse_MF}(a)-(b).
(ii) The transition point $U_{c}$ for fixed $\nu$
is usually overestimated by PMF method at large $\nu$, especially for the parameter regimes
where $U_{c}$ is large as shown in Fig.~\ref{fig:phase_diagram_large_Uc}.
This can be understood from the fact that the magnetization  is overestimated
in MF solution and the error of magnetization increases with $U$
as shown in Fig.~\ref{fig:analyse_MF}(c). (iii) When $\nu$ is
close to $1$, the MF solution of $m$ and $\tilde{\delta}=\delta-Um/2$
has a discontinuity along $U$, as can be seen in Fig.~\ref{fig:sz_mz_transition}(b)
and Fig.~\ref{fig:analyse_MF_2}(a)-(b). This discontinuity originates
from the existence of more than one local minima in the MF ground
state energy $E(M)$ in Eq. (\ref{eq:two_fermi_point_condition})
as shown by Fig.~\ref{fig:analyse_MF_2} (c). In contrast to MF
solution, the accurate VUMPS results in Fig.~\ref{fig:sz_mz_transition}(b)
exhibits continuous magnetization, suggesting that the discontinuity
of MF solution is artificial.

Finally we discuss the monotonicity of $U_{c}$ as a function of $\nu$
given by PMF method. Naively, within MF picture one expects that
increasing $\nu$ would enhance both interacting effect and total
magnetization $m$, thus $U_{c}$ would monotonically decreases. However,
$U_{c}$ obtained from PMF may increase with $\nu$ as shown in Fig.~\ref{fig:finiteU_phase_diagram}(a),
\ref{fig:phase_diagram_large_Uc}(a). This can be explained by the
noninteracting magnetization shown in Fig.~\ref{fig:analyse_MF_2}(d).
At large $\nu$ the magnetization of noninteracting ground state decreases
quickly in $(t_{\rm so},\delta)=(1,0.2)$ case and slowly in $(t_{\rm so},\delta)=(0.1,0.4)$
case. As a result, when $\nu$ is increased, stronger $U$ is required
to enhance the magnetization in the $(t_{\rm so},\delta)=(1,0.2)$ case.

\subsection{Details of $t_{\rm so}$ expansion at infinite $U$ limit}

\subsubsection{Derivations}
Before entering the tedious calculations of this method, we introduce the basic ideas. We first start from a spin polarized product state $v_0$ which is exactly the ground state at $\delta\neq 0$, $t_{so}=0$ and infinite $U$. Then particle-hole excitations resulting from SOC are introduced to state $v_0$ according to variational principle, which creates a variational many-body wavefunction $\psi_g$. The spin textures at infinite $U$ are calculated using $\psi_g$.

Our purpose is to treat $t_{so}$ as small perturbation and expand
$\langle s_{k_{0}}^{y/z}\rangle$ to lowest order of $t_{so}$. The
infinite $U$ Hamiltonian projected by Gutzwiller projector $P_{G}$
has the form
\begin{align}
\tilde{H} & =P_{G}HP_{G}\notag\\
 & =\tilde{H}_{t_{0}}+\tilde{H}_{\delta}+\tilde{H}_{t_{so}},
\end{align}
where $P_{G}$ is defined as
\begin{equation}
P_{G}=\prod_{i}(1-n_{i,\uparrow}n_{i,\downarrow}).
\end{equation}
The Hubbard interaction vanishes while terms in $H_{0}$ becomes
\begin{align}
\tilde{H}_{t_{0}}= & P_{G}H_{t_{0}}P_{G}\notag\\
= & \{-t_{0}\sum_{i}\sum_{\sigma}\sigma((1-n_{i,\bar{\sigma}})c_{i,\sigma}^{\dagger}c_{i+1,\sigma}(1-n_{i+1,\bar{\sigma}})\notag\\
 & +(1-n_{i+1,\bar{\sigma}})c_{i+1,\sigma}^{\dagger}c_{i,\sigma}(1-n_{i,\bar{\sigma}}))\}P_{G},\notag\\
\tilde{H}_{t_{so}}= & P_{G}H_{t_{so}}P_{G}\notag\\
= & \{t_{so}[\sum_{i}((1-n_{i,\downarrow})c_{i,\uparrow}^{\dagger}c_{i+1,\downarrow}(1-n_{i+1,\uparrow})\notag\\
 & -(1-n_{i+1,\downarrow})c_{i+1,\uparrow}^{\dagger}c_{i,\downarrow}(1-n_{i,\uparrow}))+h.c.]\}P_{G},\notag\\
\tilde{H}_{\delta}= & P_{G}H_{\delta}P_{G}\notag\\
= & \delta\sum_{i}(n_{i,\uparrow}-n_{i,\downarrow})\}P_{G},
\end{align}
here in the second line the spin symbol $\sigma=\pm1$ denotes a coefficient when the corresponding spin is up/down. The relevant vectors in the subspace $\text{span}(\{v_{0},\tilde{H}v_{0},(\tilde{H})^{2}v_{0}\})$ include $v_{0},v_{0}^{1},v_{1}^{1,2},v_{1}^{3},v_{1}^{4}$, which are
obtained from
\begin{align}
\tilde{H}_{t_{0}}v_{0}= & E_{kin}v_{0},\notag \\
\tilde{H}_{\delta}v_{0}= & -N\delta v_{0},\notag \\
\tilde{H}_{t_{so}}v_{0}= & t_{so}v_{0}^{1},\notag \\
\tilde{H}_{t_{0}}v_{0}^{1}= & t_{0}v_{1}^{1,2}+E_{kin}v_{0}^{1},\notag \\
\tilde{H}_{t_{so}}v_{0}^{1}= & t_{so}v_{1}^{3}+t_{so}v_{1}^{4}+O(L)t_{so}v_{0},\notag \\
\tilde{H}_{\delta}v_{0}^{1}= & \delta(2-N)v_{0}^{1}.
\end{align}
Here $E_{kin}$ is the kinetic energy of state $v_{0}$ defined by $E_{kin}=\langle v_{0}|\tilde{H}_{t_{0}}|v_{0}\rangle$.
Explicit forms of these vectors are
\begin{align}
v_{0}^{1}= & [\sum_{i}(c_{i,\downarrow}c_{i,\downarrow}^{\dagger}c_{i,\uparrow}^{\dagger}c_{i+1,\downarrow}\notag\\
 & -c_{i+1,\downarrow}c_{i+1,\downarrow}^{\dagger}c_{i+1,\uparrow}^{\dagger}c_{i,\downarrow})]v_{0},\end{align}
\begin{align}
v_{1}^{1,2}= & [\sum_{i}(c_{i-1,\downarrow}^{\dagger}c_{i+1,\downarrow}-c_{i+1,\downarrow}^{\dagger}c_{i-1,\downarrow})c_{i,\uparrow}^{\dagger}c_{i,\downarrow}\notag\\
 & +\sum_{i}(n_{i+1,\downarrow}-n_{i-1,\downarrow})c_{i,\uparrow}^{\dagger}c_{i,\downarrow}]v_{0}\notag\\
 & -\sum_{i}(1-n_{i-1,\downarrow})(2-n_{i,\downarrow})c_{i-1,\uparrow}^{\dagger}c_{i+1,\downarrow}v_{0}\notag\\
 & +\sum_{i}(1-n_{i+1,\downarrow})(2-n_{i,\downarrow})c_{i+1,\uparrow}^{\dagger}c_{i-1,\downarrow}v_{0}\notag\\
 & +\sum_{i}(1-n_{i,\downarrow})(n_{i-1,\downarrow}-n_{i+1,\downarrow})c_{i,\uparrow}^{\dagger}c_{i,\downarrow}v_{0},
\end{align}
\begin{align}
v_{1}^{3}= & [\sum_{i}((1-n_{i,\downarrow})c_{i,\uparrow}^{\dagger}c_{i+1,\downarrow}\notag\\
 & -(1-n_{i+1,\downarrow})c_{i+1,\uparrow}^{\dagger}c_{i,\downarrow})]v_{0}^{1},\notag\\
\end{align}
\begin{align}
v_{1}^{4}=- & \sum_{i}(n_{i+1,\downarrow}+n_{i-1,\downarrow})c_{i,\downarrow}^{\dagger}c_{i,\downarrow}v_{0}\notag\\
+ & \sum_{i}n_{i,\downarrow}(c_{i+1,\downarrow}^{\dagger}c_{i-1,\downarrow}+h.c.)v_{0}.
\end{align}
The meaning of these vectors are quite clear: the action of (Gutzwiller projected) spin flip term $\tilde{H}_{t_{so}}$ on the spin polarized Fermi sea $v_{0}$ defines the state $v_{0}^{1}$ where a spin is flipped. The action of $\tilde{H}_{t_{0}}$ on $v_{0}^{1}$ creates the state $v_{1}^{1,2}$ where the spin up and down particles in $v_{0}^{1}$ moves in the lattice. The action of $\tilde{H}_{t_{so}}$ on $v_{0}^{1}$ creates state $v_{1}^{3}$ if the previously flipped spin is flipped back, and creates state $v_{1}^{4}$ if one more spin is flipped. The quantum numbers of these vectors and the relevant overlaps versus lattice size $L$ are listed in the table \ref{table:scaling}.
\begin{table}
\begin{tabular}{|c|c|c|c|c|c|}
\hline
 & $v_{0}$ & $v_{0}^{1}$ & $v_{1}^{1,2}$ & $v_{1}^{3}$ & $v_{1}^{4}$\tabularnewline
\hline
$N_{\uparrow}$ & 0 & 1 & 1 & 2 & 0\tabularnewline
\hline
$N_{\downarrow}$ & N & N-1 & N-1 & N-2 & N\tabularnewline
\hline
Norm versus $L$ & 1 & $O(L^{1/2})$ & $O(L^{1/2})$ & $O(L^{1})$ & $O(L^{1})$\tabularnewline
\hline
$\langle s_{k_{0}}^{y/z}\rangle$ versus $L$ & $O(L^{0})$ & $O(L^{0})$ & $O(L^{0})$ & $O(L^{1})$ & $O(L^{1})$\tabularnewline
\hline
\end{tabular}\caption{\label{table:scaling}}
Informations about the vectors appear in $t_{so}$ expansion. Quantum numbers $N_{\uparrow},N_{\downarrow}$ denotes number of spin up, down particles in each state in first row. $N$ denotes the total particle number. The forth and fifth rows correspond to dependences of the norms and expectation values on lattice size $L$.
\end{table}
We take the approximation $N_{\uparrow}\le1$ due to weak $t_{so}$, thus $v_{1}^{3}$ is
ignored in this calculation.

Note that the above vectors are not orthonormal, and starting from these vectors we obtain a set of orthonormalized vectors $v_{0},v_{1},v_{2}$ defined as
\begin{align}
v_{1}= & v_{0}^{1}/|v_{0}^{1}|,\notag\\
\tilde{v}_{2}= & (v_{1}^{1,2}-v_{1}\langle v_{1}|v_{1}^{1,2}\rangle)\notag\\
 & +\frac{t_{so}}{t_{0}}(v_{1}^{4}-v_{0}\langle v_{0}|v_{1}^{4}\rangle),\notag\\
v_{2}= & \tilde{v}_{2}/|\tilde{v}_{2}|.
\label{eq:v1_v2_def}
\end{align}
It can be shown at small filling factor $\nu$, for $k_{0}\approx0$ the
expectation values $\langle v_{0}^{1}|n_{k_{0},\uparrow}|v_{0}^{1}\rangle$
and $\langle v_{0}^{1}|n_{k_{0},\downarrow}|v_{0}^{1}\rangle$ cancels
exactly and $\langle v_{0}^{1}|s_{k_{0}}^{z}|v_{0}^{1}\rangle=0$,
thus contribution from the higher order vector $v_{1}^{1,2}$ to expectation value $\langle s_{k_{0}}^{z}\rangle$ should be included. Since we start from state $v_{0}$, the principle of perturbation requires that $v_{0}$ vector dominates in the variational wavefunction and thus $t_{so}L^{0.5}\ll\min(t_{0},\delta)$ should be satisfied (this will be seen from combination of table \ref{table:scaling} and Eq. (\ref{eq:psi_g_appendix}) below, requiring $|c_1 v_1|\ll |v_0|$). Thus the $v_{1}^{4}$ contribution to $\langle s_{k_{0}}^{z}\rangle$ and
norms of these vector can be ignored.
On the other hand, in expression of $\langle s_{k_{0}}^{y}\rangle$
outside Fermi sea, $v_{1}^{4}$ contributes to lowest order $t_{so}$
expansion and should be retained when calculating $\langle s_{k_{0}}^{y}\rangle$.
We clarify that the condition $t_{so}L^{0.5}\ll\min(t_{0},\delta)$, which requires $t_{so}\ll L^{-0.5}$ becomes infinitely small in thermodynamic limit, is a weakness of
such perturbation treatment in many-body problems. Although the standard perturbation can deal with finitely small $t_{so}$, it is difficult to perform standard perturbation calculations due to complexity of Bethe-Ansatz wavefunction of standard Hubbard model. Nevertheless, from this analytical $t_{so}$ expansion method the existence of nontrivial many-body spin textures at weak $t_{so}$ and infinite $U$ limit can be confirmed.

The Hamiltonian matrix elements written from orthonormalized vectors
$v_{0},v_{1},v_{2}$ has the form
\begin{equation}
H=\left[\begin{array}{ccc}
H_{00} & H_{01}\\
H_{10} & H_{11} & H_{12}\\
 & H_{21} & H_{22}
\end{array}\right].
\end{equation}
The matrix elements satisfy
\begin{align}
H_{11}-H_{00}= & \Delta_{1}+2\delta,\notag\\
H_{22}-H_{00}= & \Delta_{2}+2\delta,\notag\\
H_{01}= & 4t_{so}(k_{F}-\frac{\sin2k_{F}}{2})(1-\frac{k_{F}}{\pi})L/|v_{0}^{1}|>0,\notag\\
H_{12}= & \frac{t_{0}\sqrt{|v_{1}^{1,2}|^{2}-\langle v_{1}|v_{1}^{1,2}\rangle^{2}}}{|v_{0}^{1}|}=t_{0}O(L^{0})>0.
\end{align}
The variational ground state wavefunction $\psi_{g}$
is written by an expansion of $t_{so}$ as

\begin{align}
\psi_{g}\approx & v_{0}-c_{1}v_{1}+c_{1}c_{2}v_{2}.\notag\\
c_{1}= & \frac{H_{01}}{\Delta_{1}+2\delta},\notag\\
c_{2}= & \frac{H_{12}}{\Delta_{2}+2\delta}.
\label{eq:psi_g_appendix}
\end{align}
The kinetic energy differences $\Delta_{1},\Delta_{2}$ are non-negative.
We replace the energies of $v_{1},v_{2}$ by energy of $\sum_{k}\sin kc_{k,\uparrow}^{\dagger}c_{k,\downarrow}v_{0}$
as a simple approximation, i.e., use
\begin{equation}
\Delta_{1}\approx\Delta_{2}\approx\frac{8t_{0}\sin^{3}k_{F}}{3k_{F}}
\end{equation}
to evaluate spin textures from the expression of $\psi_{g}$. Below we show spin textures inside and outside Fermi sea, respectively.

\begin{figure}[t]
\centering
\includegraphics[width=\columnwidth]{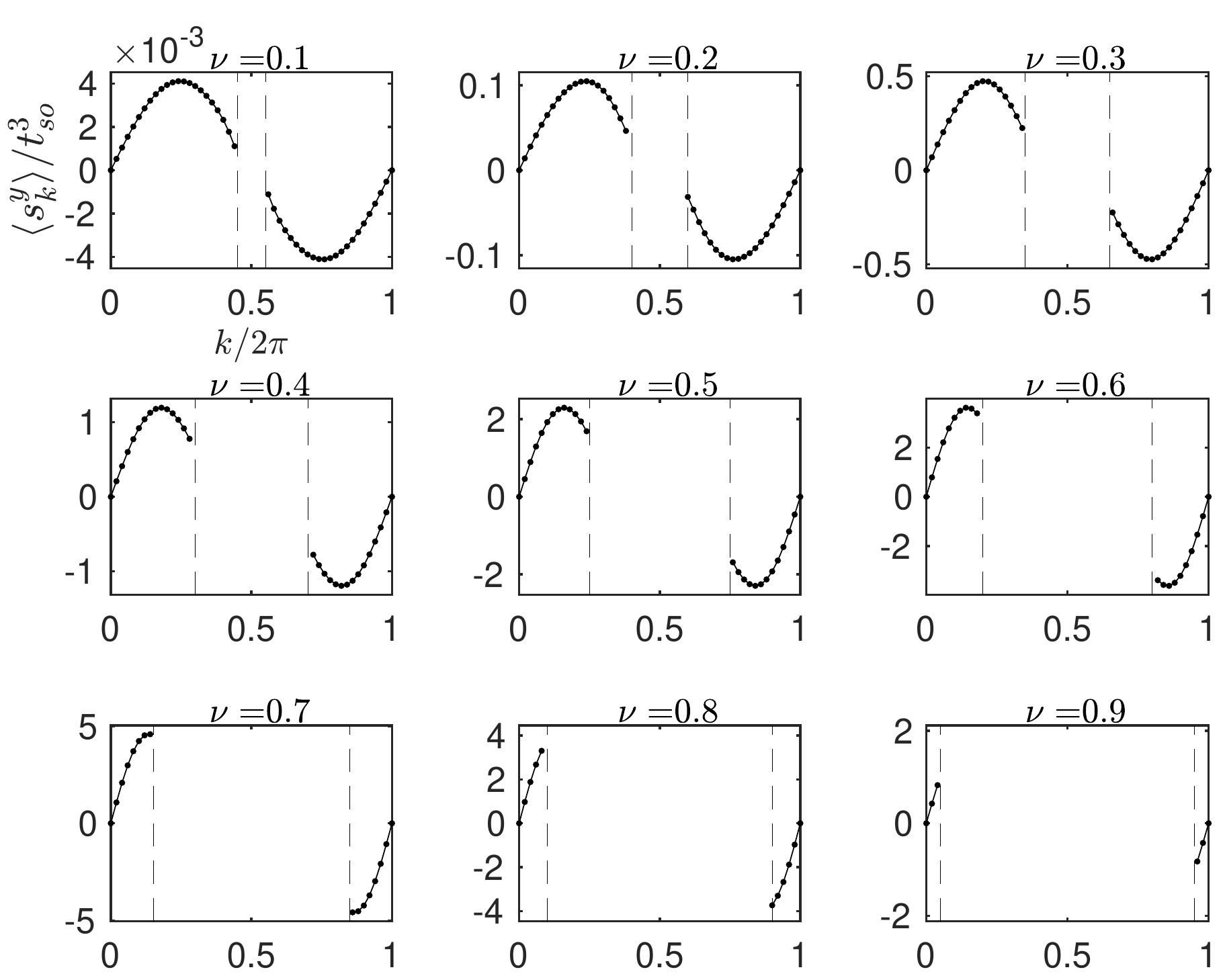}
\caption{Ground state spin textures $\langle s_{k}^{y}\rangle$ always have the same sign as the Bloch coefficient $h_{k}^{y}$ even at infinite $U$. This figure shows values of $\langle s_{k}^{y}\rangle$ for momentum points outside Fermi sea obtained from $t_{so}$ expansion ansatz in Eq. (\ref{eq:psi_g_appendix}) via numerical integration of Eq. (\ref{eq:sy_sz_appendix}), up
to lowest order $O(t_{so}^{3})$. Parameters are chosen as $(t_{0},\delta,U)=(1,1,+\infty)$.
The vertical dashed lines denote positions of Fermi momenta. The $\langle s_{k}^{y}\rangle$ for momentum points inside Fermi sea exhibits the same feature, and have simple leading order expression in Eq. (\ref{eq:inside_fermisea_appendix}).}
\label{fig:sy_largeU}
\end{figure}

\begin{figure}[t]
\centering
\includegraphics[width=\columnwidth]{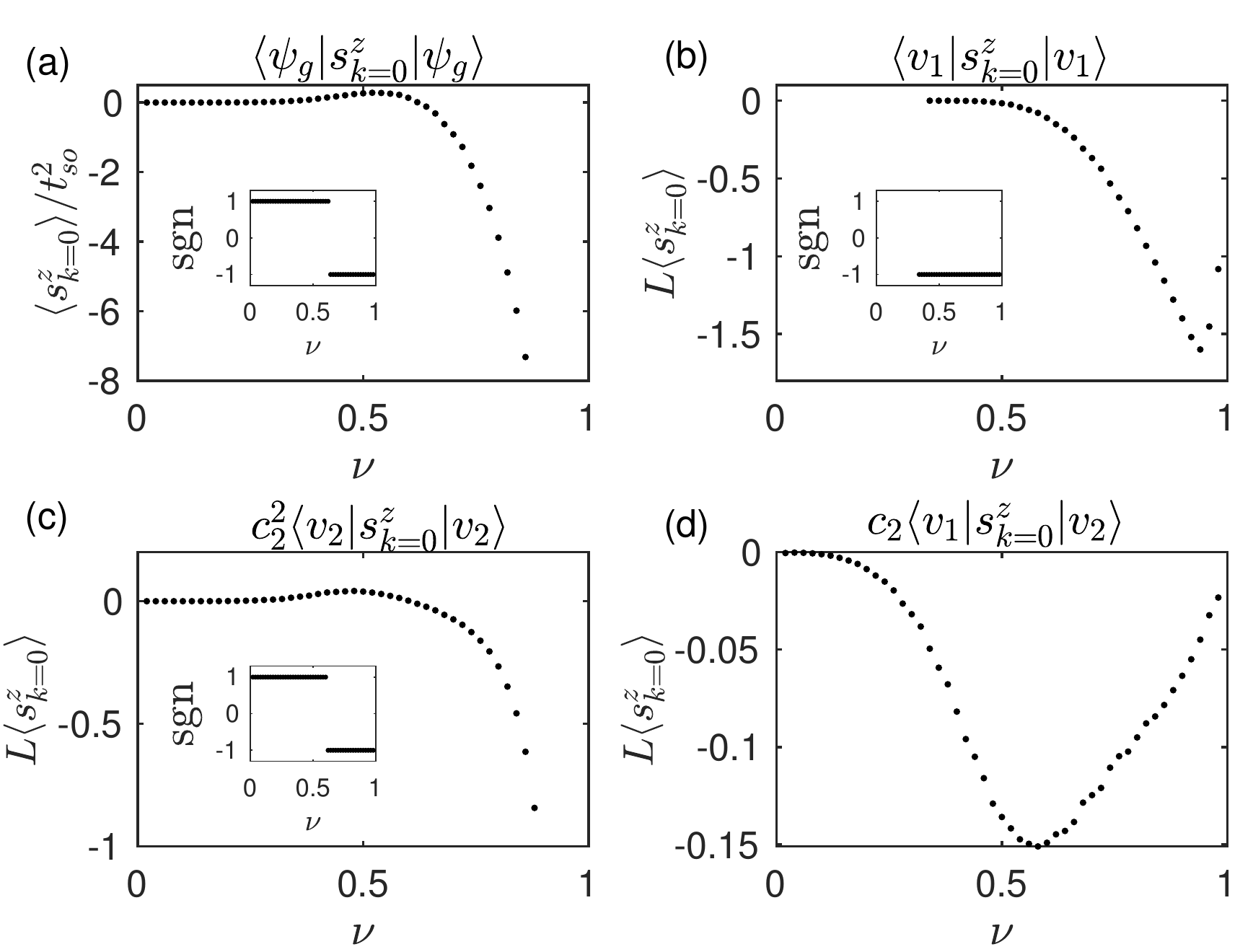}
\caption{Topological transition at infinite $U$ characterized by sign flip of $\langle s_{k=0}^{z}\rangle$, calculated from
variational ground state ansatz $\psi_{g}$ in Eq. (\ref{eq:psi_g_appendix})
and its components versus filling
$\nu$ using Eq. (\ref{eq:sy_sz_appendix}). Parameters are chosen as $(t_{0},\delta,U)=(1,1,+\infty)$. Here $L$ is lattice size. In (b), the value of $\langle s_{k=0}^{z}\rangle$ vanishes exactly for small $\nu$ and is not plotted. Therefore to obtain correct winding number, variational ansatz should be considered at least up to $v_2$.}
\label{fig:sz_largeU}
\end{figure}

For $k_{0}$ points inside Fermi sea, we obtain $\langle s_{k_{0}}^{y/z}\rangle$
at leading order of $t_{so}$:
\begin{align}
\langle s_{k_{0}}^{y}\rangle & \approx\frac{16\pi t_{so}}{\Delta_{1}+2\delta}(1-\nu)\sin k_{0},\notag\\
\langle s_{k_{0}}^{z}\rangle & \approx-1+O(t_{so}^{2}).
\label{eq:inside_fermisea_appendix}
\end{align}

For $k_{0}$ points outside Fermi sea, the expressions of $\langle s_{k_{0}}^{y/z}\rangle$
at leading order of $t_{so}$ are
\begin{align}
\langle\psi_{g}|s_{k_{0}}^{y}|\psi_{g}\rangle= & c_{1}^{2}c_{2}[-\langle v_{2}|ic_{k_{0}\downarrow}^{\dagger}c_{k_{0}\uparrow}|v_{1}\rangle \notag\\
+&c_{2}\langle v_{2}|ic_{k_{0}\downarrow}^{\dagger}c_{k_{0}\uparrow}|v_{2}\rangle]+h.c.,\notag\\
\langle\psi_{g}|s_{k_{0}}^{z}|\psi_{g}\rangle= & c_{1}^{2}[\langle v_{1}|s_{k_{0}}^{z}|v_{1}\rangle-c_{2}(\langle v_{2}|s_{k_{0}}^{z}|v_{1}\rangle+h.c.)\notag\\
 & +c_{2}^{2}\langle v_{2}|s_{k_{0}}^{z}|v_{2}\rangle].
\label{eq:sy_sz_appendix}
\end{align}
In above expressions, $c_{1},c_{2}$ coefficients, norm and expectation
values of vectors $v_{0}^{1},v_{1}^{1,2},v_{1}^{4}$ can only be integrated
in momentum space numerically in general. The patterns of $\langle s_{k_{0}}^{y}\rangle$
in above expression are shown in Fig.~\ref{fig:sy_largeU}. At
small $\nu$ limit, the expression of $\langle s_{k_{0}}^{y}\rangle$
outside Fermi sea can be simplified to an analytical expression
\begin{equation}
\langle s_{k_{0}}^{y}\rangle\approx\frac{64\pi^{4}t_{so}^{3}\nu^{4}}{3(\Delta_{1}+2\delta)^{3}}\sin k_{0}.
\end{equation}
The values of $\langle\psi_{g}|s_{k=0}^{z}|\psi_{g}\rangle$ and its
three components $\langle v_{1}|s_{k=0}^{z}|v_{1}\rangle$, $c_{2}\langle v_{2}|s_{k=0}^{z}|v_{1}\rangle$,
$c_{2}^{2}\langle v_{2}|s_{k=0}^{z}|v_{2}\rangle$ are shown in Fig.~\ref{fig:sz_largeU}.
These three terms are independent on $t_{so}$ since only leading order contributions from $t_{so}$ are considered and $\langle\psi_{g}|s_{k=0}^{z}|\psi_{g}\rangle\propto t_{so}^{2}$ due to the $c_{1}^{2}$ factor. The numerical integration results show that the signs of the expectation values as functions of filling factor $\nu$ are
\begin{align}
\mathrm{sgn}(\langle v_1|s_{k=0}^{z}|v_1\rangle) & =-\Theta(\nu-\nu_{v_1}) ,\notag\\
\mathrm{sgn}(\langle v_2|s_{k=0}^{z}|v_2\rangle) & =\Theta(\nu_{v_2}-\nu)-\Theta(\nu-\nu_{v_2})
\label{eq:v1_v2_expectation}
\end{align}
as shown in Fig.~\ref{fig:sz_largeU}. Here $\Theta(x)$ denotes the step function where $\Theta(x)=0$ for $x \le 0$ and $\Theta(x)=1$ for $x>0$. The changes of signs in above $\langle s_{k=0}^z\rangle$ expressions across the critical filling factors $\nu_{v_1},\nu_{v_2}$ result from the fact that the wavefunctions $v_1 , v_2$ depend on filling factor $\nu$. To be specific, for $v_{1}$ state one can see $\langle v_{1}|s_{k=0}^{z}|v_{1}\rangle$ exactly vanishes at small $\nu$ and is negative at large $\nu$; while for $v_{2}$ state one can see $\langle v_{2}|s_{k=0}^{z}|v_{2}\rangle$ is positive at small $\nu$ and negative at large $\nu$. The above signs in Eq. (\ref{eq:v1_v2_expectation})) do not depend on other Hamiltonian parameters in this leading order calculation, since the contributions of $v_1^4$ term in Eq. (\ref{eq:v1_v2_def})) is only at leading order for $s_k^z$ outside Fermi sea. Further numerical integration of Eq. (\ref{eq:psi_g_appendix}) at generic Hamiltonian parameters shows that spin texture $\langle\psi_{g}|s_{k=0}^{z}|\psi_{g}\rangle$ of the variational ground state $\psi_g$ has a sign flip as
\begin{align}
\mathrm{sgn}(\langle s_{k=0}^{z}\rangle)  =\Theta(\nu_{c}-\nu)-\Theta(\nu-\nu_{c}),
\end{align}
hence the winding number has a transition at critical $\nu_c$.

\subsection{Exact ground state spin textures in special parameter limits}
In order to understand the pattern of many-body spin textures better, here we consider two special limits where exact properties of observable many-body spin texture can be obtained. The first limit is the large $U$ limit at half filling where expression of spin textures at lowest $U^{-1}$ order can be expressed explicitly. The second limit is the flat band limit with $t_{0}=t_{\rm so},\delta=0$, where the exact spin direction in momentum space can be obtained. These two cases supports the fact that the common characteristics of spin textures is that the sign of $\langle s_{k}^{y}\rangle$ is always the same as the Bloch Hamiltonian coefficient $h_{k}^{y}$.

\subsubsection{The half filling and large $U$ limit with $0<\delta<2t_{0}$, }
We choose the on-site terms
\begin{equation}
\tilde{H}_{0}=\delta\sum_{i}(n_{i,\uparrow}-n_{i,\downarrow})+U\sum_{i}n_{i,\uparrow}n_{i,\downarrow},
\end{equation}
as unperturbed Hamiltonian, and choose the hopping term
\begin{align}
\tilde{H}_{1}=  \sum_{i} \psi_{i}^{\dagger}(it_{\rm so}\tau_{y}-t_{0}\tau_z) \psi_{i+1} +h.c.
\end{align}
as perturbation. Due to nonzero $\delta$, the unperturbed ground
state at half filling is non-degenerate and written as
\begin{align}
\psi^{(0)} & =\prod_{i}c_{i\downarrow}^{\dagger}|vac\rangle\notag\\
 & =\prod_{k}c_{k\downarrow}^{\dagger}|vac\rangle.
\end{align}
We consider perturbed wavefunction up to order of $U^{-1}$, i.e.,
the first order correction
\begin{equation}
\psi^{(1)}= \frac{t_{\rm so}}{-2\delta-U}\sum_{i}(c_{i,\uparrow}^{\dagger}c_{i+1,\downarrow}-c_{i+1,\uparrow}^{\dagger}c_{i,\downarrow})\psi^{(0)}.
\end{equation}
The expectation of hopping matrix $\langle c_{i,\sigma}^{\dagger}c_{j,\sigma'}\rangle$
up to order $U^{-1}$ are calculated with the first order perturbed
wave function:
\begin{align}
\langle c_{i,\uparrow}^{\dagger}c_{j,\downarrow}\rangle= & -\delta_{i-j,1}\frac{t_{\rm so}}{2\delta+U}+\delta_{i-j,-1}\frac{t_{\rm so}}{2\delta+U},\notag\\
\langle c_{i,\uparrow}^{\dagger}c_{j,\uparrow}\rangle= & O(\frac{1}{U^{2}})\delta_{i,j},\notag\\
\langle c_{i,\downarrow}^{\dagger}c_{i,\downarrow}\rangle= & 1-\delta_{i,j}O(\frac{1}{U^{2}}).
\end{align}
We then obtain the momentum space spin texture in large $U$ limit
\begin{align}
\langle s_{k}^{y}\rangle= & \frac{4t_{\rm so}}{U+2\delta}\sin k,\notag\\
\langle s_{k}^{z}\rangle= & -1+O(\frac{1}{U^{2}}).
\end{align}

\subsubsection{The Flat band $t_{0}=t_{\rm so},\delta=0$ case}

The Hamiltonian in $t_{0}=t_{\rm so},\delta=0$ case becomes
\begin{align}
H=&-2t_{0}\sum_{k}\psi_{k}^{\dagger}(\cos k\tau_{z}+\sin k\tau_{y})\psi_{k} \notag\\
&+U\sum_{i}n_{i,\uparrow}n_{i,\downarrow},
\end{align}
where the spin direction in momentum space is exactly known. The single
particle band is flat with the dispersion $\epsilon_{k}^{\pm}=\pm2t_{0}$.
For a lattice with size $L$, consider the following operation. First,
rotate all the spin in $y-z$ plane along $x$ axis by the angle $2\pi/L$,
denoted by
\begin{equation}
R=e^{i\frac{2\pi}{L}\sum_{i}s_{i}^{x}/2}=e^{i\frac{2\pi}{L}\sum_{k}s_{k}^{x}/2},
\end{equation}
the Hamiltonian is transformed to
\begin{align}
RHR^{-1}= & -2t_{0}\sum_{k}\psi_{k}^{\dagger}[\cos(k+\frac{2\pi}{L})\tau_{z} \\
&+\sin(k+\frac{2\pi}{L})\tau_{y}]\psi_k \notag\\
 & +U\sum_{i}n_{i,\uparrow}n_{i,\downarrow}.
\end{align}
Then we perform gauge transformation which shifts the momentum by
$\frac{2\pi}{L}$, denoted by
\begin{equation}
Uc_{i\sigma}^{\dagger}U^{-1}=e^{i\frac{2\pi i}{L}}c_{i\sigma}^{\dagger},
\end{equation}
the Hamiltonian then becomes
\begin{align}
URHR^{-1}U^{-1}= & -2t_{0}\sum_{k}\psi_{k}^{\dagger}(\cos k\tau_{z}+\sin k\tau_{y})\psi_{k} \notag\\
 &+ U\sum_{i}n_{i,\uparrow}n_{i,\downarrow}.
\end{align}
One can see that the Hamiltonian is invariant under $UR$ symmetry,
thus spin textures satisfy
\begin{align}
\langle\psi|s_{k}^{y}|\psi\rangle= & \langle\psi|R^{-1}U^{-1}URs_{k}^{y}R^{-1}U^{-1}UR|\psi\rangle\notag\\
= & \langle\psi|URs_{k}^{y}R^{-1}U^{-1}|\psi\rangle\notag\\
= & \langle\psi|U\cos\frac{2\pi}{L}s_{k}^{y}+\sin\frac{2\pi}{L}s_{k}^{z}U^{-1}|\psi\rangle\notag\\
= & \langle\psi|\cos\frac{2\pi}{L}s_{k+\frac{2\pi}{L}}^{y}+\sin\frac{2\pi}{L}s_{k+\frac{2\pi}{L}}^{z}|\psi\rangle,\notag\\
\langle\psi|s_{k}^{z}|\psi\rangle= & \langle\psi|\cos\frac{2\pi}{L}s_{k+\frac{2\pi}{L}}^{z}-\sin\frac{2\pi}{L}s_{k+\frac{2\pi}{L}}^{y}|\psi\rangle.
\end{align}
In above calculation if the ground state is degenerate, the expectation
value should be defined from their average. From above expression
we immediately obtain
\begin{align}
\theta_{k+\frac{2\pi}{L}}= & \theta_{k}+\frac{1}{L},
\end{align}
and spin length $S_{k}$ is constant in FBZ. The spin direction $\theta_{k}$
at $k=0$ can only be $0$ or $1$ due to parity symmetry. One naively
expect that $\theta_{k=0}=0$ has lower single-particle energy. Numerical
calculations shows $\theta_{k=0}=0$, thus the spin texture in FBZ
is
\begin{equation}
\theta_{k}=\frac{k}{2\pi}.
\end{equation}
This $UR$ symmetry ensures that winding number of spin texture is
always $1$ no matter how large $U$ is. For the more general $t_{0}\neq t_{\rm so},\delta=0$
case, although numerical results suggests winding number is also always
$1$, we have no proof since the $UR$ symmetry used above is no more
satisfied.

\subsection{Spin textures from high temperature expansion}
Here we present detailed high temperature expansion calculations. We first present lowest order analytical expressions for spin textures and detailed analysis of spin texture patterns, and then provide derivations of winding number transition points in four special limits.

In the HTE method the Hamiltonian is separated as $H=  \tilde{H}_{0}+\tilde{H}_{1}$ where
\begin{align}
\tilde{H}_{0}= & U\sum_{i}n_{i\uparrow}n_{i\downarrow}+\delta\sum_{i}(n_{i,\uparrow}-n_{i,\downarrow})-\mu\sum_{i\sigma}n_{i\sigma},\notag\\
\tilde{H}_{1}= & \sum_{i} \psi_{i}^{\dagger}(it_{\rm so}\tau_{y}-t_{0}\tau_z) \psi_{i+1} +h.c.,
\end{align}
here the on-site terms are taken as unperturbed part and hoppings are taken as perturbation.
In HTE method, the partition function of $\tilde{H}_{0}$ for each site is independent, and they
are connected by perturbation term $\tilde{H}_{1}$. Since the expansion
parameter is $t_{0}\beta$ and $t_{\rm so}\beta$, finite order expansion
is a good approximation at high temperature. We define spin-dependent
chemical potential $\mu_{\sigma}=\mu-\sigma\delta$ and single site
weights $x=e^{\beta\mu_{\uparrow}},y=e^{\beta\mu_{\downarrow}},d=e^{\beta(2\mu-U)}$
for various configurations. The single site probabilities take the
form
\begin{align}
\kappa_{e} & =\frac{1}{1+x+y+d},\notag\\
\kappa_{\uparrow} & =\frac{x}{1+x+y+d},\notag\\
\kappa_{\downarrow} & =\frac{y}{1+x+y+d},\notag\\
\kappa_{d} & =\frac{d}{1+x+y+d}.
\end{align}
We here show the spin texture at first order expansion:
\begin{align}
\langle s_{k}^{y}\rangle_{0th}= & 0,\notag \\
\langle s_{k}^{y}\rangle_{1st}= & 4t_{\rm so}\sin k\frac{1}{(1+x+y+d)^{2}}\notag \\
 \times &[\frac{e^{\beta\mu_{\bar{\sigma}}}-e^{\beta\mu_{\sigma}}}{\mu_{\bar{\sigma}}-\mu_{\sigma}}+\frac{e^{\beta(\mu_{\sigma}+2\mu_{\bar{\sigma}}-U)}-e^{\beta(2\mu_{\sigma}+\mu_{\bar{\sigma}}-U)}}{\mu_{\bar{\sigma}}-\mu_{\sigma}}\notag \\
  +&\frac{e^{2\beta\mu_{\bar{\sigma}}}-e^{\beta(\mu_{\sigma}+\mu_{\bar{\sigma}}-U)}}{\mu_{\bar{\sigma}}-\mu_{\sigma}+U}+\frac{e^{\beta(\mu_{\bar{\sigma}}+\mu_{\sigma}-U)}-e^{2\beta\mu_{\sigma}}}{\mu_{\bar{\sigma}}-\mu_{\sigma}-U}],\notag \\
\langle s_{k}^{z}\rangle_{0th}= & \kappa_{\uparrow}-\kappa_{\downarrow},\notag \\
\langle s_{k}^{z}\rangle_{1st}= & 2t_{0}\cos k \notag \\
\times&\frac{\beta(x+y)+4xy\frac{1-e^{-\beta U}}{U}+\beta e^{-\beta U}xy(x+y)}{(1+x+y+d)^{2}}.\label{eq:HTE_1st}
\end{align}
The momentum distribution $n_{k}=n_{k,\uparrow}+n_{k,\downarrow}$
expanded at first order is
\begin{align}
\langle n_{k,\sigma}\rangle_{0th}= & \kappa_{\sigma}+\kappa_{d},\notag\\
\langle n_{k,\sigma}\rangle_{1st}= & 2\sigma t_{0}\cos k \notag\\
\times &\frac{\beta e^{\beta\mu_{\sigma}}+2e^{2\beta\mu}\frac{1-e^{-\beta U}}{U}+\beta e^{\beta(2\mu-U)}e^{\beta\mu_{\bar{\sigma}}}}{(1+x+y+d)^{2}},
\end{align}
and the filling factor $\nu$ is determined by chemical potential $\mu$ via
\begin{equation}
\nu=  \kappa_{\uparrow}+\kappa_{\downarrow}+2\kappa_{d}.
\end{equation}
All four terms in bracket of $\langle s_{k}^{y}\rangle_{1st}$ are
positive and independent on momentum, thus only at $k=0,\pi$ the $\langle s_{k}^{y}\rangle_{1st}$
is zero. Therefore winding number at finite temperature is also determined
by sign of $\langle s_{k}^{z}\rangle$ at $k=0,\pi$:
\begin{align}
\langle s_{k=0}^{z}\rangle= & \kappa_{\uparrow}-\kappa_{\downarrow} \notag\\
+&2t_{0}\frac{\beta(x+y)+4xy\frac{1-e^{-\beta U}}{U}+\beta e^{-\beta U}xy(x+y)}{(1+x+y+d)^{2}},\notag\\
\langle s_{k=\pi}^{z}\rangle= & \kappa_{\uparrow}-\kappa_{\downarrow}\notag\\
-&2t_{0}\frac{\beta(x+y)+4xy\frac{1-e^{-\beta U}}{U}+\beta e^{-\beta U}xy(x+y)}{(1+x+y+d)^{2}}.
\end{align}
Note that $\langle s_{k=\pi}^{z}\rangle$ is always negative since
we have set $\delta>0$. The winding number is now determined by sign
of $\langle s_{k=0}^{z}\rangle$. The corrections of $t_{\rm so}$ on
$\langle s_{k}^{z}\rangle$ is at least at second order and is unimportant
at high temperature compared to $t_0$ contribution.

To understand winding number phase diagram better, below we evaluate
the transition points of winding number in four limiting $(U,\beta)$
parameter regimes. We set $t_{0}=1$ and restrict $t_{\rm so},\delta$ to be of order $O(1)$.

\subsubsection{Noninteracting case: $U=0$, $\beta\ll1$ }

The noninteracting spin texture can be obtained exactly from Eq. (\ref{eq: H0_diag}).
The spin direction at each momentum point is the same as that of lower
band, thus the transition point is exactly $\delta=2t_{0}$ independent
on $\beta$ and $\nu$. In HTE calculation, $d=xy$ due to $U=0$,
and the spin texture at first order expansion has the form

\begin{align}
\langle s_{k}^{z}\rangle= & \frac{x-y}{(1+x)(1+y)}\notag \\
+ & 2\beta t_{0}\cos k(\frac{x}{(1+x)^{2}}+\frac{y}{(1+y)^{2}}),\notag \\
\langle s_{k}^{y}\rangle= & 2t_{\rm so}\sin k\frac{y-x}{\delta}\frac{1}{(1+x)(1+y)}.\label{eq:HTE_U0}
\end{align}
This agrees with first order $\beta t_{0}$ and $\beta t_{\rm so}$ Taylor
expansion of exact expression of spin texture. The transition point
of winding number obtained from Eq. (\ref{eq:HTE_U0}) has $\beta$
dependence due to the fact that it is $\beta$'s first order approximation.
However, at high temperature $\beta\delta\ll1,\beta t_{0}\ll1$ regime,
one can expand $e^{\pm\beta\delta}$ as $1\pm\beta\delta$ in Eq.
(\ref{eq:HTE_U0}). Let $\langle s_{k=0}^{z}\rangle=0$ using $\langle s_{k=0}^{z}\rangle$
in Eq. (\ref{eq:HTE_U0}), one can obtain the correct condition
\begin{equation}
2t_{0}=\delta
\end{equation}
for transition of winding number. This simple case illustrates validity of the HTE method at high temperature. Note that in above calculation $e^{\beta\mu}$
should not be approximated by Taylor expansion: when $\beta\rightarrow0$,
$\mu$ should tends to infinity to ensure the filling factor
\begin{equation}
\nu=\frac{2e^{\beta\mu}}{(1+e^{\beta\mu})^{2}}
\end{equation}
unchanged by $\beta$.
We also have checked that expression of spin texture obtained from HTE up to third order is the same as that from Taylor expansion of exact result when $U=0$.

\subsubsection{Infinite temperature limit: $\beta\ll\frac{1}{U}\ll1$}

In this case, temperature is much larger than any other parameters
including $U$. Since $\beta\delta,\beta U$ are small quantities,
terms in Eq. (\ref{eq:HTE_1st}) should be expanded as
\begin{align}
e^{\beta\delta} & \approx1+\beta\delta+(\beta\delta)^{2}/2,\notag\\
e^{-\beta U} & \approx1-\beta U+(\beta U)^{2}/2,\notag\\
\frac{1-e^{-\beta U}}{U} & \approx\beta-\beta^{2}U/2.
\end{align}
Critical condition $\langle s_{k=0}^{z}\rangle=0$ expanded
at first order of $\beta$ becomes

\begin{align}
2t_{0}\frac{\beta e^{\beta\mu}+2e^{2\beta\mu}\beta+\beta e^{3\beta\mu}}{(1+2e^{\beta\mu}+e^{2\beta\mu})^{2}}-\frac{\beta\delta e^{\beta\mu}}{(1+2e^{\beta\mu}+e^{2\beta\mu})} & =0,
\end{align}
which gives
\begin{equation}
2t_{0}=\delta.
\end{equation}
From this formula one can know if temperature $T=1/\beta$ is much
larger than any other parameters, at $0<\delta<2t_{0}$ regime the
winding number will eventually becomes $1$ as if there is no interaction.

\subsubsection{Infinite interaction limit: $\frac{1}{U}\ll\beta,\frac{1}{U}\ll1$}

In this case, $U$ is much larger than any other parameters including
high temperature $T=1/\beta$ as the case studied in Ref. \cite{PhysRevE.89.063301}.
Terms involving $U$ like $e^{-\beta U},\frac{1-e^{-\beta U}}{U},\frac{e^{2\beta\mu_{\bar{\sigma}}}-e^{\beta(\mu_{\sigma}+\mu_{\bar{\sigma}}-U)}}{\mu_{\bar{\sigma}}-\mu_{\sigma}+U}$
in Eq. (\ref{eq:HTE_1st}) can be discarded directly (this is equivalent
to considering Gutzwiller projected Hamiltonian). Spin textures and
filling factor become
\begin{align}
\langle s_{k}^{y}\rangle= & 4t_{\rm so}\sin k\frac{1}{(1+x+y)^{2}}\frac{e^{\beta\mu_{\bar{\sigma}}}-e^{\beta\mu_{\sigma}}}{\mu_{\bar{\sigma}}-\mu_{\sigma}},\notag\\
\langle s_{k}^{z}\rangle= & \frac{x-y}{1+x+y}+2t_{0}\cos k\frac{\beta(x+y)}{(1+x+y)^{2}},\notag\\
\nu= & \frac{x+y}{1+x+y}.
\end{align}
At high temperature $\frac{1}{U}\ll\beta\ll1$ limit, using $e^{\pm\beta\delta}\approx1\pm\beta\delta$
in expression of $x,y$, following the preceding calculation, filling
$\nu$ becomes
\begin{equation}
\nu=\frac{2e^{\beta\mu}}{1+2e^{\beta\mu}},
\end{equation}
and the critical condition condition $\langle s_{k=0}^{z}\rangle=0$
at first order of $\beta$ becomes
\begin{equation}
2t_{0}(1-\nu)=\delta.
\end{equation}
At intermediate temperature, we show $\beta$ and $t_{\rm so},\delta$
dependence of $\nu_{c}$ in Fig.~\ref{fig:phase_diagram_infiniteU}
through numerical calculation of third order HTE. In subfigure (a) it
is shown that $\nu_{c}$ has no evident $t_{\rm so}$ dependence. In subfigure (b)
one can see $\nu_{c}$ is a monotonically decreasing function of $\delta$, while whether
$\nu_{c}$ increases or decreases with increasing $\beta$ depends
on magnitude of $\delta$.

\subsubsection{Half filling case: $\nu=1$, $\beta\ll1$ }

The half filling condition $\nu=1$ implies particle-hole symmetry
$U=2\mu$, such that $d=e^{\beta(2\mu-U)}=1$ and
\begin{equation}
\nu=\frac{x+y+2d}{1+x+y+d}=1.
\end{equation}
In this case, through relation $U=2\mu$ and $e^{\pm\beta\delta}\approx1\pm\beta\delta$
at high temperature one can expand critical condition $\langle s_{k=0}^{z}\rangle=0$
and simplify it to the form
\begin{equation}
2t_{0}[\beta U+e^{\beta U/2}(1-e^{-\beta U})]-\delta(\beta U)(1+e^{\beta U/2})=0.
\end{equation}
One can easily find if $t_{0}/\delta$ is fixed, the solution of $\beta U$
in above equation is also fixed. Therefore in this case with fixed
$t_{0}/\delta$ ratio the critical $U_{c}$ satisfies
\begin{equation}
U_{c}\propto\frac{1}{\beta}=T.
\end{equation}
Note that in the derivation $e^{\beta U}$ should not be expanded since $U=2\mu$ and $\beta U$ is not a small quantity.

\subsection{The conventional gapped SPT phase protected by parity symmetry at $\nu=1$}
Here we investigate the conventional gapped SPT phase protected by parity symmetry $P$ at half filling. We first derive a topological invariant from the unique gapped ground state condition and then illustrate the mechanism of protected entanglement spectrum degeneracy. The topological invariant introduced below are used to determine gap closing point in the numerical simulations.

The basic idea is to transform the fermionic Hamiltonian to a spin model via Jordan-Wigner transformation, and then derive the SPT topological invariant using MPS (VUMPS) formalisms following Refs. \cite{perez2008string,pollmann2010entanglement}, with the Jordan-Wigner transformed parity symmetry $P$ introduced in Appendix A. Recall that $P$ has the form
\begin{align}
P & =\tilde{I}(\otimes_{i}u_{i}),
\end{align}
where $\tilde{I}$ is the bond-centered inversion in Jordan-Wigner
representation, $u_i=\text{diag}([1,1,-1,1])$ is a on-site unitary acting
on the local bases $[|0\rangle,|\uparrow\rangle,|\downarrow\rangle,|\uparrow\downarrow\rangle]$.

We denote $A_{\alpha,\beta}^{s}$ as tensor of translation invariant
injective MPS that approximates the unique gapped ground state well. We choose the gauge where $A_{\alpha,\beta}^{s}$ is
in left canonical form, and the transfer matrix
\begin{equation}
E_{(\alpha,\alpha'),(\beta,\beta')}(A,A)=\sum_{s}(A_{\alpha,\beta}^{s})^{*}A_{\alpha',\beta'}^{s}
\end{equation}
has positive definite diagonal matrix $R$ as right dominant eigenvector.
Define unitary $\tilde{L}$ as the left dominant eigenvector of transfer
matrix $E_{(\alpha,\beta'),(\beta,\alpha')}(A,R^{\frac{1}{2}}(\sum_{s'}u^{s,s'}A^{s'})^{T}R^{-\frac{1}{2}})$,
the symmetry $P$ of ground state requires the magnitude of dominant
eigenvalue to be $1$ and tensor $A$ is transformed under $P$ as
\begin{equation}
(\sum_{s'}u^{s,s'}A^{s'})^{T}=e^{i\phi}R^{-\frac{1}{2}}\tilde{L}^{-1}A^s\tilde{L}R^{\frac{1}{2}}.
\end{equation}
Through injectivity of MPS and commutation relation $[\tilde{L},R^{\frac{1}{2}}]=0$,
one obtains $\tilde{L}^{*}\tilde{L}=e^{i\Phi}$, where $\Phi$ can't
be gauged away. Hence $\tilde{L}^{T}=e^{-i\Phi}\tilde{L}$ and $\tilde{L}=e^{-2i\Phi}\tilde{L}$,
one find $\Phi=0$ or $\pi$. The nontrivial case $\Phi=\pi$ implies
$\tilde{L}$ is a skew-symmetric unitary matrix, and $[\tilde{L},R^{\frac{1}{2}}]=0$
implies eigenvalues of $R$ must be even-fold degenerate. Since in
such MPS gauge condition the bipartite entanglement spectrum of MPS corresponds
to diagonal elements of $R$, the bipartite entanglement spectrum of
MPS are even-fold degenerate. Note that $\Phi$ only takes discrete
values $0$ or $\pi$, continuous change of MPS wavefunction will
not change value of $\Phi$. The only possibility for change of $\Phi$
is closing of the gap, thus $\Phi$ can be taken as a $Z_2$ topological invariant. At $U=0$ and $0<\delta<2t_{0}$ we calculated
the $\tilde{L}$ matrix using VUMPS, and found $\tilde{L}^{*}\tilde{L}=-1$.
Since the property $\tilde{L}^{*}\tilde{L}=-1$ and even-fold degeneracy
of entanglement spectrum are protected by gap and $P$ symmetry, the
gapped interacting ground state which is smoothly connected to noninteracting
topological insulator can be classified as topological phase protected
by parity symmetry $P$.

\noindent

\end{document}